\documentclass[a4paper,12pt]{article}
\usepackage{jheppubm}

\usepackage[utf8]{inputenc}
\usepackage[english]{babel}
\usepackage{amsmath}
\usepackage{amsthm}
\usepackage{amsfonts}
\usepackage{amssymb}
\usepackage{hyperref}
\usepackage{textcomp}
\usepackage{bm}
\usepackage{amsbsy}
\usepackage{graphicx}
\usepackage{mathrsfs}
\usepackage{epsfig}
\usepackage{color}
\usepackage{euscript}
 \usepackage{multirow}
 \usepackage{ulem}

\newcommand{\nn}{\nonumber}
\newcommand{\be}{\begin{equation}}
\newcommand{\ee}{\end{equation}}
\newcommand{\bea}{\begin{eqnarray}}
\newcommand{\eea}{\end{eqnarray}}

\newcommand{\fb}{\mathfrak{b}}

\newcommand{\fg}{\mathfrak{g}}

\newcommand{\fm}{\mathfrak{m}}

\newcommand{\fF}{\mathfrak{F}}

\newcommand{\cA}{\cal A}
\newcommand{\cS}{\cal S}

\numberwithin{equation}{section}

\title{Energy Loss in Holographic Anisotropic Model for Heavy
  Quarks in External Magnetic Field}

\author{Irina Ya. Aref'eva$^a$, Kristina Rannu$^a$ and Pavel Slepov$^a$}

\affiliation{$^a$Steklov Mathematical Institute, Russian Academy of
  Sciences,\\ Gubkina str. 8, 119991, Moscow, Russia
}

\emailAdd{arefeva@mi-ras.ru}
\emailAdd{rannu-ka@rudn.ru}
\emailAdd{slepov@mi-ras.ru}

\abstract{Energy loss in anisotropic hot dense QGP in external
  magnetic field is studied within holographic approach. Energy loss
  is calculated by estimation of behaviour of the spatial Wilson
  loops using the effective potential technique. We examine the
  dependence of the effective potential on the spatial Wilson loops
  orientation in fully aniso\-tropic background. For this purpose we
  obtain general formulas for the effective potential and study
  appearance of the effective potential dynamical wall. We consider
  particular fully anisotropic model \cite{2011.07023} supported by
  Einstein-Dilaton-three-Maxwell action. The effective potential
  strongly depends on the parameters of anisotropy and magnetic field,
  therefore the energy loss depends on physical parameters -- $T$,
  $\mu$, $c_B$ and orientation. Orientation is determined by angles
  between the moving heavy quark velocity, the axis of heavy ions
  collision and their impact parameter vector.}


\begin{document}

\maketitle


\section{Introduction}

Nowadays holographic approach to study quark-gluon plasma (QGP) is
actively developed \cite{Solana, IA, DeWolf, Arefeva:2016rob,
  Arefeva:2018bnh, A-DF} (and refs therein). Holography is one of the
most effective tools to investigate the
ultrarelativistic heavy-ion collisions (HIC). Different experimental
data such as transport coefficients, thermalization time,
multiplicity, direct-photon spectra etc. can be described within
holographic QCD (HQCD). It is also expected that holography predicts
more detailed form of QCD phase diagram. Different isotropic and
anisotropic models holographic models to describe QGP were considered
in \cite{2011.07023, Solana, IA, DeWolf, Arefeva:2016rob,
  Arefeva:2018bnh,A-DF, 0604204, 0611304, 0812.0792, 0903.2859,
  1004.1880, 1006.0055, 1006.5461, 1112.4402, 1201.0820, 1202.4436,
  1206.2824, 1301.0385, 1303.6929, 1406.1865, AG, 1411.5332,
  1505.07894, 1506.05930, 1511.02721, 1512.04062, 1512.06493,
  1606.03995, 1610.04618, 1610.09814, 1611.06339, 1612.06248,
  1703.09184, 1705.07587, 1707.00872, 1708.05691, 1710.06417,
  1802.05652, 1803.06764, 1808.05596, 1810.12525, 1810.07019,
  1811.11724, 1906.12316, 1907.01852, 1908.02000, ARS-2019qfthep, SS,
  1909.06917, 1910.02269, 2002.00075, 2003.05847, 2004.01965,
  2005.00500, Ballon-Bayona:2020xls, 2007.06980, 2009.05562,
  2009.05694, 2010.04578, 2010.06762, 2011.09474}. \\

One of HQCD models' classes is based on the Einstein-dilaton-Maxwell
action, where the Maxwell field supports chemical potential. Choice of
the warp factor in the metrics strongly influences the phase
transition structure of HQCD. This choice should be done with
correspondence to lattice results (Columbia plot) \cite{Brown:1990ev,
  1912.04827}. \\

It is important to add anisotropy into the holographic theory
\cite{1202.4436} as QGP is an anisotropic media just after the HIC. To
deal with anisotropic HQCD one considers Einstein-dilaton-two Maxwell
model with additional Maxwell field to support the anisotropy in
metrics. Such anisotropic model was considered in \cite{1802.05652,
  1808.05596} to reconstruct heavy quarks scenario and in
\cite{2009.05562} for the light quarks case. In these models
anisotropy is defined by a parameter $\nu$. The value $\nu = 4.5$
gives an accordance with the experimental data for the energy
dependence of the total multiplicity of particles created in
HIC. Isotropic holographic models had not been able to reproduce this
experimental dependence (\cite{AG} and refs therein). The model
\cite{1802.05652} describes smeared confinement/deconfinement phase
transitions, as the position of the confinement/deconfinement phase
transition line depends on orientation of quark pair in respect to the
HIC line \cite{1808.05596}. That model also indicates the relations of
the fluctuations of the multiplicity, i.e. the entanglement entropy,
with the background phase transitions \cite{2003.05847}. The
anisotropy of the background metric also influences on corresponding
jet quenching \cite{1202.4436, 1606.03995,1710.06417}. \\

In this paper we calculated spatial Wilson loops (SWL) in a fully
anisotropic holographic model for heavy quarks. It is known that for
isotropic models the string tension of SWL is proportional to the
corresponding drag force \cite{Zahed,1707.05045}. This relation with a
small modification takes place also for anisotropic models
\cite{A-DF}. Therefore we can calculate SWL to estimate the drag
forces for a quark moving in QGP. Drag forces defining energy loss are
the subject of intensive studies in holographic approach
\cite{0605158, 0605182, 0605199, 0611272, 0906.1890, 1202.3696,
  1312.7474, 1410.7040, 1412.8433, 1505.07379, 1507.06556, 1605.06061,
  1606.01598, 1707.05045, 1711.08943, 1901.09304}. In this paper the
SWLs are calculated in the model set up by
Einstein-dilaton-three-Maxwell action and describe two different types
of anisotropy: anisotropy for producing the multiplicity dependence on
energy and anisotropy associated with the magnetic field
\cite{2011.07023}. \\

This paper is organised as follows. In Subsect.\ref{SubSect: AO} the
Born-Infeld action for SWL in fully anisotropic background is
presented. In Subsect.\ref{Sect:BI} the large $\ell$ asymptotic of
the Born-Infeld action for SWL are calculated. In
Subsect.\ref{SubSect:DW} explicit forms of dynamical wall (DW)
equations are defined and asymptotics for particular cases of
orientations of SWL are presented. In Subsect.\ref{Subsect:Model} Einstein-dilaton-three-Maxwell holographic anisotropic model
\cite{2011.07023} and its thermodynamic properties are described.  
In Sect.\ref{Sect:solution} SWL are considered in 
the metric supported by Einstein-dilaton-three-Maxwell action
 and effective potentials' dependence on magnetic
field and HIC anisotropy are studied. The conclusion is given in Sect.\ref{Sect:conclusion}. Some formulas
describing the model \cite{2011.07023} are presented in Appendix
\ref{appendixA}.

\newpage
\section{Setup. Spatial Wilson loops in fully anisotropic background}
\label{Sect:SWL}

\subsection{Arbitrary Orientation}\label{SubSect: AO}

We consider fully anisotropic background
\bea
  ds^2 = G_{\mu\nu}dx^{\mu}dx^{\nu} = \cfrac{L^2 \fb(z)}{z^2} \left[
    - \, g(z) dt^2 + \fg_1 dx^2 + \fg_2  dy_1^2 + \fg_3  dy_2^2 +
    \cfrac{dz^2}{g(z)} \right] \label{Gbackgr}
\eea 
and are going to calculate differently oriented SWLs in this background.

Following the holographic approach we have to calculate the value of
the Nambu-Goto action for test string in the background
\eqref{Gbackgr}:
\be
  S = \frac{1}{2 \pi \alpha'} \int d\xi^{1} d\xi^{2} \sqrt{- \det
    h_{\alpha\beta}},
\ee
where the induced metric
\be
  h_{\alpha\beta} = G_{\mu\nu} \partial_{\alpha}
  X^{\mu} \partial_{\beta} X^{\nu}.
\ee
depends on the orientation. Here we use the same parametrization of
the orientation of Wilson rectangle as specified the entanglement
rectangular parallelepiped in \cite{2003.05847}. In 3-dimensional
space the parameters specifying orientations of both objects, the
rectangle in 3D space and rectangular parallelepiped, were the
same. There is obvious difference in parametrizations of the world
surface for the Wilson rectangle and world volume of rectangular
parallelepiped.

To get the parametrization of the wold sheet we use the representation
of the rotation matrix $M(\phi, \theta,\psi )$ in 3-dimensional space
\bea
x^i&=& \sum _{j=1,2,3} a_{ij}(\phi, \theta,\psi )\,\zeta^j,
\qquad i=1,2,3, \label{Emb}
\eea
in terms of the Euler angles $\phi, \theta,\psi $:
\bea
M(\phi, \theta,\psi )={\begin{pmatrix}a_{11}(\phi, \theta,\psi ) &
    a_{12}(\phi, \theta,\psi ) & a_{13}(\phi, \theta,\psi ) \\
    a_{21}(\phi, \theta,\psi ) & a_{22}(\phi, \theta,\psi ) &
    a_{23}(\phi, \theta,\psi ) \\
    a_{31}(\phi, \theta,\psi ) & a_{32}(\phi, \theta,\psi ) &
    a_{33}(\phi, \theta,\psi ) 
\end{pmatrix}}
\label{Eulmat}
\eea
where
\bea
\begin{array}{lll}
&a_{11}(\phi,\theta,\psi) =\cos \phi \cos\psi -\cos \theta \sin \phi \sin\psi,\,\,\,  \\
&a_{12}(\phi,\theta,\psi) =-\cos \psi \sin\phi -\cos\phi \cos \theta \sin\psi,\,\,\,\, \\
&a_{13}(\phi,\theta,\psi) =\sin\theta \sin \psi, \\
&a_{21}(\phi,\theta,\psi) =\cos \theta \cos \psi \sin\phi+\cos \phi \sin\psi, \\
&a_{22}(\phi,\theta,\psi) =\cos \phi \cos \theta \cos\psi-\sin \phi \sin\psi,\\ 
&a_{23}(\phi,\theta,\psi) =-\cos \psi \sin \theta, \\
&a_{31}(\phi,\theta,\psi) =\sin\phi \sin \theta, \\
&a_{32}(\phi,\theta,\psi) =\cos\phi\sin \theta , \\
&a_{33}(\phi,\theta,\psi) =\cos \theta.
 \end{array}\\\ \nn \label{EP}
\eea
Here $\phi$ is the angle between $\zeta^1$-axis and the node line (N),
$\theta$ is the angle between $\zeta^3$ and $x^3$-axes, $\psi$ is the
angle between the node line N and $x^1$-axis.

To describe the nesting of the 2-dimensional world sheet in
5-dimensional space time we use 
\bea \label{GPar}
X^0(\xi) &=& const,\nn \\
X^i(\xi) &=& \sum _{\alpha=1,2} a_{i\alpha }(\phi, \theta,\psi )\,\xi^\alpha,
\qquad i=1,2,3, \qquad \alpha=1,2, \label{Emb} \\
X^4(\xi) &=& z(\xi^1),\nn 
\eea
where $x^i$ are spatial coordinates and $a_{ij}(\phi,\theta,\psi)$ are
entries of the rotation matrix.

We write the line element of the induced metric for the arbitrary
oriented spatial Wilson loop as
\bea
ds^2&=&g_{\alpha \beta }\,d\xi^\alpha d\xi^\beta, \qquad
\alpha,\beta=1,2
\eea
and substitute the differentials $dx^M$ following from the embedding
relations \eqref{Emb}:
\bea
ds^2&=&\frac{L^2 b_s(z)}{z^2}\left(\sum
_{i=1,2,3}\fg_i(z)d(x^i)^2+\frac{d(x^4)^2}{g}\right)\nn\\
&=&\frac{L^2 b_s(z)}{z^2}\left( \sum _{i=1,2,3}\fg_i(z)\,\Big(\sum
  _{j=1,2}a_{ij}(\phi, \theta,\psi ) \, d\xi^{j}\Big)^2 
+z'^{2} \ \frac{d(\xi^1)^2}{g(z)}\right).
\eea
We have
\bea g_{\alpha\beta}&=&\frac{L^2 b_s(z)}{z^2}\,\bar{g}_{\alpha\beta},\\
\bar g_{11}(z,\phi, \theta,\psi )&=&\fg_{1}a_{11}^2+\fg_{2}a_{21}^2+\fg_{3}a_{31}^2+\frac{z'^{2}}{g},\nn
\\
\bar g_{22}(z,\phi, \theta,\psi )&=&\fg_{1}a_{12}^2+\fg_{2}a_{22}^2+\fg_{3}a_{32}^2,\nn\\
\bar g_{12}(z,\phi, \theta,\psi )&=&\fg_{1}a_{11} a_{12}+\fg_{2}a_{21}a_{22}+\fg_{3}a_{13}a_{32},\nn \\
\bar g_{21}&=&\bar g_{12}.
\label{barg}
\eea
Determinant of the induced metric for SWL is
\bea
\det g_{\alpha\beta} &=& \left(\frac{L^2 b_s}{z^2}\right)^2 \Biggl(
\left(\fg_{1}a_{11}^2 + \fg_{2}a_{21}^2 + \fg_{3}a_{31}^2 +
  \frac{z'^{2}}{g}\right) (\fg_{1}a_{12}^2 + \fg_{2}a_{22}^2 +
\fg_{3}a_{32}^2) - \nn \\  
&-&(\fg_{1}a_{11} a_{12} + \fg_{2}a_{21}a_{22} +
\fg_{3}a_{13}a_{32})^2 \Biggr).
\eea

The Nambu-Goto action for SWL is given by integration over the world
sheet~${\cal W}$
\be
{\cal S}_{SWL} = \int _{{\cal W}} \left(\frac{L^2 b_s}{z^2}\right)
\sqrt{\left(\fg_{1}\fg_{2}a_{33}^2 + \fg_{1}\fg_{3}a_{23}^2 +
    \fg_{2}\fg_{3}a_{13}^2 + \fg_{3}^2a_{32}^2(a_{31}^2-a_{13}^2) +
    \frac{z'^{2}}{g} \, \bar g_{22}\right)} \ d\xi^{1}d\xi^{2}.
\label{S_swl}\ee
\normalsize

The effective potential is 
%
\be
{\cal V}(z(\xi)) = \left(\frac{L^2 b_s}{z^2}\right) \,
\sqrt{\fg_{1}\fg_{2}a_{33}^2 + \fg_{1}\fg_{3}a_{23}^2 +
  \fg_{2}\fg_{3}a_{13}^2 +
  \fg_{3}^2a_{32}^2(a_{31}^2-a_{13}^2)}. \label{V_swl}
\ee

The effective potential and the action depend on the angles and
anisotopy. This result can be compared with the action and the effective
potential for holographic entanglement entropy  (HEE) \cite{2003.05847}:
\be
{\cal S}_{HEE} = \int _{{\cal P}}\left(\frac{L^2b_s}{z^2}\right)^{3/2}
\sqrt{\left(\fg_{1}\fg_{2}\fg_{3} + \frac{z'^{2}}{g}\,(\bar g_{22}\bar
    g_{33}-\bar g_{23}^2)\right)} \ d\xi^{1}d\xi^{2}d\xi^{3},
\label{S_gen}
\ee
\bea
{\cal V}_{HEE}(z) = \left(\frac{L^2 b_s}{z^2}\right)^{3/2}
\sqrt{\fg_{1}\fg_{2}\fg_{3}}, \label{GV}
\eea
where $g$, $\fg_1$, $\fg_2$, $\fg_3$ are functions of $z$ and $\bar
g_{22}$, $\bar g_{33}$, $\bar g_{23}$ are functions of $z$ and the
Euler angles. Note that the effective potential for HEE does not
depend on the angles.

\subsection{Born-Infeld type action and large $\ell$ asymptotics}
\label{Sect:BI}

The considered actions for SWL, temporal WL and HEE \cite{Arefeva:2018bnh,SS,2003.05847} are the particular
cases of the BI action:
\be
\label{BI}
{\cal S}=\int _{-\ell/2}^{\ell/2} M(z(\xi))\sqrt{{\cal {\cal
      F}}(z(\xi))+(z^{\prime}(\xi))^ 2} \, d\xi.
\ee
This action defines the dynamical system with a dynamic variable $z =
z(\xi)$ and time~$\xi$. The effective potential is
\be
\label{EfPot}
{\cal V}(z(\xi))\equiv M(z(\xi))\sqrt{{\cal F}(z(\xi))}.
\ee
This system has the  first integral:
\bea
\label{FI}
\frac{M(z(\xi)){\cal F}(z(\xi))}{\sqrt{{\cal
      F}(z(\xi))+(z'(\xi))^2}}={\cal I}. 
\eea
From \eqref{FI} we can find the ``top'' point ${z_{*}}$ (the closed
position of the minimal surface to the horizon), where ${z'(\xi) =
  0}$:
\be
M(z_{*})\sqrt{F(z_{*})}={\cal I}.
\ee
Finding  $z^{\prime}$ from \eqref{FI} one gets representations for
the length $\ell$ and the action ${\cal S}$ \eqref{BI}:
\bea\label{ell1}
\frac\ell2 &=&
\int_0^{z_*}\frac{1}{\sqrt{{\cal F}(z)}} \
\frac{dz}{\sqrt{\frac{{\cal V}^2(z)}{{\cal V}^2(z_*)} -1 }}, \\
\frac{{\cal S}}{2} &=&
\int_\epsilon^{z_*}\frac{M(z)dz}{\sqrt{1-\frac{{\cal V}^2(z_*)}{{\cal
        V}^2(z)}}}.
\label{calS1}
\eea

We have  two options to have $\ell\to \infty$.
\begin{itemize}
\item The existence of  a stationary point of ${\cal V}(z)$
  \be
  {\cal V}^\prime\Big|_{z_{DW}} = 0.
  \ee
  This point is called a dynamical wall (DW) point. One takes the top
  point $z_*$ equal to the dynamical wall position. Since near
  the top point
  \be
  \sqrt{\frac{{\cal V}^2(z)}{{\cal V}^2(z_{DW})}-1}= \sqrt{\frac{
      {\cal V}''(z_{DW})}{{\cal V}(z_{DW})}}(z-z_*)+{\cal O}((z-z_*)^2),
  \ee
  one gets
  \bea
  \ell &\underset{z\to z_*}{\sim }& \frac{1}{\sqrt{F(z_{DW})}} \,
  \sqrt{\frac{{\cal V}(z_{DW})}{{\cal V}''(z_{DW})}} \, \log (z-z_*),
  \\
  \cS&\underset{z\to z_*}{\sim}& M(z_{DW}) \, \sqrt{\frac{{\cal
        V}(z_{DW})}{{\cal V}''(z_{DW})}} \log (z-z_*).
  \eea
  Hence
  \bea
  \cS&\sim&  M(z_{DW})\cdot\sqrt{F(z_{DW})} \cdot \ell,
  \eea
  \bea
  \sigma _{DW} &=& M(z_{DW}) \,\sqrt{F(z_{DW})}. \label{ten}
  \eea
\item There is no stationary point of ${\cal V}(z)$ in the region $0 < z <
  z_h$, and we suppose it to be near horizon
  \be
  F(z) = \fF(z_h) (z_h-z)+{\cal O}((z_h-z)^2),
  \ee
  i.e. near horizon (the sting streach on the horizon). In this case
  we take $z_*=z_h$ and there are the following options: 
  \begin{itemize}
  \item
    if $M(z_h)\neq \infty$, we have
    \bea
    \ell&\to& \infty,\\
    S&\to&0;
    \eea
  \item  if $M(z) \underset{z\to z_h}{\to}\infty$ as
    \be
    M(z)\underset{z\sim z_h}{\sim} \frac{\fm (z_h)}{\sqrt{z-z_h}},
    \ee
    we have
    \bea
    \ell &\underset{z\to z_h}{\sim }& \frac{1}{\sqrt{\fF (z_h)}} \
    \frac{1}{\sqrt{-\frac{2 {\cal V}'(z_h)}{{\cal V}(z_h)}}} \ \log
    (z-z_h), \\
    \cS&\underset{z\to z_*}{\sim}& \fm(z_h) \
    \frac{1}{\sqrt{-\frac{2{\cal V}'(z_h)}{{\cal V}(z_h)}}} \ \log
    (z-z_h)
    \eea
    and therefore
    \be
    \sigma _h=\fm(z_h)\,\fF^{1/2}(z_h).
    \ee
  \end{itemize}
  
\end{itemize}

\subsection{Particular cases and DW equations}\label{SubSect:DW}

Let us consider particular cases of \eqref{V_swl} and \eqref{S_swl} to 
understand  the picture with SWLs in fully anisotropic background more
instructively. \\
1)~$\phi=0$,  $\theta=0$, $\psi=0$; $a_{11}=a_{22}=a_{33}=1$, 
$a_{12}=a_{21}=a_{31}=a_{31}=a_{32}=a_{23}=0$:\\
\be
{\cal S}_{xY_{1}}=\int _{{\cal P}}\left(\frac{L^2 b_s}{z^2}\right)
\sqrt{\left(\fg_{1}\fg_{2}+\frac{z'^{2}}{g}\fg_{2}\right)} \
d\xi^{1}d\xi^{2},
\ee
\be
{\cal V}_{xY_{1}}(z(\xi)) = \left(\frac{L^2 b_s}{z^2}\right)\
\sqrt{\fg_{1}\fg_{2}};
\ee
2)~$\phi=\pi/2$, $\theta=0$, $\psi=0$; $a_{21}=a_{33}=-a_{12}=1$,
$a_{11}=a_{13}=a_{22}=a_{23}=a_{31}=a_{32}=0$:\\
 \be
{\cal S}_{Xy_{1}}=\int _{{\cal P}}\left(\frac{L^2 b_s}{z^2}\right)\sqrt{\left(\fg_{1}\fg_{2}+\frac{z'^{2}}{g}\fg_{1}\right)}d\xi^{1}d\xi^{2},
\ee
\be
{\cal V}_{Xy_{1}}(z(\xi))=\left(\frac{L^2 b_s}{z^2}\right)\
\sqrt{\fg_{1}\fg_{2}};
\ee
3)~$\phi=0$,  $\theta=\pi/2$, $\psi=0$; $a_{11}=-a_{23}=a_{32}=1$,
$a_{12}=a_{13}=a_{21}=a_{22}=a_{31}=a_{33}=0$:\\
\be
{\cal S}_{xY_{2}}=\int _{{\cal P}}\left(\frac{L^2 b_s}{z^2}\right)
\sqrt{\left(\fg_{1}\fg_{3}+\frac{z'^{2}}{g}\fg_3\right)}d\xi^{1}d\xi^{2},
\ee
\be
{\cal V}_{xY_{2}}(z(\xi))=\left(\frac{L^2 b_s}{z^2}\right)\
\sqrt{\fg_{1}\fg_{3}};
\ee
4)~$\phi=\pi/2$, $\theta=\pi/2$, $\psi=-\pi/2$;
$a_{22}=a_{31}=-a_{13}=1$,
$a_{11}=a_{12}=a_{21}=a_{23}=a_{32}=a_{33}=0$:\\
\be
{\cal S}_{y_{1}Y_{2}}=\int _{{\cal P}}\left(\frac{L^2 b_s}{z^2}\right)
\sqrt{\left(\fg_{2}\fg_{3}+\frac{z'^{2}}{g}\fg_2\right)}d\xi^{1}d\xi^{2},
\ee
\be
{\cal V}_{y_{1}Y_{2}}(z(\xi))=\left(\frac{L^2 b_s}{z^2}\right)\
\sqrt{\fg_{2}\fg_{3}}.
\ee
These results correspond to \cite{A-DF}. The general form of the  DW
equation (if DW exists): ${\cal V}'(z)=0$ \cite{Arefeva:2016rob,
  0604204}.

The equations for the DW for SWL in particular cases for different
potentials:
\bea
{\cal DW}_{xY_1} = {\cal DW}_{Xy_1}&\equiv&
\frac{2 b_s'(z)}{b_s(z)} + \frac{\fg_1'(z)}{\fg_1(z)} +
\frac{\fg_2'(z)}{\fg_2(z)} - \frac{4}{z}\Bigg|_{z = z_{DW}}
\hspace{-15pt} = 0, \\
{\cal DW}_{xY_2} &\equiv&
\frac{2 b_s'(z)}{b_s(z)} + \frac{\fg_1'(z)}{\fg_1(z)} +
\frac{\fg_3'(z)}{\fg_3(z)} - \frac{4}{z}\Bigg|_{z = z_{DW}}
\hspace{-15pt} = 0,\\
{\cal DW}_{y_1Y_2} &\equiv&
\frac{2 b_s'(z)}{b_s(z)} + \frac{\fg_2'(z)}{\fg_2(z)} +
\frac{\fg_3'(z)}{\fg_3(z)} - \frac{4}{z}\Bigg|_{z = z_{DW}}
\hspace{-15pt} = 0.
\eea

Let us take the metric in the string frame that supported by
Einstein-dilaton-three Maxwell action that was obtained in
\cite{2011.07023}:
\begin{gather}
  ds^2 = \cfrac{L^2 b_s(z)}{z^2}  \left[
    - \ g(z) dt^2 + dx^2 + \left(
      \cfrac{z}{L} \right)^{2-\frac{2}{\nu}} \hspace{-5pt} dy_1^2
    + e^{c_B z^2} \left( \cfrac{z}{L} \right)^{2-\frac{2}{\nu}}
    \hspace{-5pt} dy_2^2
    + \cfrac{dz^2}{g(z)} \right], \label{eq:2.03} \\
  b_s(z) = e^{{2{\cA}(z)}+\sqrt{\frac2{3}} \phi(z,z_0)}, \label{eq:2.04}
\end{gather}
where $L$ is the AdS-radius, $b_s(z)$ is the warp-factor,
$\phi(z,z_0)$ is the dilaton field, $z_0$ is the point at which
$\phi(z_0)=0$ (the boundary condition for dilaton field), $g(z)$ is
the blackening function, $\nu$ is the parameter of HIC  anisotropy and
$c_B$ is the coefficient of secondary anisotropy related to the
external magnetic field $F_{\mu\nu}^{(B)}$. Note that choice of
${\cA}(z)$ determines the heavy/light quarks description of the model,
so we follow previous works and consider ${\cA}(z) = - \, c z^2/4$ for
heavy quarks \cite{1802.05652} and ${\cA}(z) = - \, a \, \ln (b z^2 +
1)$ for light quarks \cite{2009.05562}. Therefore for the solution
\cite{2011.07023} we can get $\fg_1 = 1$, $\fg_2 = (z/L)^{2-2/\nu}$,
$\fg_3 = (z/L)^{2-2/\nu}e^{c_Bz^2}$. For this particular case:
\bea\label{sigmaxY1}
\sigma_{xY_1}=\sigma_{Xy_1}=\left(\frac{L^2 b_s(z)}{z^2}\right)\
\sqrt{\fg_{1}\fg_{2}}=\left(\frac{L^{1+1/\nu}
    b_s(z)}{z^{1+1/\nu}}\right),\\
\sigma_{xY_2}=\left(\frac{L^2 b_s(z)}{z^2}\right)\
\sqrt{\fg_{1}\fg_{3}}=\left(\frac{L^{1+1/\nu}
    b_s(z)}{z^{1+1/\nu}}\right)e^{c_Bz^2/2},\label{sigmaxY2}\\
\sigma_{y_1Y_2}=\left(\frac{L^2 b_s(z)}{z^2}\right)\
\sqrt{\fg_{2}\fg_{3}}=\left(\frac{L^{2/\nu}
    b_s(z)}{z^{2/\nu}}\right)e^{c_Bz^2/2},\label{sigmayY2}
\eea
where $z = z_h$ or $z = z_{DW}$ (if the dynamical wall exists). 

Note  that drag forces for metric \eqref{Gbackgr} with $\fg_1=1$
 have been calculated directly in \cite{A-DF} with the result 
\bea
 p_x&=& v_x\,\frac{b_s(z)}{z^2} \\
  p_{y_1}&=& v_{y_1}\,\frac{b_s(z)}{z^2}\,\fg_2 (z) \\
   p_{y_2}&=& v_{y_2}\,\frac{b_s(z)}{z^2}\,\fg_3(z),
\eea
that reproduce \eqref{sigmaxY1}, \eqref{sigmaxY2} and \eqref{sigmayY2}
for  $\fg_1=1$ if we identify 
\bea
v_x&=&v \,\sqrt{\fg_2},\\
v_{y_1}&=&v \,\frac{\sqrt{\fg_3}}{\fg_2},\\
v_{y_2}&=&v \,\frac{\sqrt{\fg_2}}{\sqrt{\fg_3}}\eea
with some constant $v$.

\subsection{Model and its thermodynamical properties}
\label{Subsect:Model}

In the next sections we calculate the string tensions
\eqref{sigmaxY1}-\eqref{sigmayY2} and their dependence on the
thermodynamic parameters and magnetic field. Most of the quantities
we dealt with in the previous sections depend explicitly on $z_h$,
while we need the temperature dependence. As we deal with the model
considered in \cite{2011.07023}, we remind here its main features and
specifically the behavior of temperature as function of $z_h$, $c_B$
and $\mu$.

We take the action in Einstein frame and the metric ansatz 
\begin{gather}
  \begin{split}
    S &= \cfrac{1}{16\pi G_5} \int d^5x \ \sqrt{-g} \ \times \\
    &\times \left[ R - \cfrac{f_1(\phi)}{4} \ F^{_{(1)}2} 
      - \cfrac{f_2(\phi)}{4} \ F^{_{(2)}2}
      - \cfrac{f_B(\phi)}{4} \ F^{_{(B)}2}
      - \cfrac{1}{2} \ \partial_{\mu} \phi \partial^{\mu} \phi
      - V(\phi) \right],
  \end{split}\label{eq:2.01} \\
  \begin{split}
    F_{\mu\nu}^{(1)} = \partial_{\mu} A_{\nu} - \partial_{\nu} A_{\mu},\
&  {\mbox{i.e.}} \quad A_{\mu}^{(1)} = A_t (z) \delta_\mu^0, \\
    F_{\mu\nu}^{(2)} = q \ dy^1 \wedge dy^2, \ &  {\mbox{i.e.}} \quad
    F_{y_1 y_2}^{(2)} = q,\\
    F_{\mu\nu}^{(B)} = q_B \ dx \wedge dy^1, \ &  {\mbox{i.e.}} \quad
   F_{x y_1}^{(B)} = q_B,
  \end{split}\label{eq:2.02} \\
  ds^2 = \cfrac{L^2}{z^2} \ \fb(z) \left[
    - \ g(z) dt^2 + dx^2 + \left(
      \cfrac{z}{L} \right)^{2-\frac{2}{\nu}} \hspace{-5pt} dy_1^2
    + e^{c_B z^2} \left( \cfrac{z}{L} \right)^{2-\frac{2}{\nu}}
    \hspace{-5pt} dy_2^2
    + \cfrac{dz^2}{g(z)} \right], \label{eq:2.03} \\
  \fb(z) = e^{2{\cA}(z)}, \label{eq:2.04}
\end{gather}
where $\phi = \phi(z)$ is the scalar field, $f_1(\phi)$, $f_2(\phi)$
and $f_B(\phi)$ are the coupling functions associated with the Maxwell
fields $A_{\mu}$, $F_{\mu\nu}^{(2)}$ and $F_{\mu\nu}^{(B)}$
correspondingly, $q$ and $q_B$ are constants and $V(\phi)$ is the
scalar field potential, $L$ is the AdS-radius, $\fb(z)$ is the
warp factor in Einstein frame and ${\cA}(z)$ is related with $\fb(z)$
according \eqref{eq:2.04}, $g(z)$ is the blackening function, $\nu$ is
the parameter of primary anisotropy, caused by non-symmetry of
heavy-ion collision (HIC), and $c_B$ is the coefficient of secondary
anisotropy related to the external magnetic field
$F_{\mu\nu}^{(B)}$. Choice of ${\cA}(z)$ determines the heavy/light
quarks description of the model, so we follow previous works and
consider ${\cA}(z) = - \, c z^2/4$ and $f_1 = z^{-2+\frac{2}{\nu}}$
for heavy quarks \cite{1802.05652}. The EOM solution can be found in
Appendix \ref{appendixA}.

Temperature and entropy can be written as:
\begin{gather}
  \begin{split}
    T &= \cfrac{|g'|}{4 \pi} \, \Bigl|_{z=z_h}
    = \cfrac{1}{2 \pi} \left| - \ e^{\frac{1}{4}(3c-2c_B)z_h^2} \ (2
      c_B - c) \ z_h^{1+\frac{2}{\nu}} \left\{
        \cfrac{\mu^2 \ e^{\frac{1}{4}(c-2c_B)z_h^2}}{4 L^2 \left( 1 -
            e^{\frac{1}{4}(c-2c_B)z_h^2} \right)^2} \right. \right. +
    \\
    &+ \ \left( \cfrac{3}{4} \right)^{1+\frac{1}{\nu}}
    \cfrac{(2 c_B - c)^{\frac{1}{\nu}}}{\Gamma\left(1 + \frac{1}{\nu}
        \ ; 0\right) - \Gamma\left(1 + \frac{1}{\nu} \ ; \frac{3}{4}
        (2 c_B - c) z_h^2 \right)} \ \times \\
    &\left. \times \left. \left[
          1 - \cfrac{\mu^2 \ (2 c_B - c)^{-\frac{1}{\nu}}}{4 L^2
            \left( 1 - e^{\frac{1}{4}(c-2c_B)z_h^2} \right)^2} \left(
            \Gamma\left(1 + \frac{1}{\nu} \ ; 0\right) - \Gamma\left(1
              + \frac{1}{\nu} \ ; (2 c_B - c) z_h^2 \right) \right)
        \right] \right\} \right|,
  \end{split}\label{eq:4.01} \\
  s = \cfrac{1}{4} \left( \cfrac{L}{z_h} \right)^{1+\frac{2}{\nu}}
  e^{-\frac{1}{4}(3c-2c_B)z_h^2}. \label{eq:4.02}
\end{gather}

\begin{figure}[t!]
  \centering
  \includegraphics[scale=0.42]{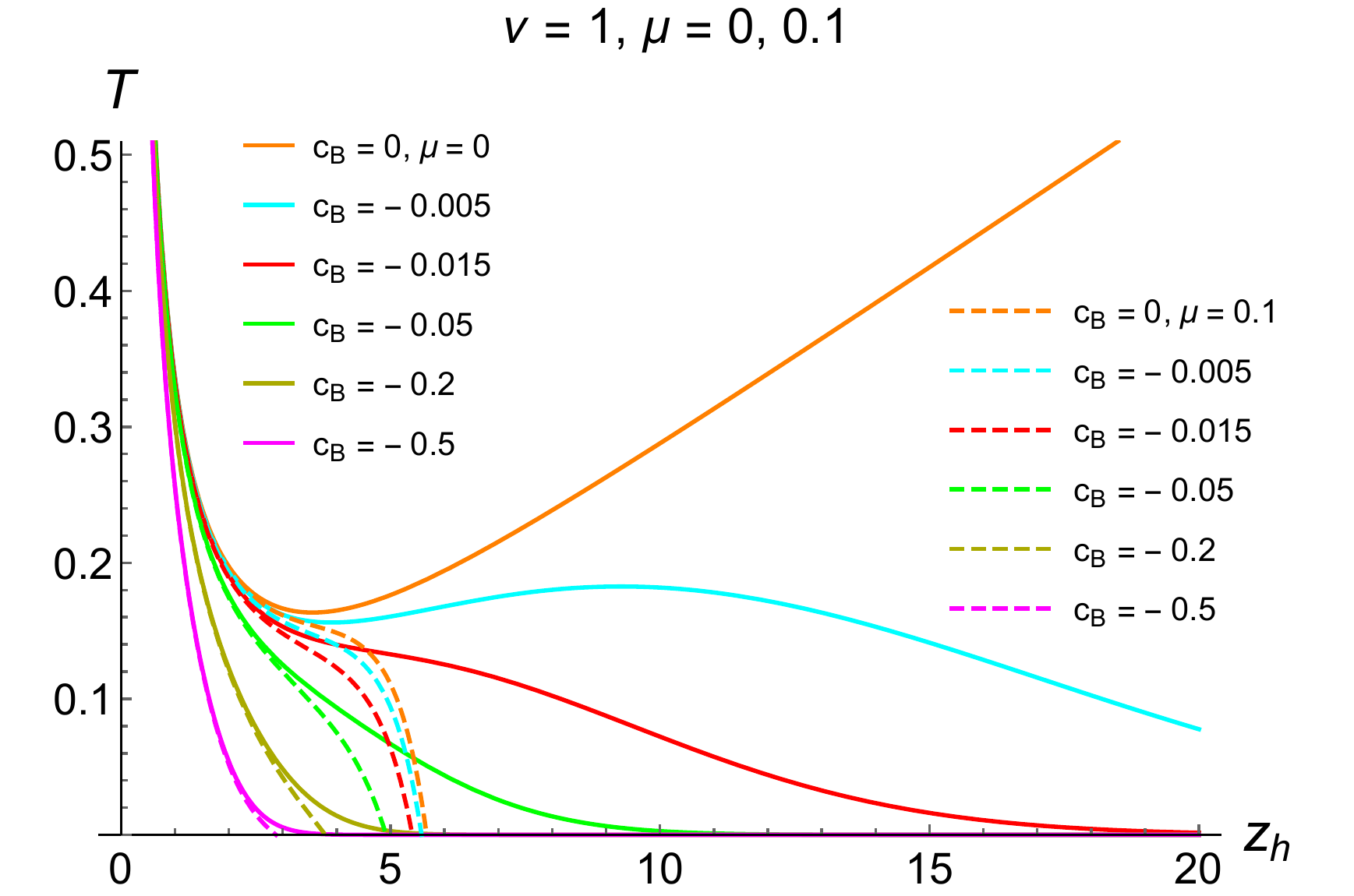}
  \includegraphics[scale=0.42]{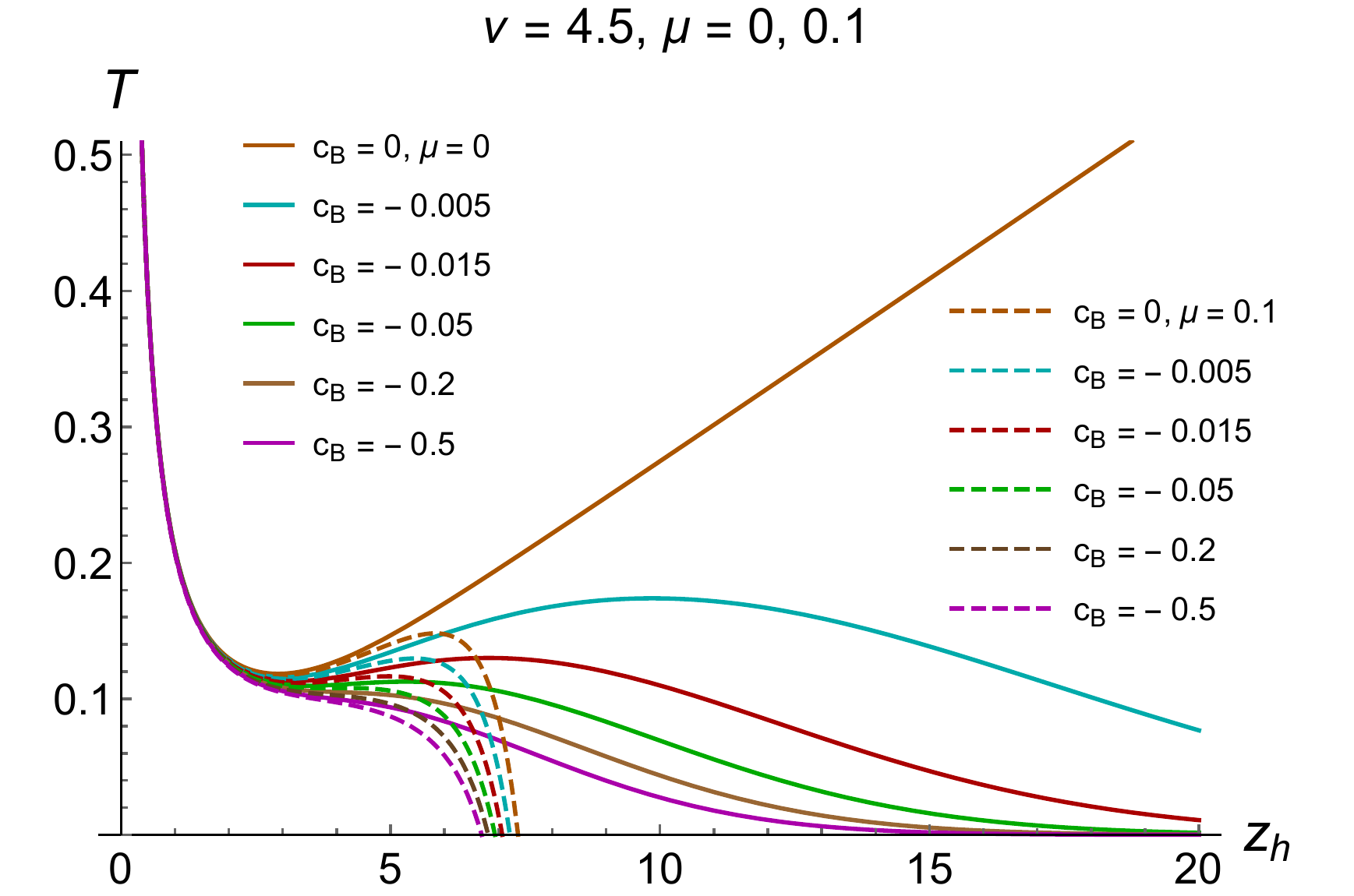} \\
  A \hspace{70mm} B \\ \ \\
  \includegraphics[scale=0.42]{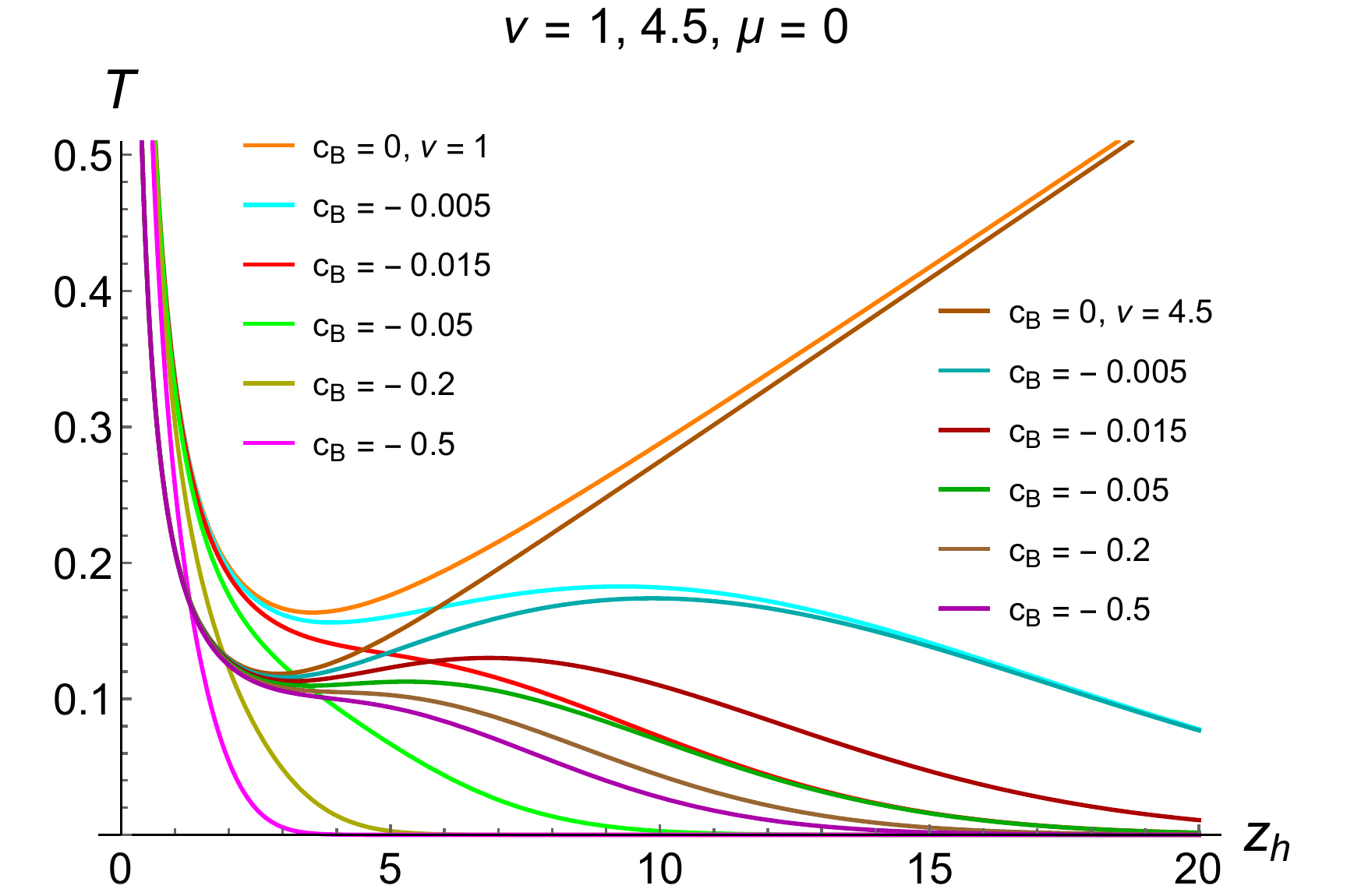} 
  \includegraphics[scale=0.42]{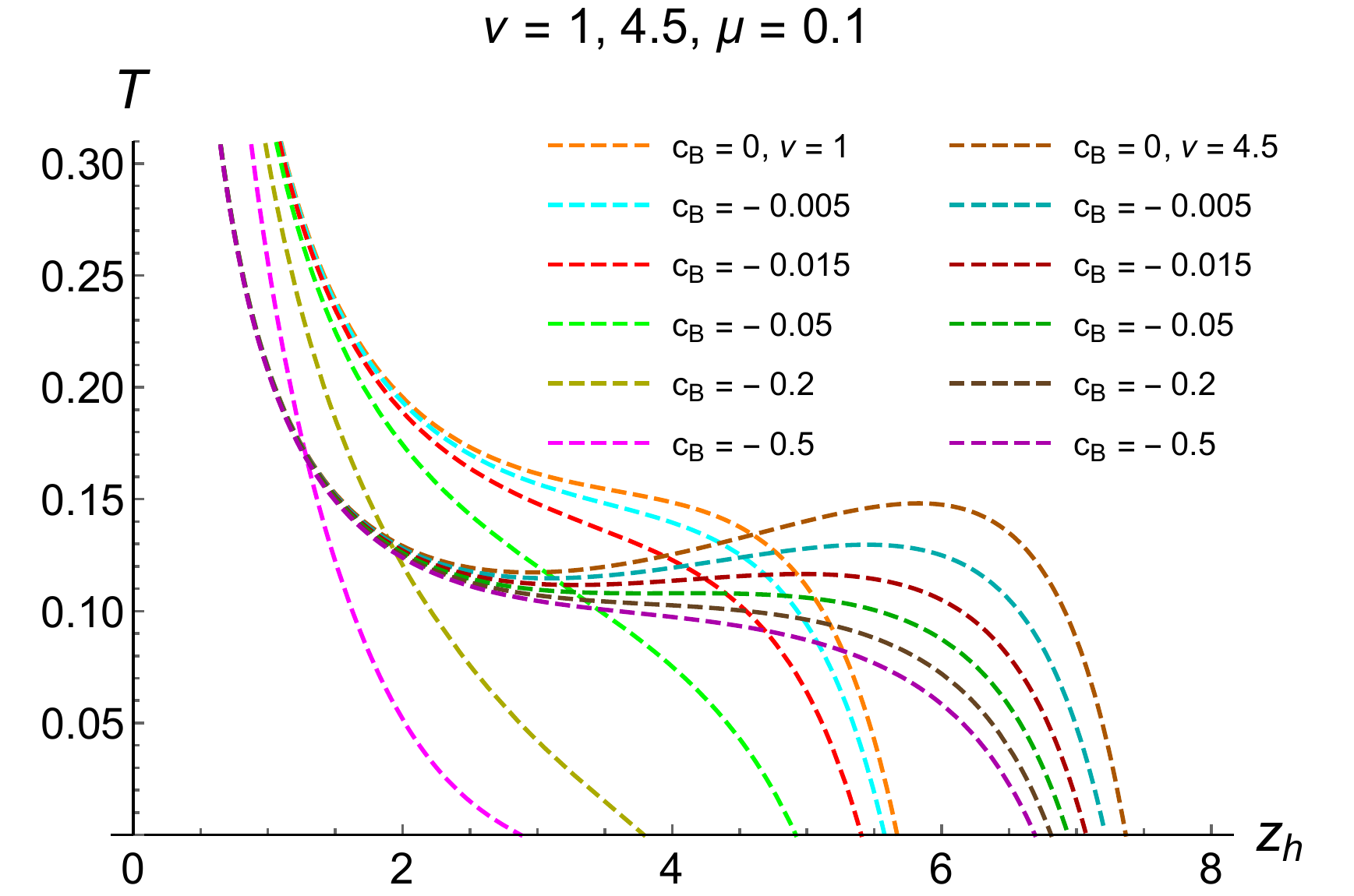} \\
  C \hspace{70mm} D
  \caption{Temperature dependence on $z_h$ for different $c_B$, $\mu =
    0, \ 0.1$, $\nu = 1$ (A) and $\nu = 4.5$ (B); for $\nu = 1, \ 4.5$,
    $\mu = 0$ (C) and $\mu = 0.1$ (D).
  }
  \label{Fig:Temperature}
\end{figure}

\begin{figure}[h!]
  \centering
  \includegraphics[scale=0.42]{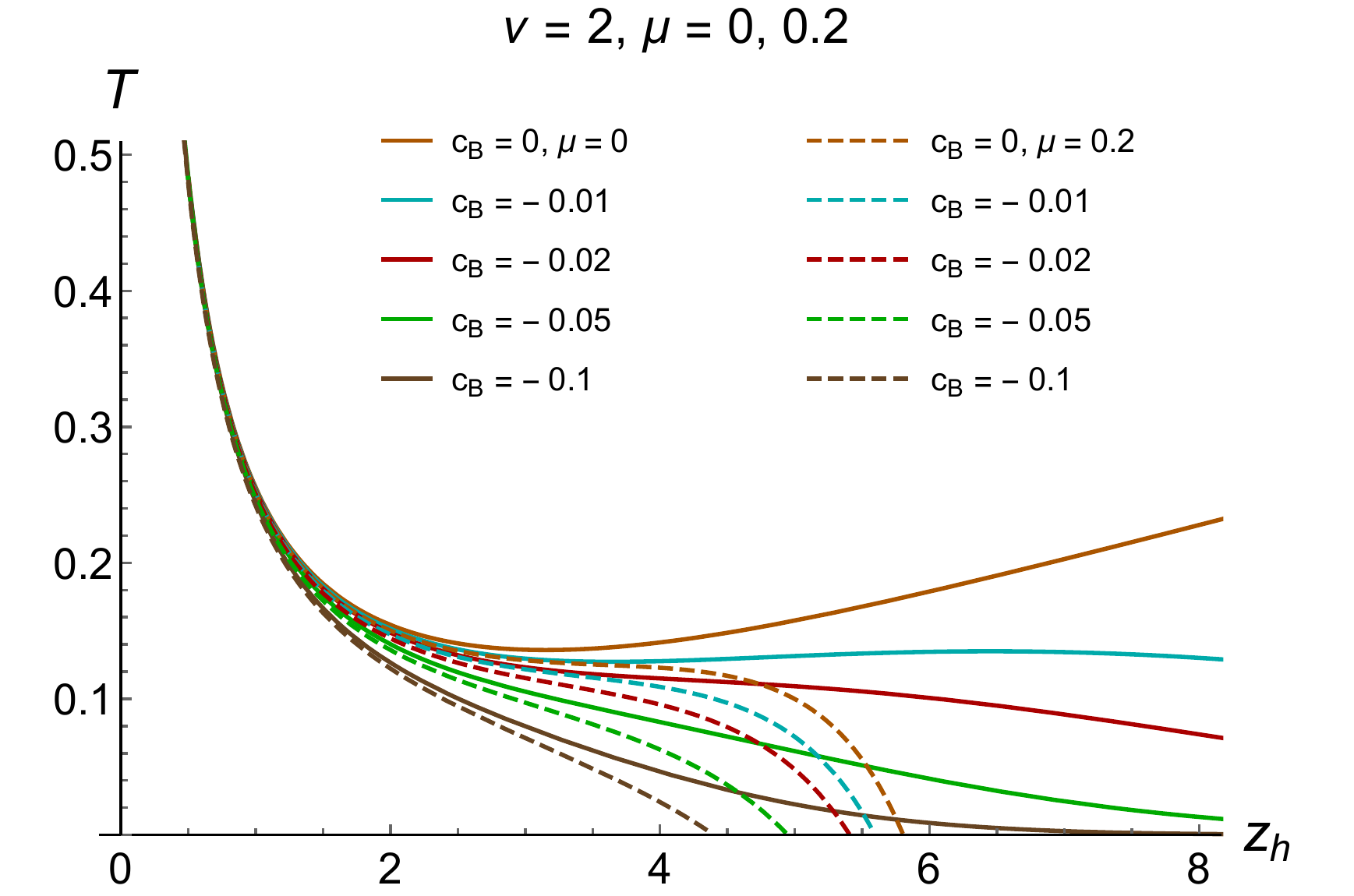}\\
  \caption{Temperature dependence on $z_h$ for different $c_B$, $\nu =
    2$, $\mu = 0, \ 0.2$.}
  \label{Fig:Tnu2}
\end{figure}

For zero chemical potential and zero magnetic field in holographic
models describing heavy quarks behavior \cite{1802.05652, 1808.05596}
temperature is a two-valued function of horizon having a local minimum
(Fig.\ref{Fig:Temperature}, \ref{Fig:Tnu2}). Increasing branch with
larger $z_h$ values is unstable, so the phase transition (collapse) to
the decreasing (stable) branch with smaller $z_h$ values along any
isotherm $T > T_{min}$ is possible. Because of the local minimum
temperatures $0 \le T < T_{min}$ can't be reached thus limiting the
possibility of the system cool-down.

Turning magnetic field or chemical potential on makes temperature a
three-valued function due to a local maximum appearance
(Fig.\ref{Fig:Temperature}A,B). Thus magnetic field and chemical
potential reinforce each other's effect on the temperature
behavior. Here the BH collapse and therefore the 1-st order phase
transition are possible for $T_{\min} < T < T_{max}$ where $T(z_h)$ is
three-valued. Both decreasing branch with larger $z_h$ (right) and
increasing branch (middle) are unstable, and the collapse occurs from
decreasing unstable branch to the decreasing stable one (left) 
bypassing through the increasing unstable interval between the local
minimum and the local maximum. For $T > T_{max}$ and $0 \le T <
T_{min}$ phase transition doesn't happen but any temperatures are
available.
 

As chemical potential and magnetic field have similar effect on the
temperature there are rather narrow intervals of $\mu$ and $c_B$ for
which collapse is still possible. As we can see on plots of
Fig.\ref{Fig:Temperature}A, even for $c_B = - 0.015$ or $\mu = 0.1$
temperature is monotonic already. So the BH collapse and the 1-st
order phase transition corresponding to it become very fragile in the
magnetic field, and non-zero chemical potential only exacerbates this
situation.

The other factor that becomes significant is the primary spatial
anisotropy parametrized by the coefficient $\nu$. It makes temperature
extremes more pronounced and inhibits their smoothing longer
(Fig.\ref{Fig:Temperature}C). This tendency keeps for non-zero
$\mu$ (Fig.\ref{Fig:Temperature}D), thus making phase transition more
resistant to stronger coupling with magnetic field and to increase in
chemical potential. On Fig.\ref{Fig:Tnu2} intermediate case of small 
primary anisotropy is presented.

\begin{figure}[t!]
  \centering
  $\nu = 1$ \hspace{70 mm} $\nu=4.5$\\ \ \\
  \includegraphics[scale=0.45]{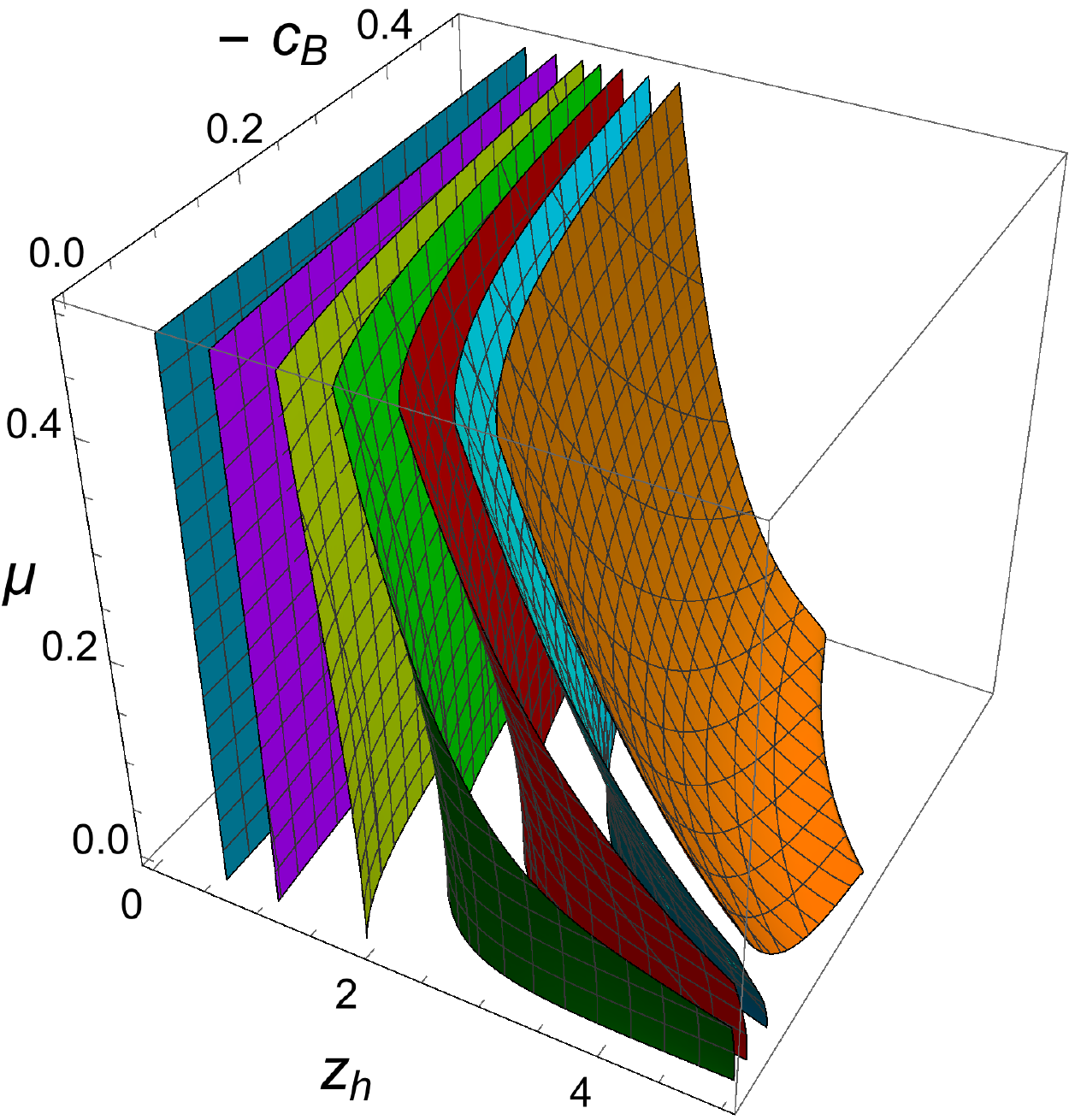} \quad
  \includegraphics[scale=0.5]{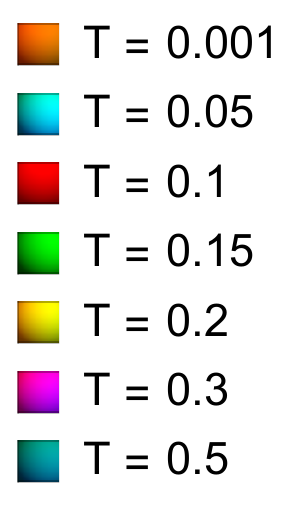} \quad
  \includegraphics[scale=0.45]{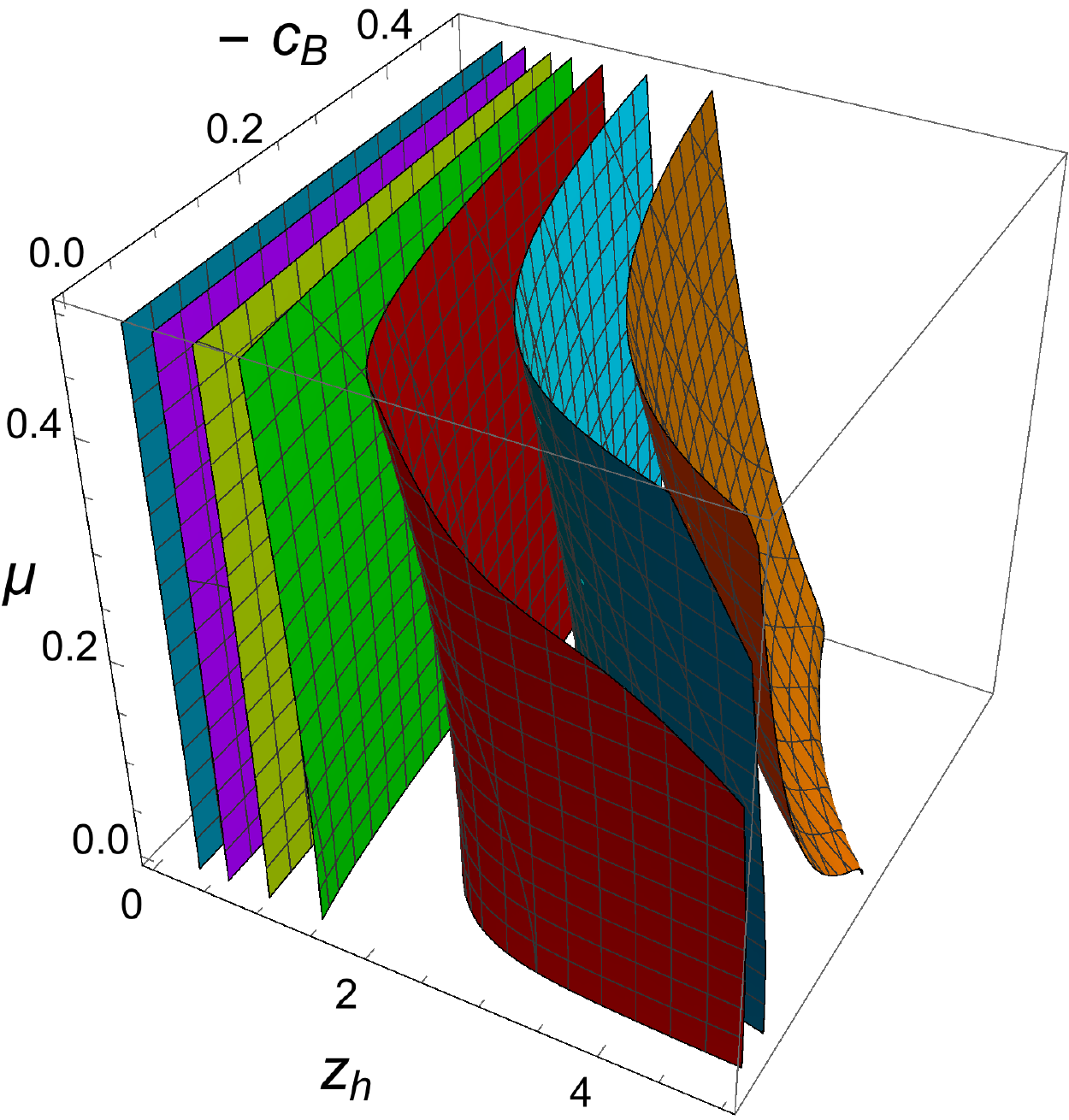} \\ \ \\
  A \hspace{80mm} B \\ \ \\
  \includegraphics[scale=0.45]{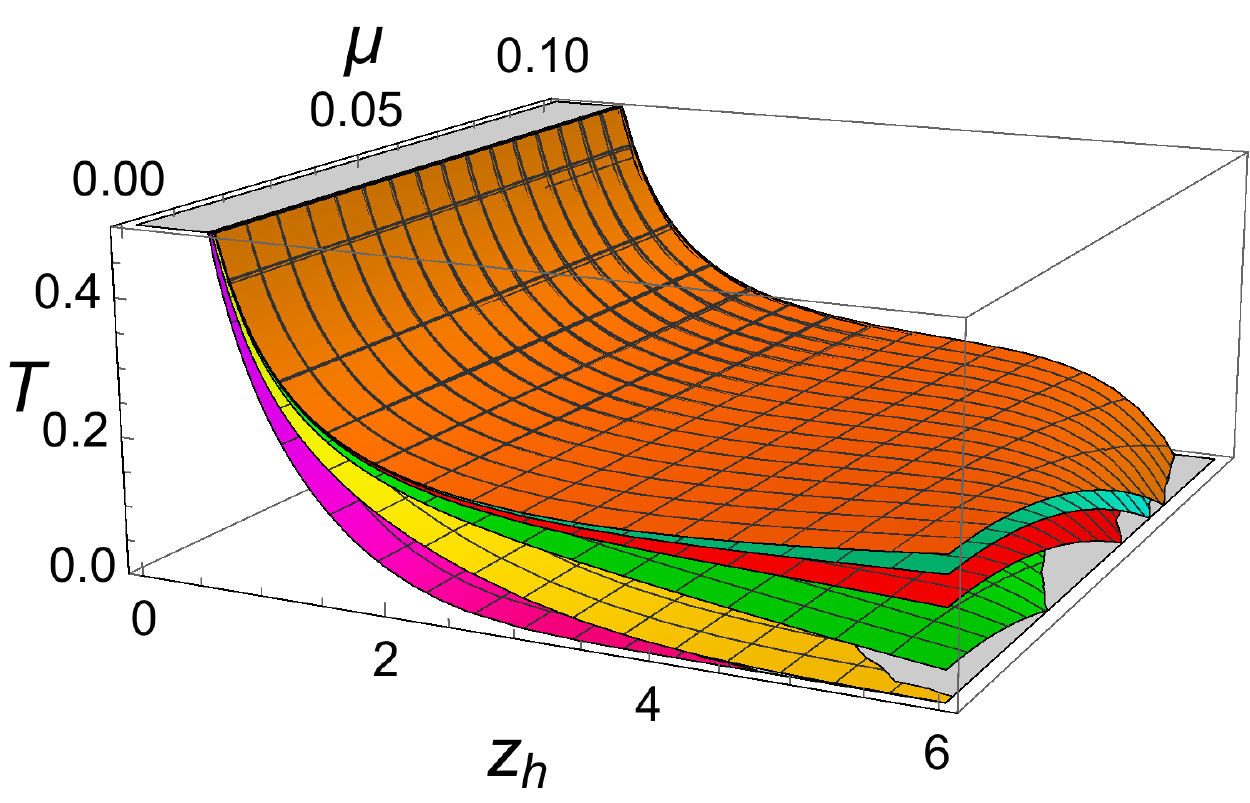} \quad
  \includegraphics[scale=0.5]{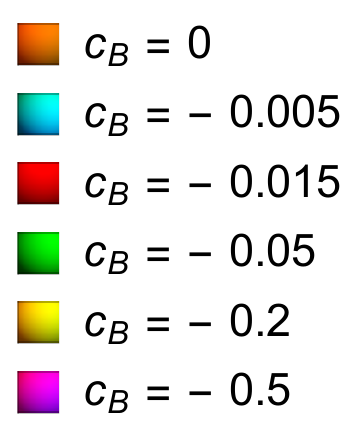} \quad
  \includegraphics[scale=0.45]{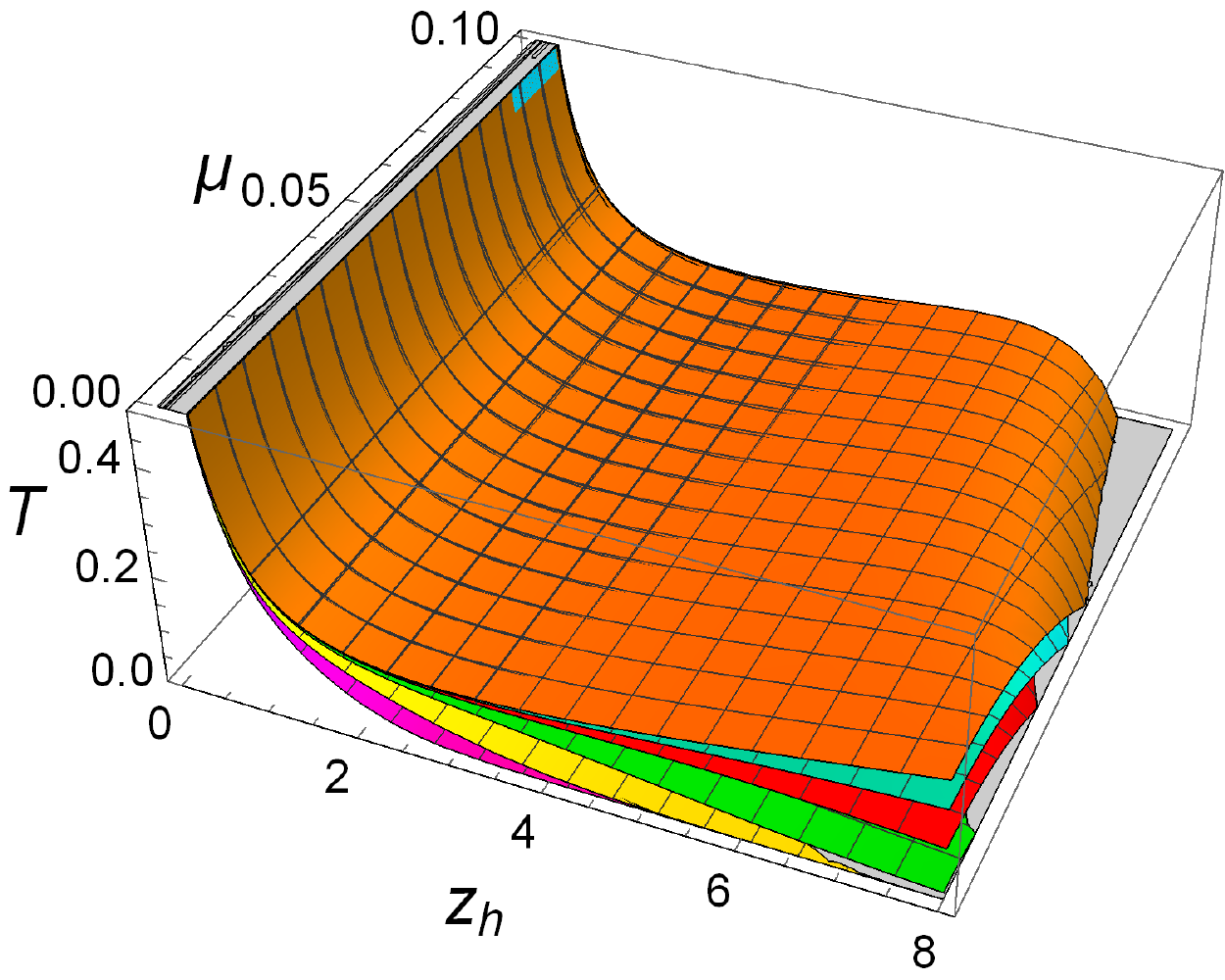}\\
  C \hspace{80 mm} D\\ \ \\
  \caption{Contour plots of temperature depending on $z_h$, $- \, c_B$
    and $\mu$ for $\nu = 1$ (A) and $\nu = 4.5$ (B). Dependence of
    temperature on $z_h$ and $\mu$ for different $c_B$ shown by
    different colors for $\nu = 1$ (C) $\nu = 4.5$ (D). 
  }
  \label{Fig:Temperature3D}
\end{figure}

\begin{figure}[t!]
  \centering
  \includegraphics[scale=0.45]{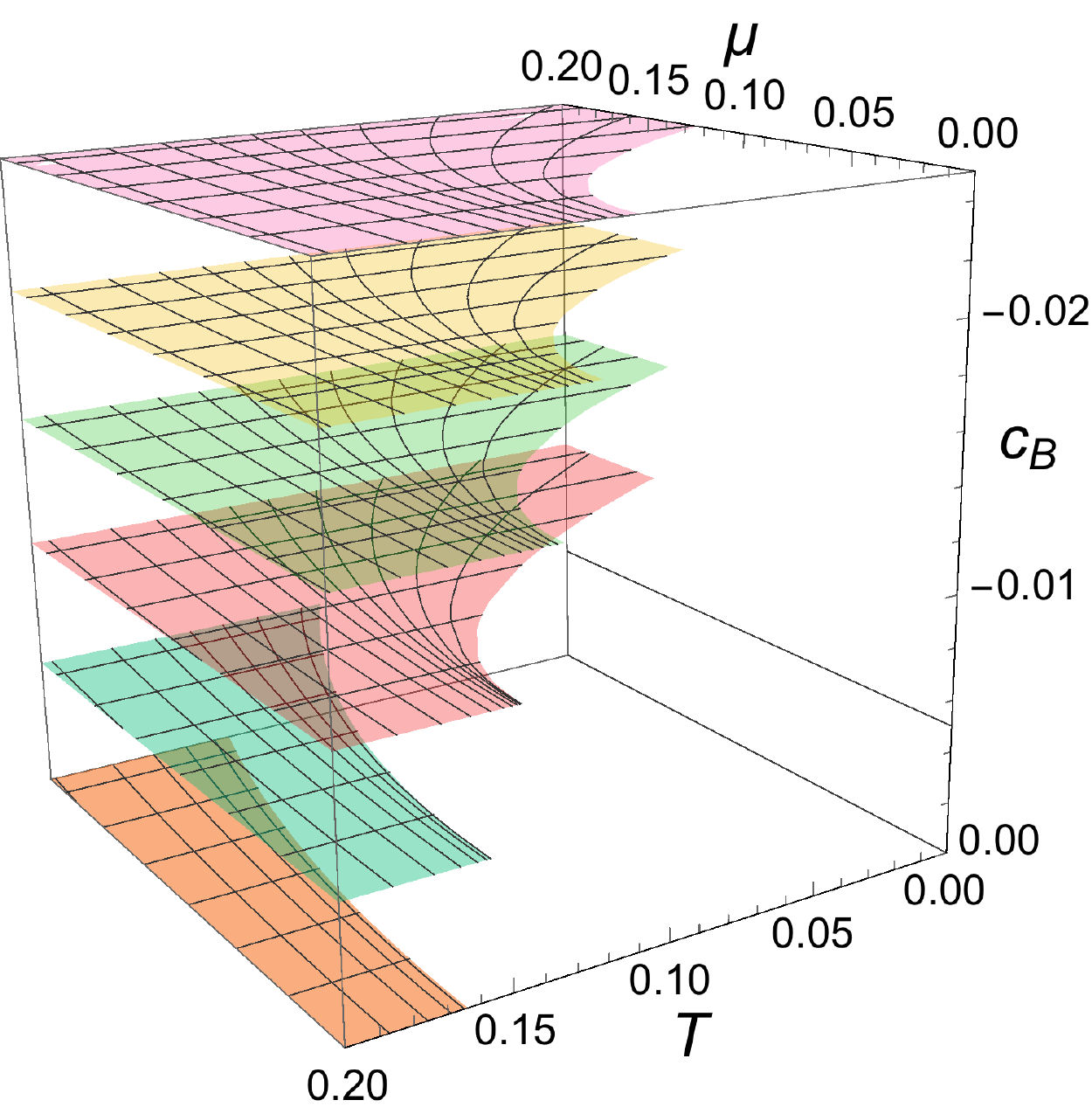} \quad
  \includegraphics[scale=0.5]{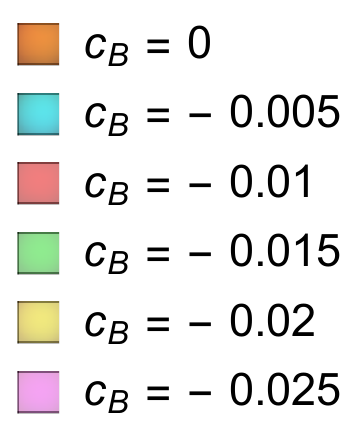} \quad
  \includegraphics[scale=0.45]{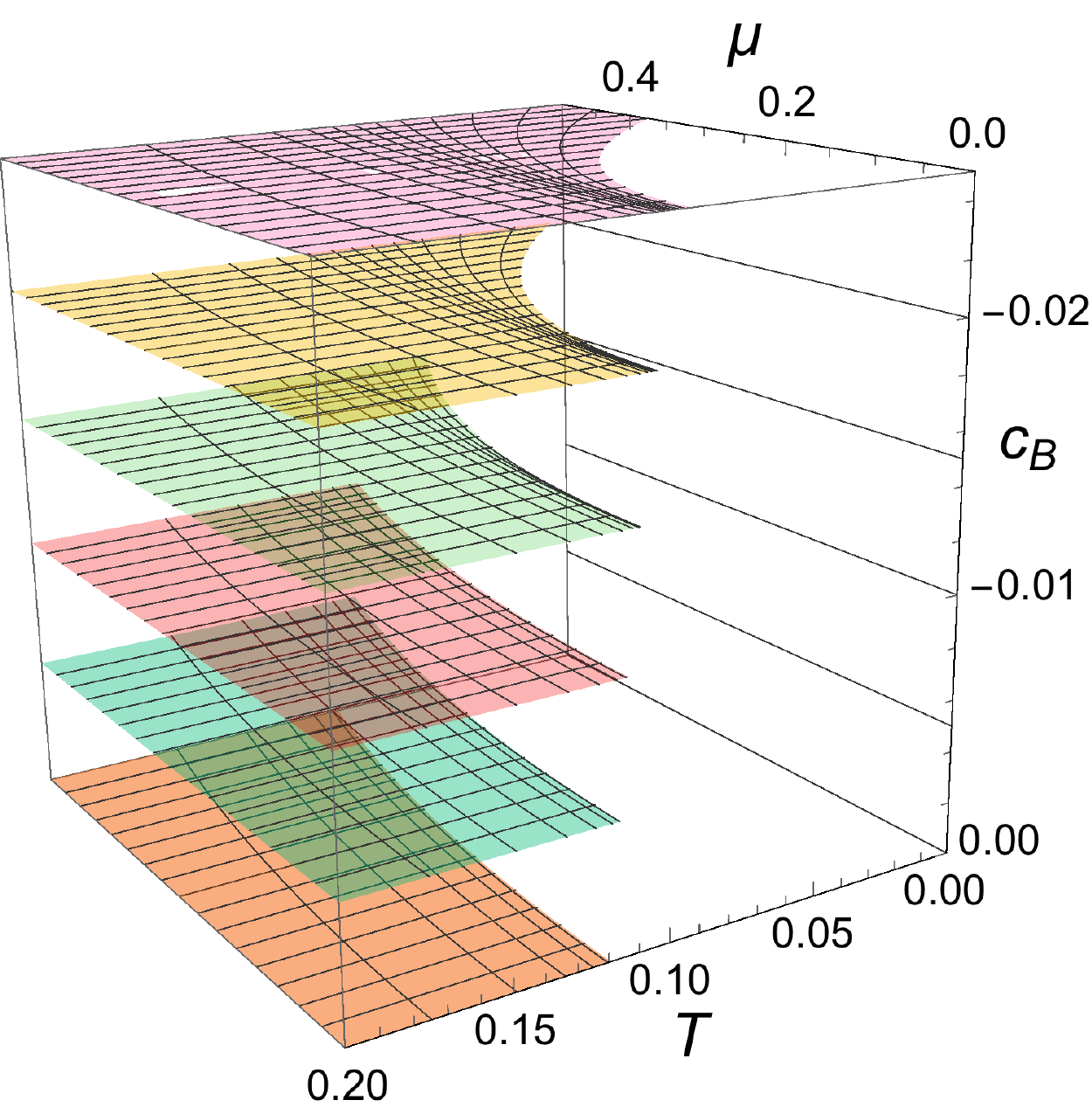}\\
  A \hspace{80mm} B
  \caption{Domains of physical parameters of AHM (anisotropic
    holographic matter). Dependence of the temperature on $z_h$ for
    $0 \le \mu \le 0.2$ and $- 0.025 \le c_B \le 0$ for $\nu = 1$ (A)
    and for $\nu = 4.5$ (B). Fixed $c_B$ values are shown by different
    colors. }
\label{Fig:3D-domain}
\end{figure}

On Fig.\ref{Fig:3D-domain} temperature as a function of $z_h$, $\mu$
and $c_B$ is presented. At small values temperature is rather
sensitive to chemical potential and magnetic field presence
(Fig.\ref{Fig:3D-domain}.A). The larger absolute value of $\mu$ or
$c_B$ leads to smaller $z_h$ for the fixed temperature. But their
influence significantly weakens with increasing temperature. Isotherm
surfaces for large $T$ are almost flat and parallel to
$\mu$ -- $c_B$ plane, i.e. they are determined mainly by horizon, so that
contribution of magnetic field and chemical potential are minor. For
$\nu = 4.5$ (Fig.\ref{Fig:3D-domain}.B) this process is faster than
for $\nu = 1$ as influence of $\mu$ and $c_B$ is weakened from the
start.

On Fig.\ref{Fig:3D-domain}C,D $T(z_h,\mu)$ for different $c_B$ and
$\nu = 1, \, 4.5$ is presented. Here we see the effects discussed
above in relation to Fig.\ref{Fig:Temperature}, \ref{Fig:Tnu2}. Each
colored surface corresponds to some $c_B$ value, and the surfaces
corresponding to larger absolute values of $c_B$ lie below. For $\nu =
1$ (Fig.\ref{Fig:3D-domain}C) surface for fixed $c_B$ is flatter than
for $\nu = 4.5$ (Fig.\ref{Fig:3D-domain}D). 

On Fig.\ref{Fig:3D-domain} possible temperature values depending on
magnetic field influence and chemical potential are shown. Planes
corresponding to some fixed (descrete) values of $c_B$ are highlighted
as levels for clarity. Empty spaces mean regions with unattainable
temperature values.

\newpage
\section{Numerical Results}\label{Sect:solution}

The main ingredient of our calculations is a finding position of the
dynamical wall for effective potentials. The effective potentials
depend on the orientation. In particular cases for orientations
$xY_1$, $xY_2$ and $y_1Y_2$ the effective potentials are given by
equations  \eqref{sigmaxY1}, \eqref{sigmaxY2} and \eqref{sigmayY2}. We
denote the corresponding potentials as ${\cal V}_1$,  ${\cal V}_2$ and
${\cal V}_3$ for simplicity. In the next subsections we present the
forms of these potentials. The effective potentials depend on $c$,
$z_0$, $z_h$ and  $c_B$. Note that effective potential does not depend
on $\mu$.

\subsection{Wilson loop $W_{xY_{1}}$}

\begin{figure}[h!]
  \centering
  \includegraphics[scale=0.46]{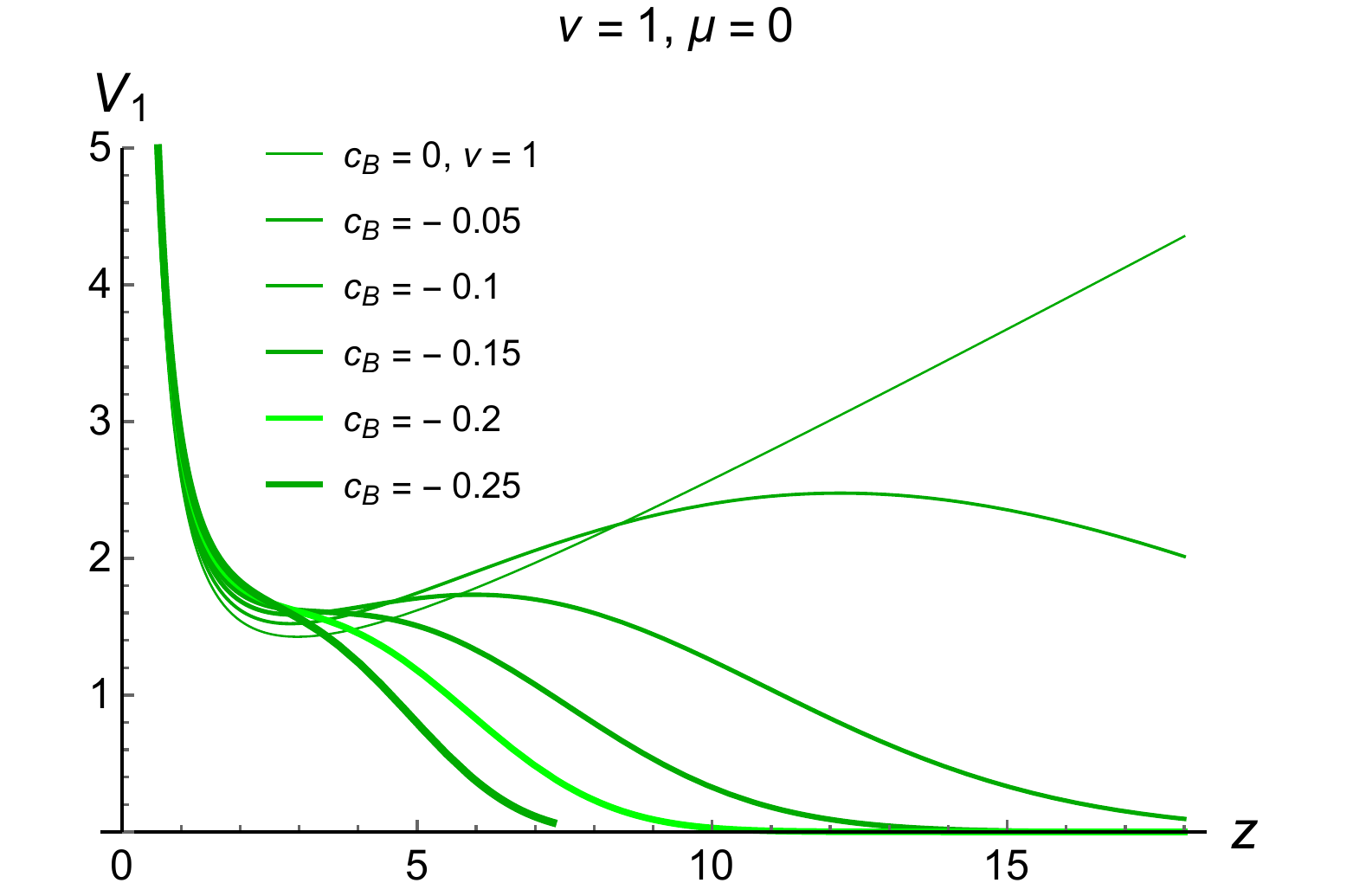}
  \includegraphics[scale=0.46]{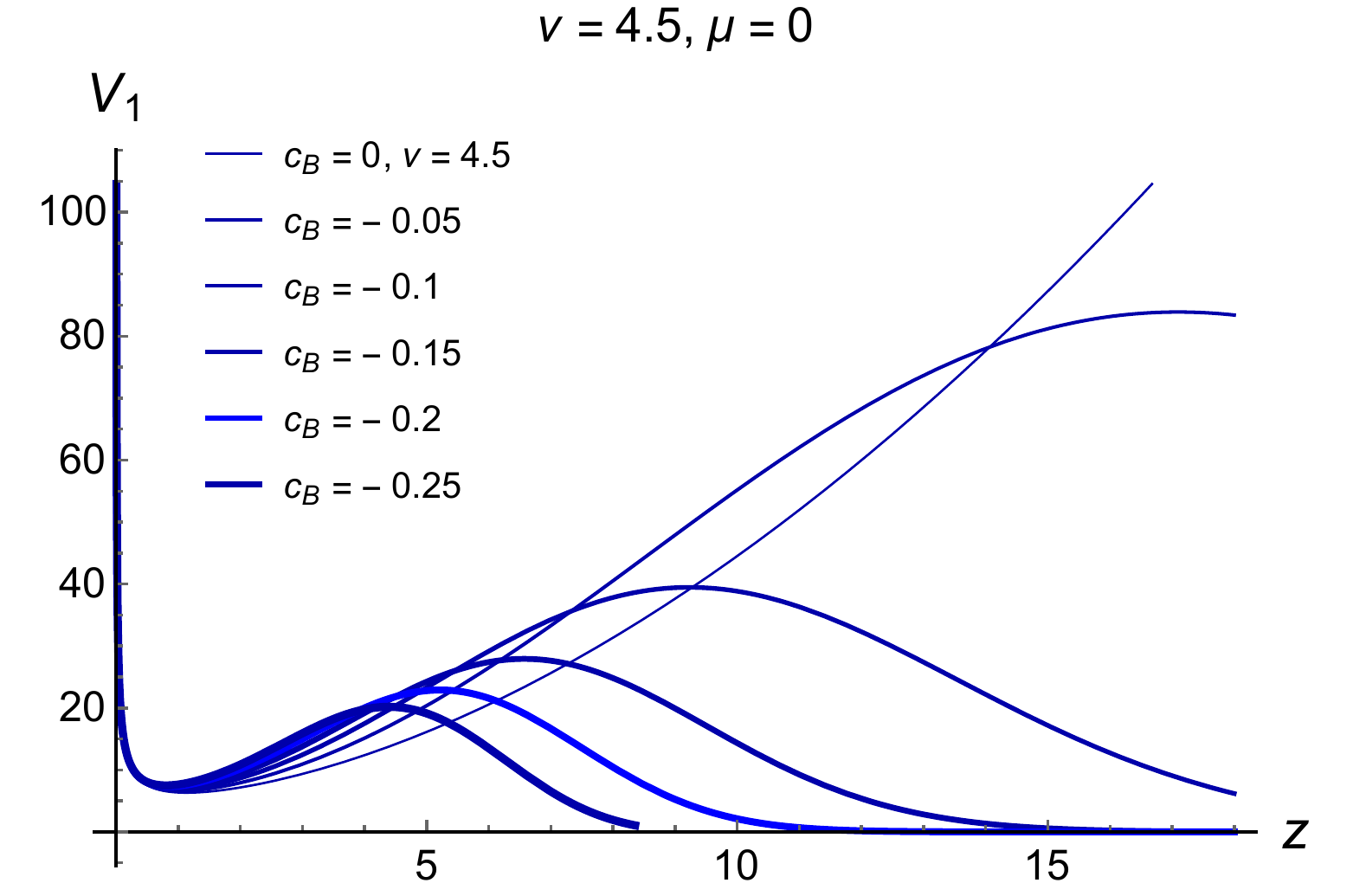} \\
  A \hspace{70mm} B
  \caption{The dependence of the effective potential ${\cal V}_1$ on
    the magnetic field for $\nu = 1$ (A) and $\nu = 4.5$ (B).
  }
  \label{Fig:V1}
\end{figure}

The behavior of the effective potential ${\cal V}_1$ that corresponds
to the Wilson loop $W_{xY_{1}}$ on the magnetic field is presented on
Fig.\ref{Fig:V1}.

We see that ${\cal V}_1$ has local minimums, i.e. has dynamical walls
where ${\cal V}'(z) = 0$ for $0 < c_B < c_{B,cr}$. At $c_{B,cr} = 0.2$
for $\nu = 1$ the effective potential has an inflection point and for
$c_B > c_{B, cr}$ the dynamical walls disappear. Locations of the
dynamical walls depends on the value of $c_B$ and are given in the
Tables~\ref{tab:V1nu1} and~\ref{tab:V1nu45}.

\begin{table}[b!]
\centering
\begin{tabular}{|l|c|l|l|l|}
\hline
\multicolumn{1}{|c|}{${\cal V}_1$} & \multicolumn{4}{c|}{$\nu=1$} \\ \hline
$-\,c_B$ & 0 & 0.05 & 0.1 & 0.138 \\ \hline
$z_{DW}$ & \multicolumn{1}{l|}{2.969} & 2.899 & 2.849 & 3.579 \\ \hline
\end{tabular}
\caption{Locations of DW for ${\cal V}_1$ at $\nu = 1$.}
\label{tab:V1nu1}
\end{table}

\begin{table}[h]
\centering
\begin{tabular}{|l|c|l|l|l|l|l|l|}
\hline
\multicolumn{1}{|c|}{${\cal V}_1$} & \multicolumn{7}{c|}{$\nu=4.5$} \\ \hline
$-\,c_B$ & 0 & 0.02 & \multicolumn{1}{c|}{0.05} & \multicolumn{1}{c|}{0.1} & \multicolumn{1}{c|}{0.15} & \multicolumn{1}{c|}{0.2} & \multicolumn{1}{c|}{0.25} \\ \hline
$z_{DW}$ & \multicolumn{1}{l|}{1.131} & 1.083 & 1.031 & 0.952 & 0.896 & 0.841 & 0.798 \\ \hline
\end{tabular}
\caption{Locations of DW for ${\cal V}_1$ at $\nu = 4.5$.}
\label{tab:V1nu45}
\end{table}

\begin{figure}[h!]
  \centering
  \includegraphics[scale=0.35]{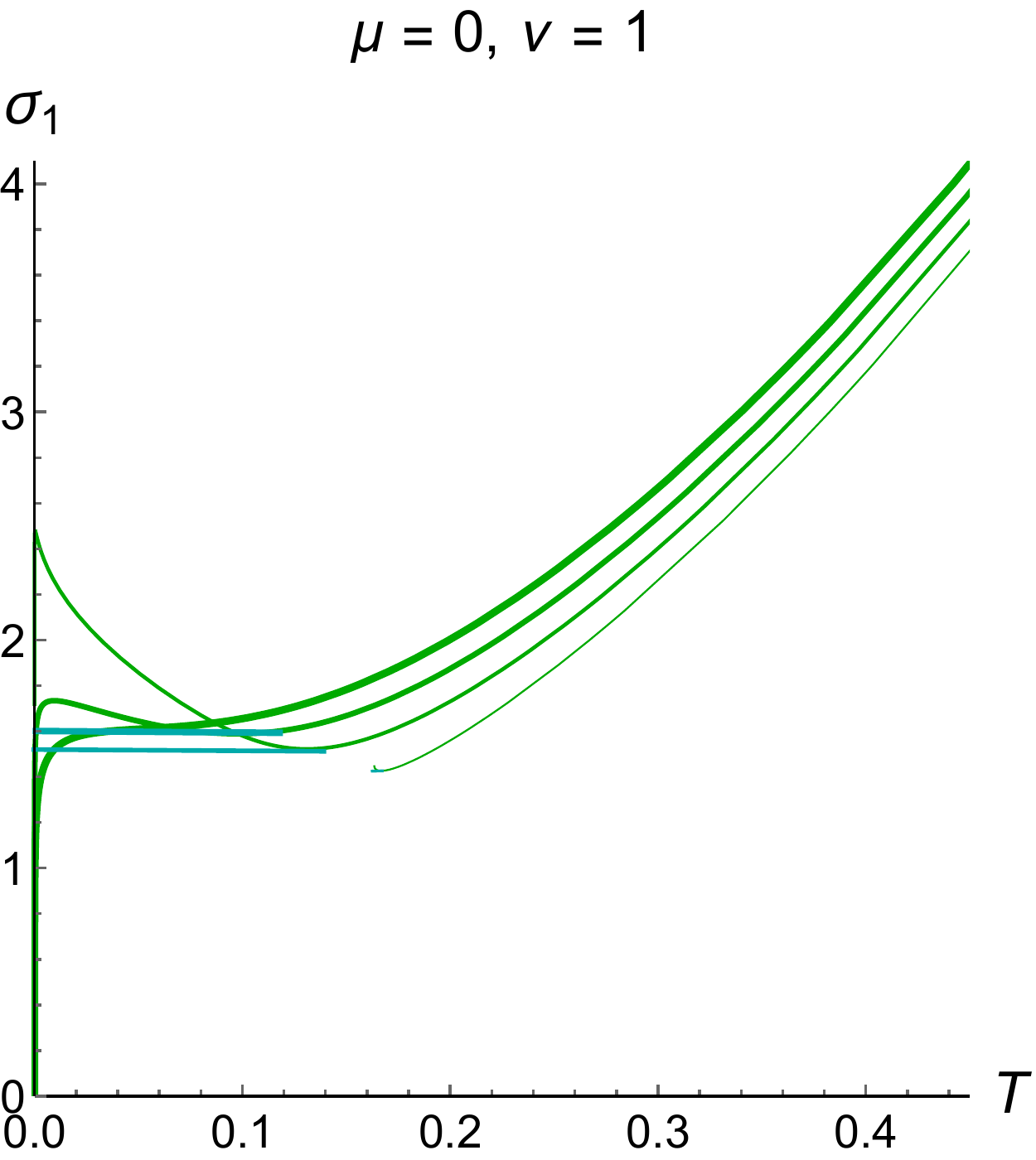}\quad
  \includegraphics[scale=0.35]{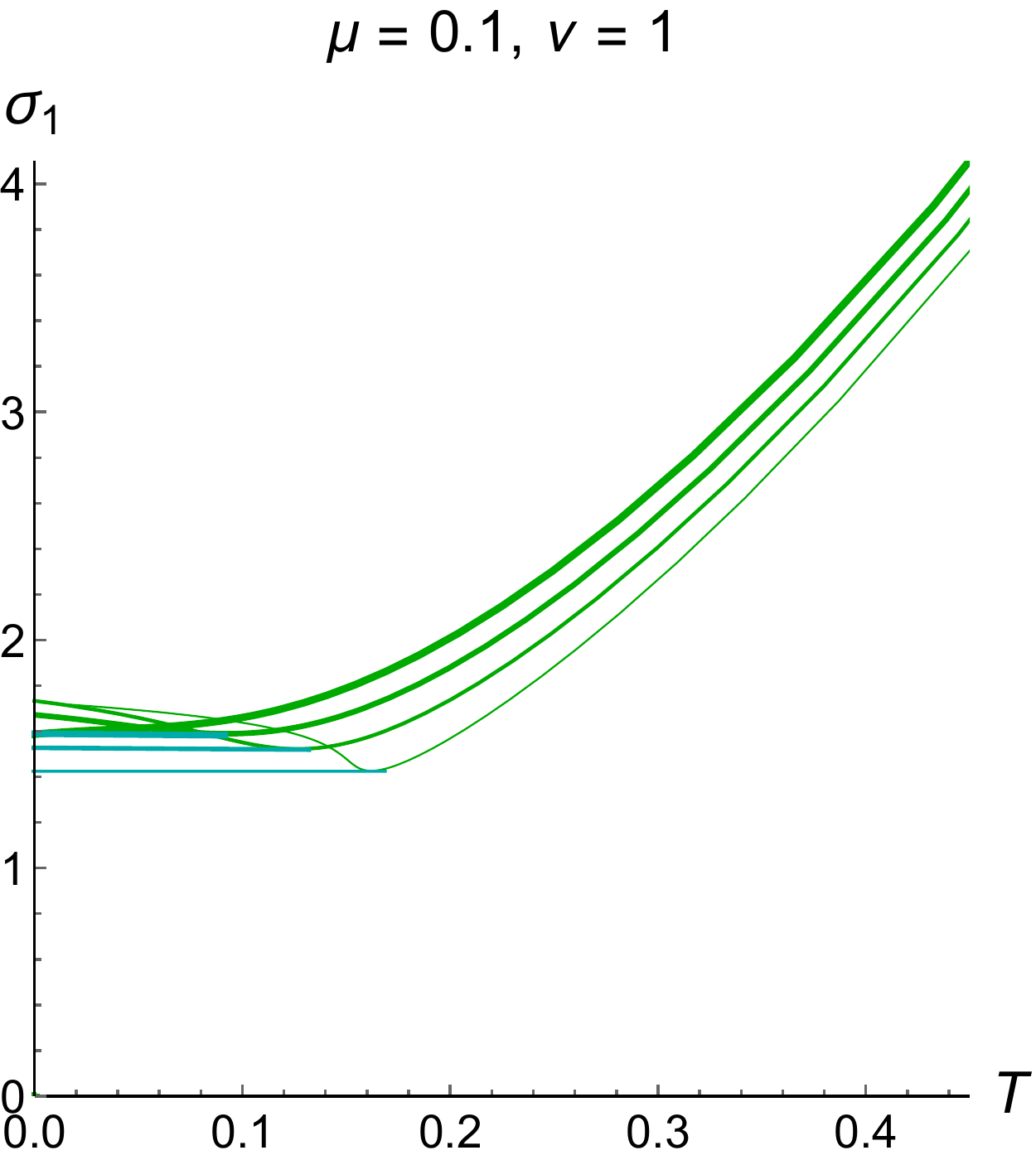}\qquad
  \includegraphics[scale=0.35]{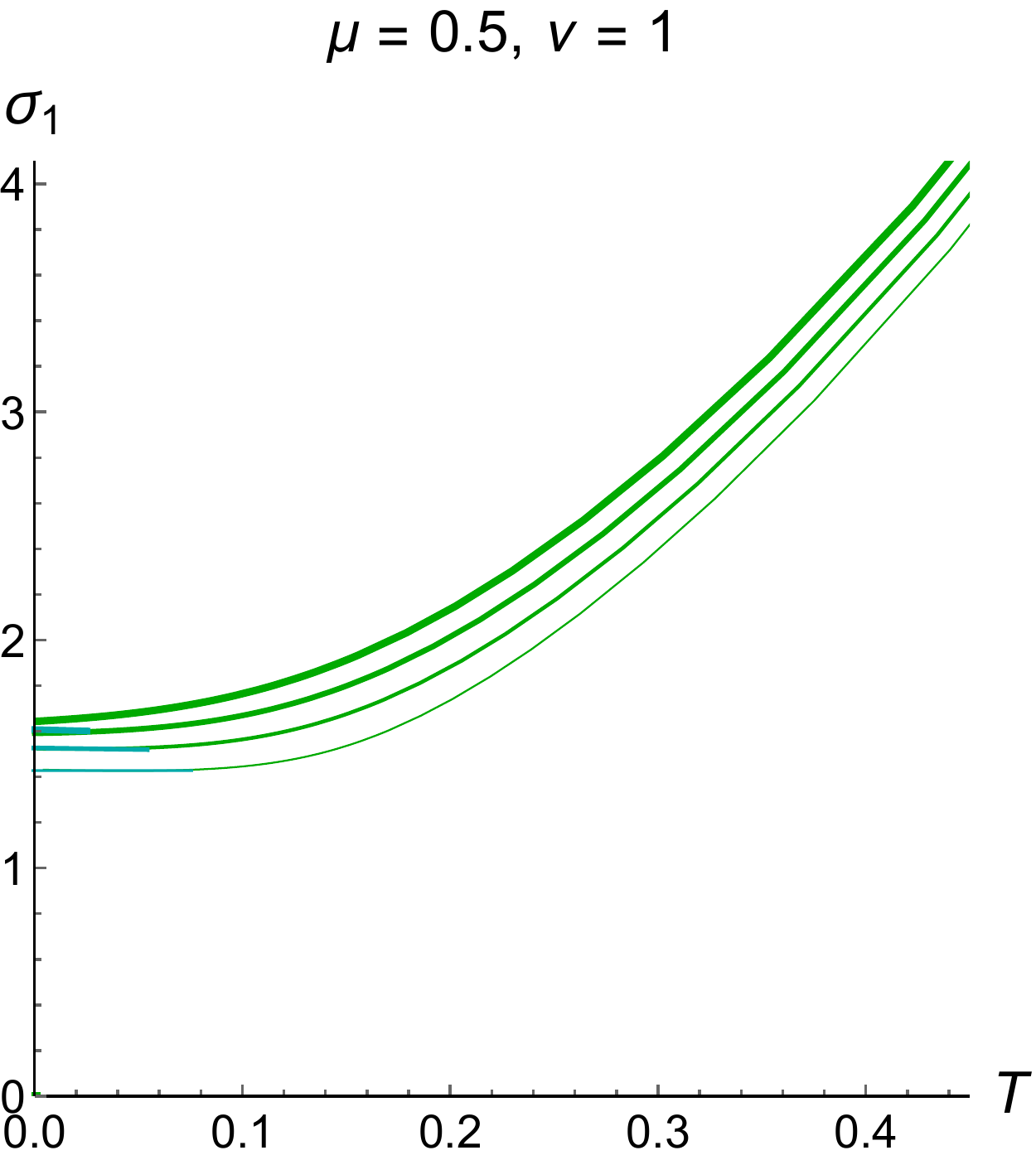}\\
  A \hspace{40 mm} B \hspace{40 mm} C
  \includegraphics[scale=0.5]{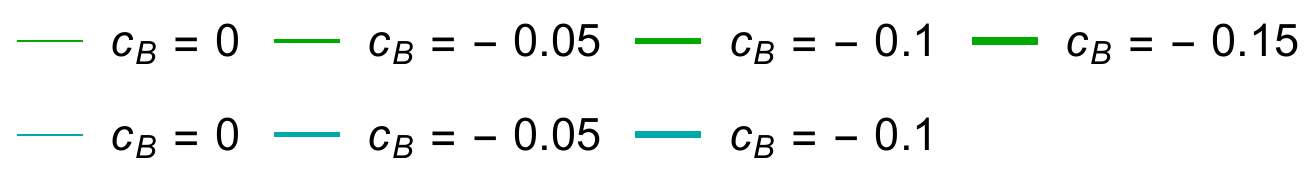}
  \caption{Dependence of $\sigma_1$ on temperature for different
    values of $c_B$ for $\mu = 0$ (A), $\mu = 0.1$ (B) and $\mu = 0.5$
    (C); $\nu = 1$.
  }
  \label{Fig:sigma-1-nu1}
\end{figure}

\begin{figure}[h!]
  \centering
  \includegraphics[scale=0.3]{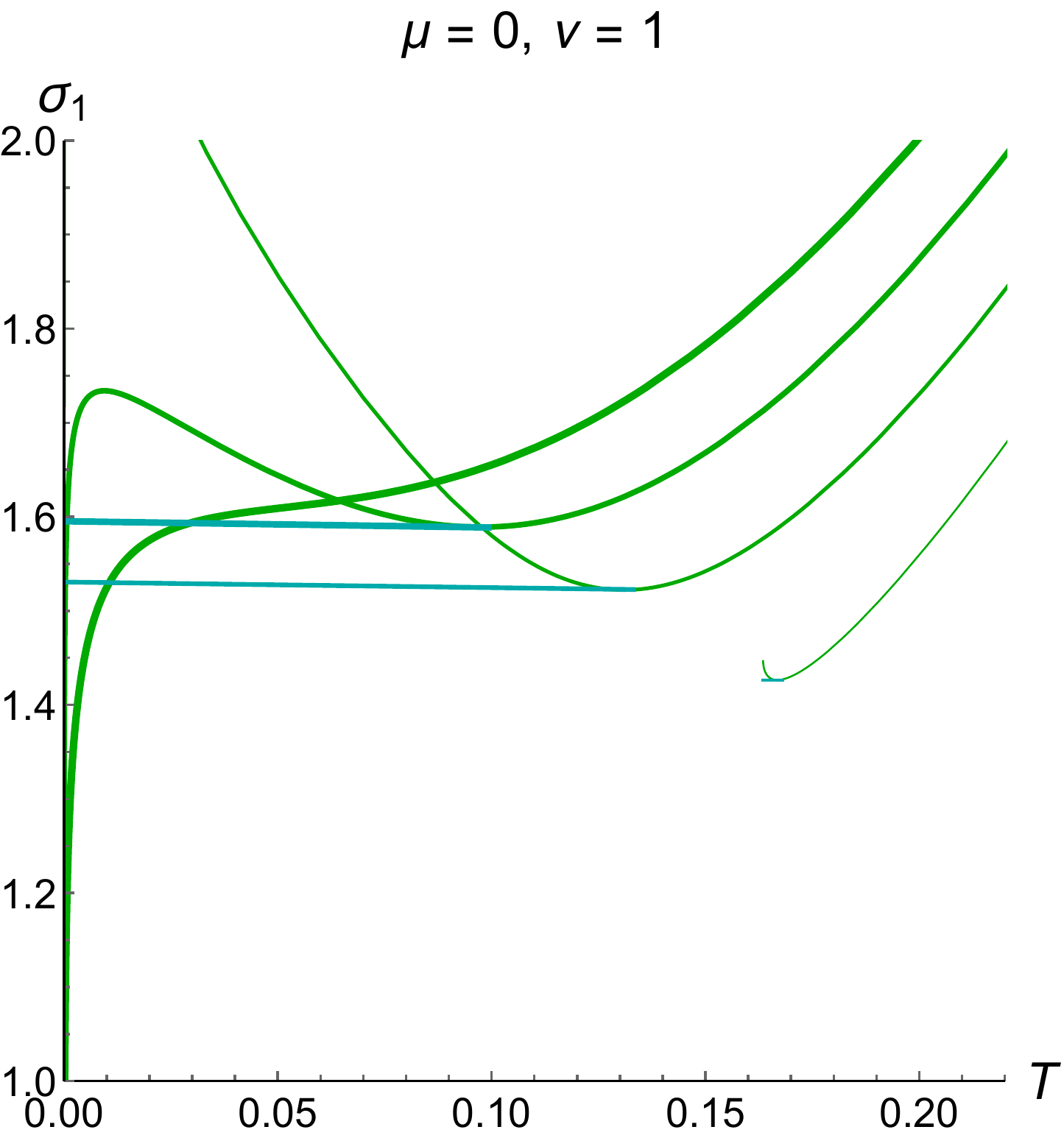}\quad
  \includegraphics[scale=0.3]{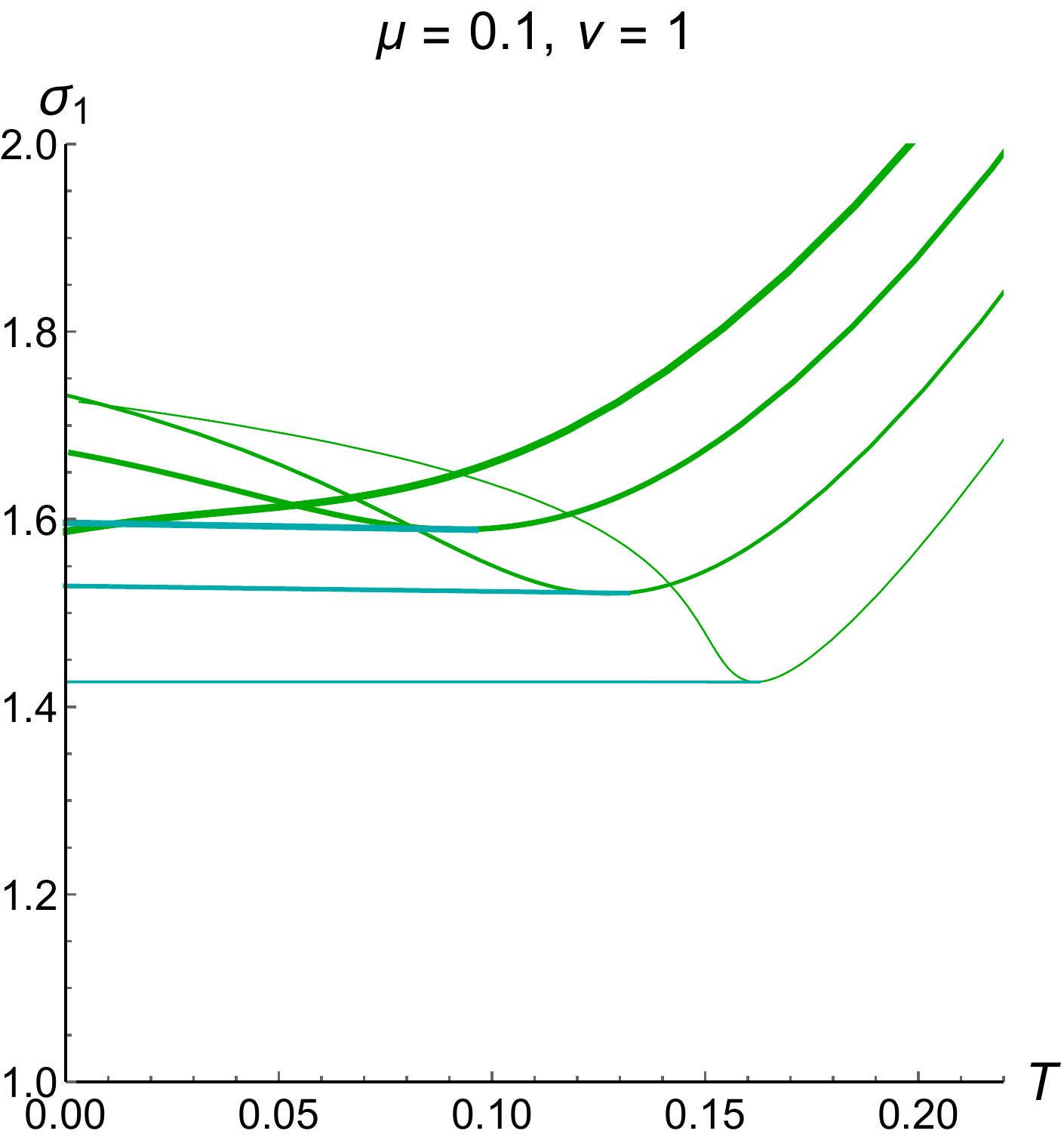}\quad
  \includegraphics[scale=0.3]{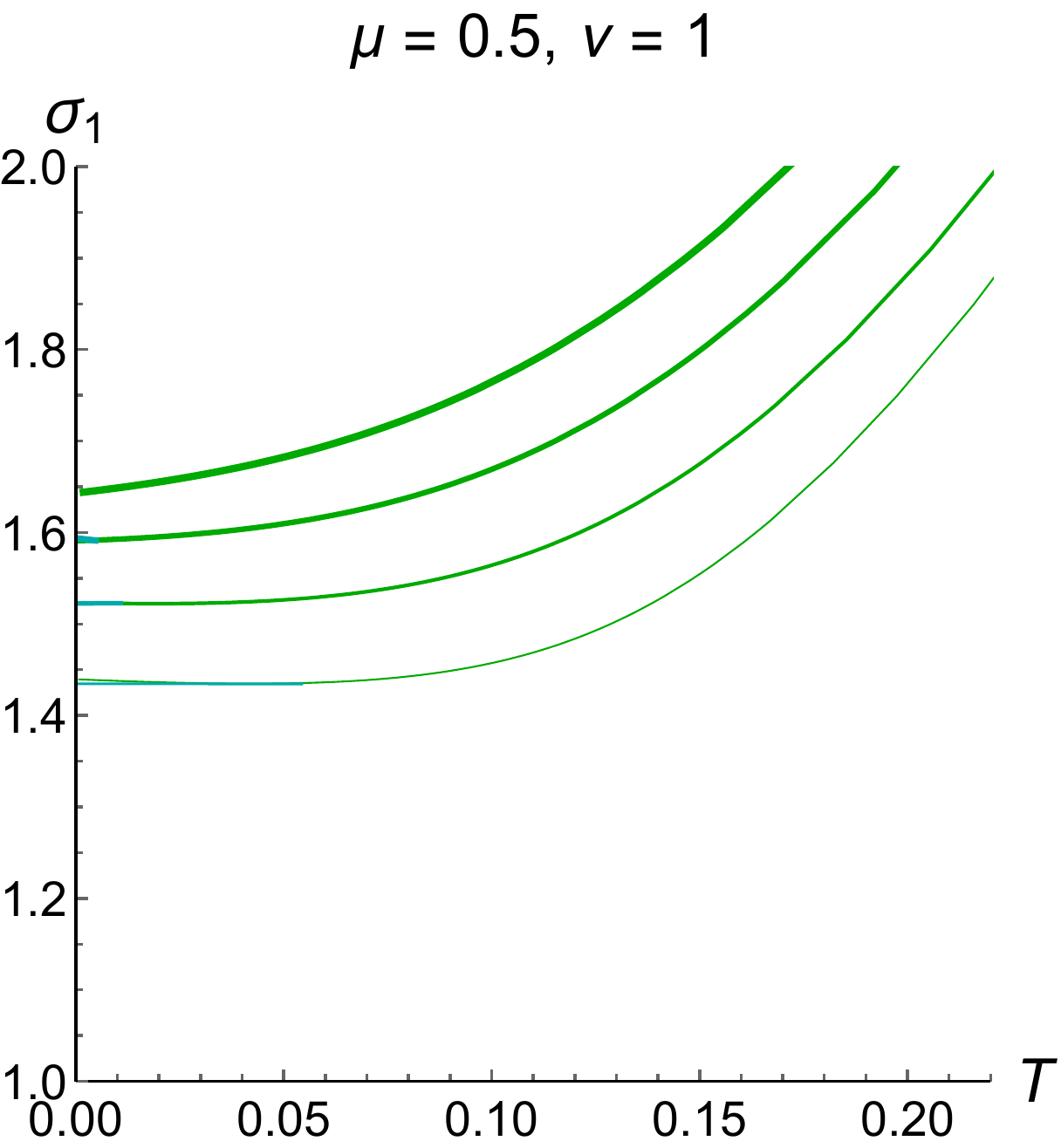}\\
  A \hspace{40 mm} B \hspace{40 mm} C \\ \ \\
  \includegraphics[scale=0.7]{Plots/sigma1-nu1-leg.pdf}
  \caption{Dependence of $\sigma_1$ from Fig.\ref{Fig:sigma-1-nu1} in
    larger scale. Plot legends are the same for all plots.
  }
  \label{Fig:sigma-1-nu1-large}
\end{figure}

\begin{figure}[h!]
  \centering
  \includegraphics[scale=0.35]{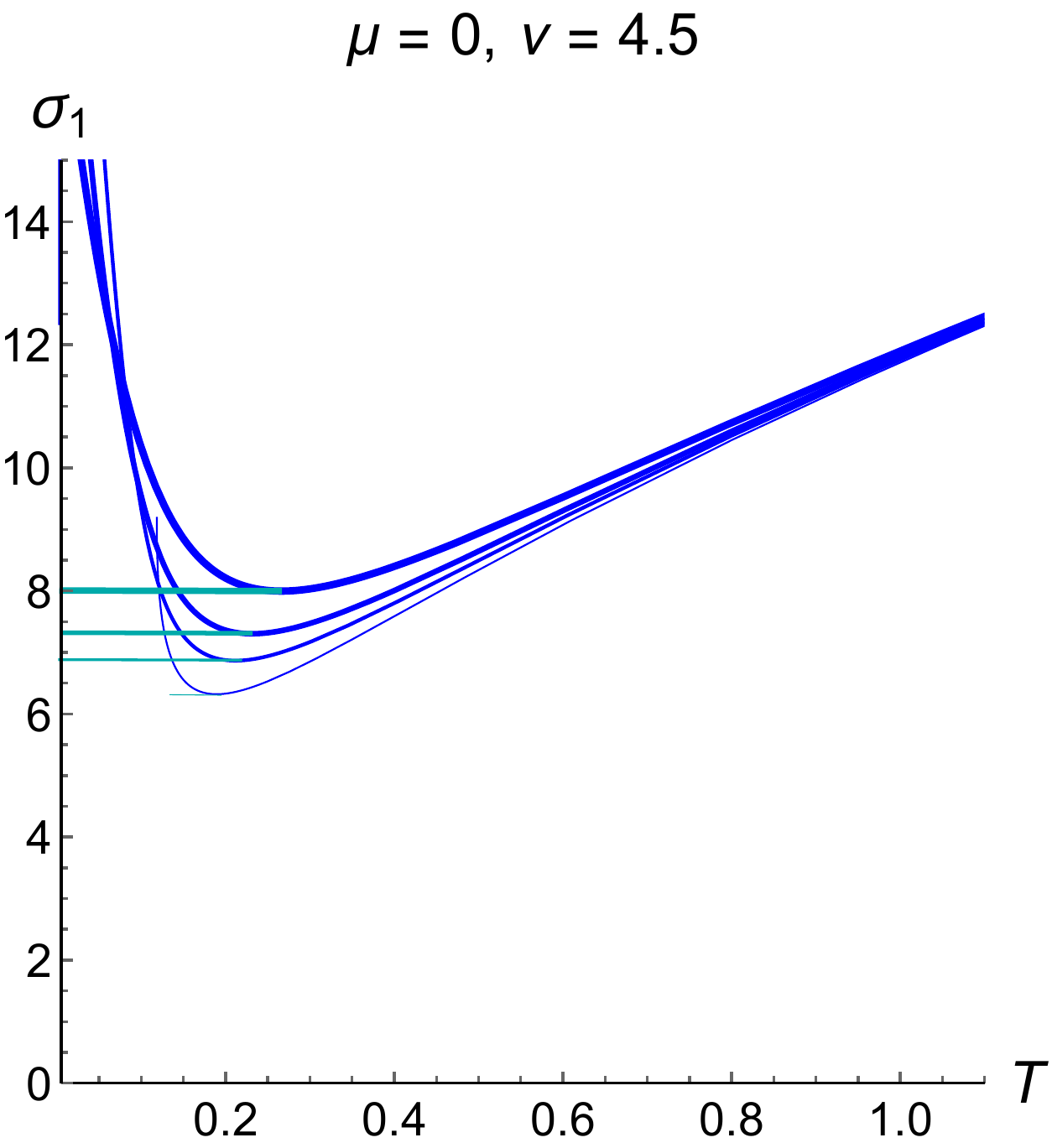}\quad
  \includegraphics[scale=0.35]{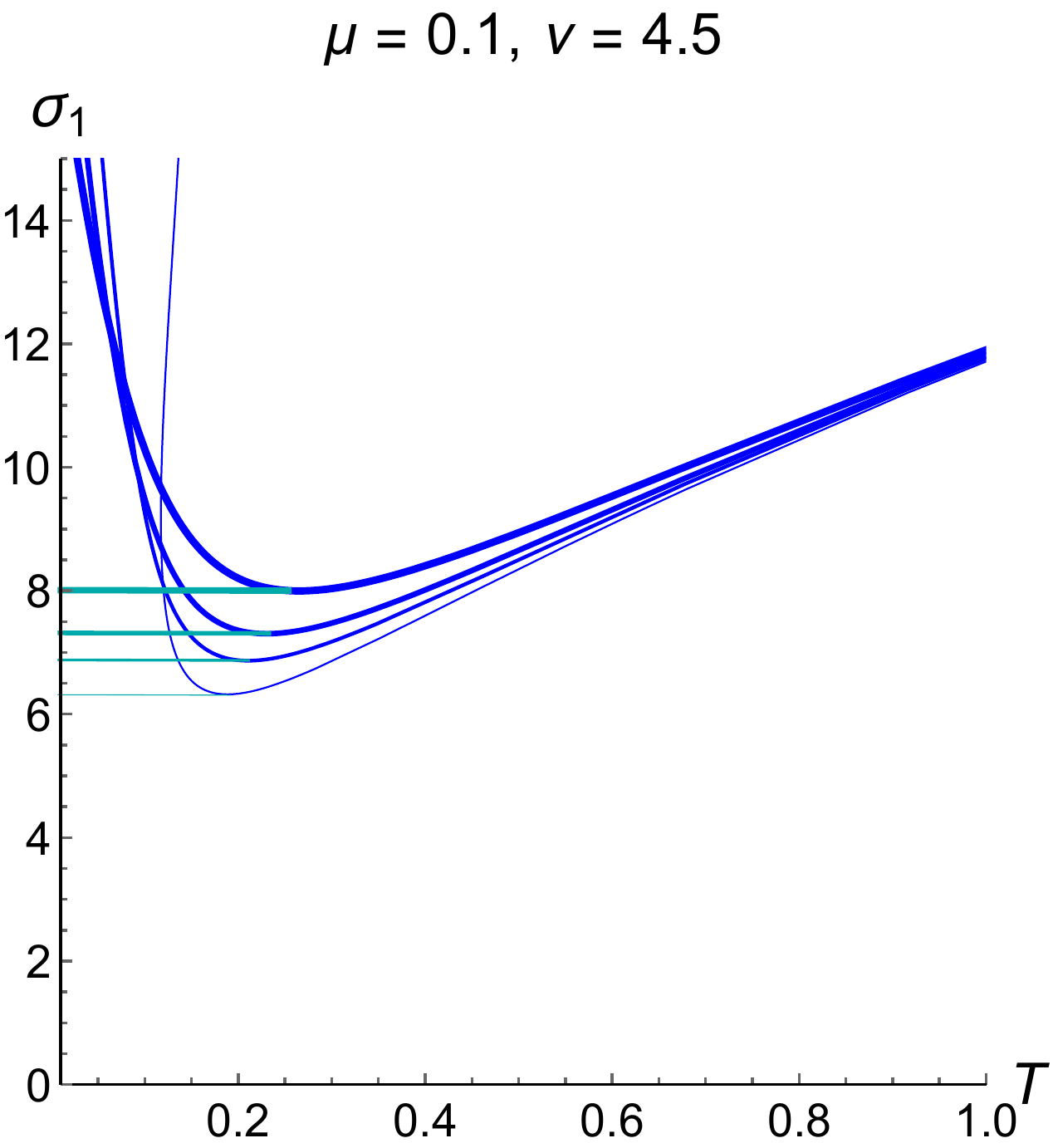}\qquad
  \includegraphics[scale=0.35]{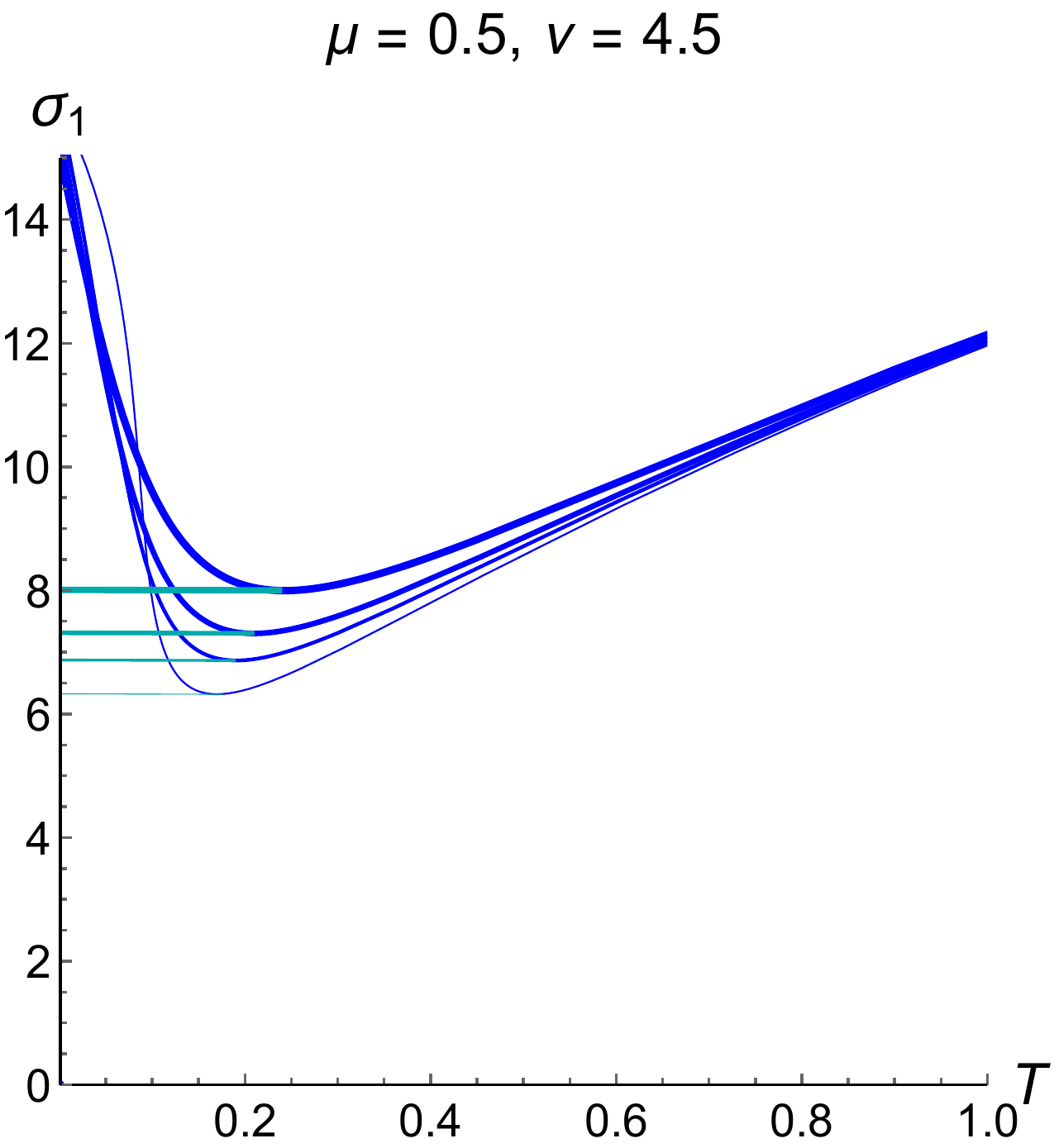}\\
  A \hspace{40 mm} B \hspace{40 mm} C
  \caption{Dependence of $\sigma_1$ on temperature and magnetic field
    for $\mu = 0$ (A), $\mu = 0.1$ (B) and $\mu = 0.5$ (C); $\nu =
    4.5$.
  }
  \label{Fig:sigma-1-nu45}
\end{figure}

\begin{figure}[h!]
  \centering
  \includegraphics[scale=0.32]{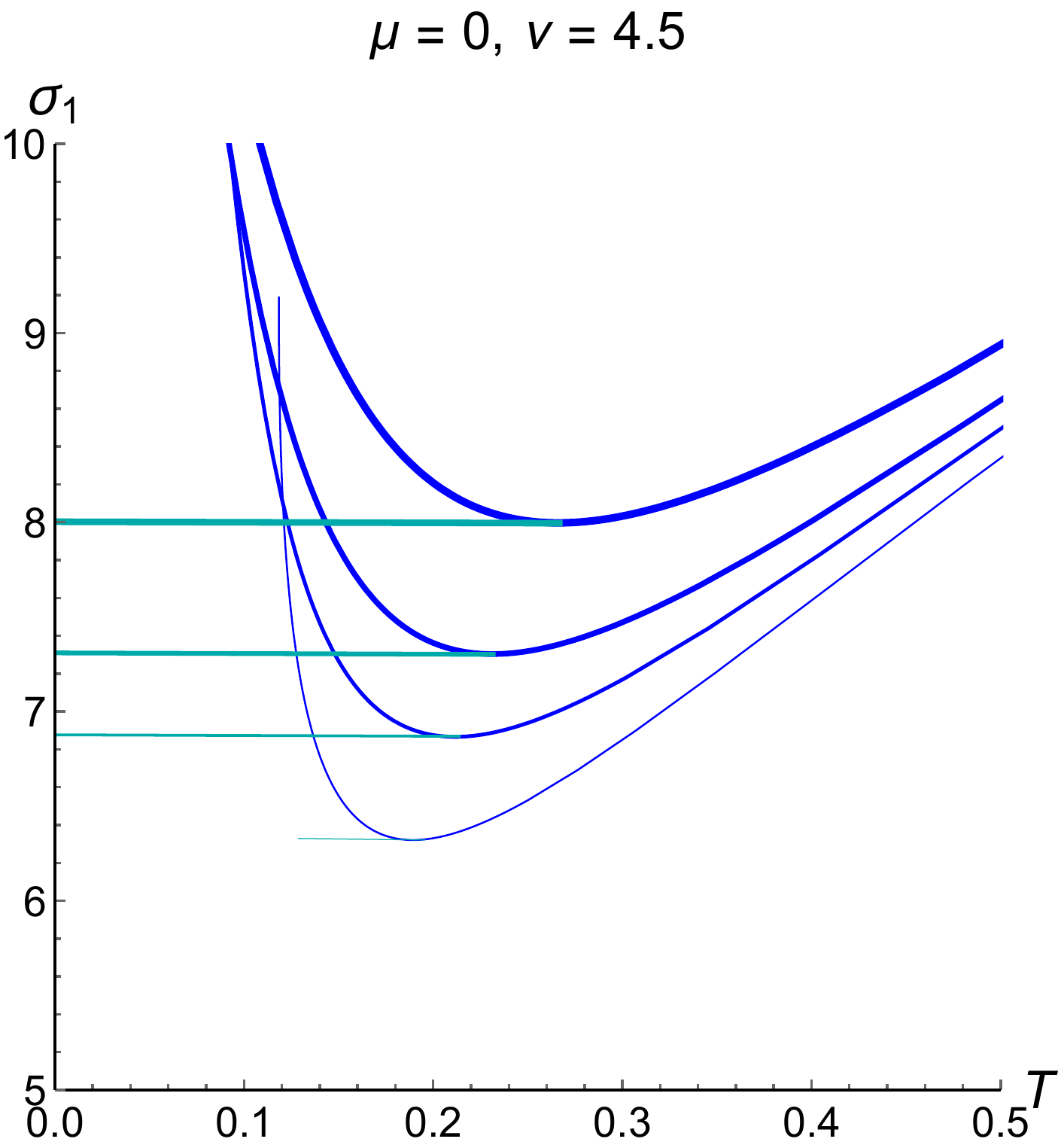}\quad
  \includegraphics[scale=0.35]{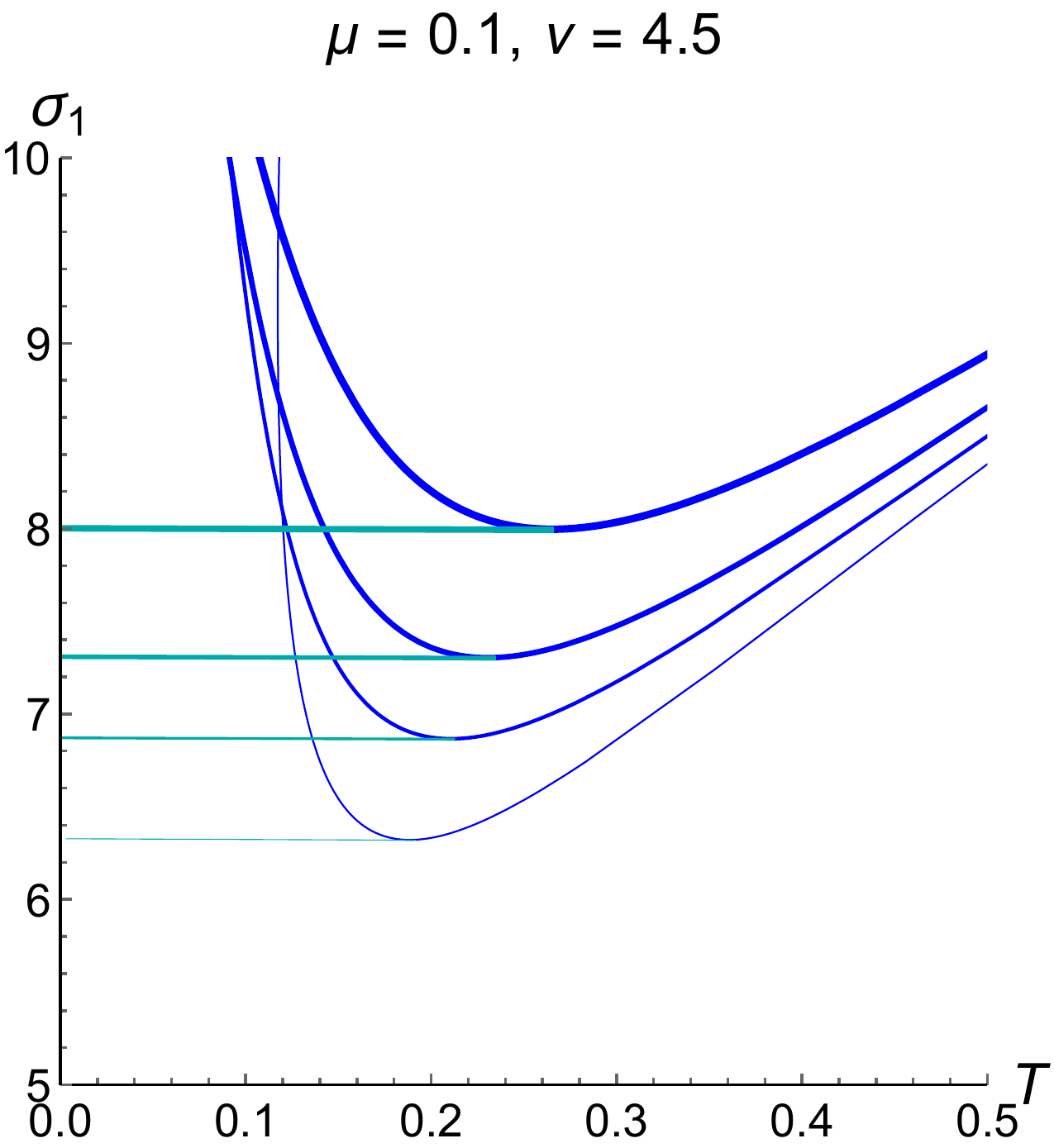}\quad
  \includegraphics[scale=0.37]{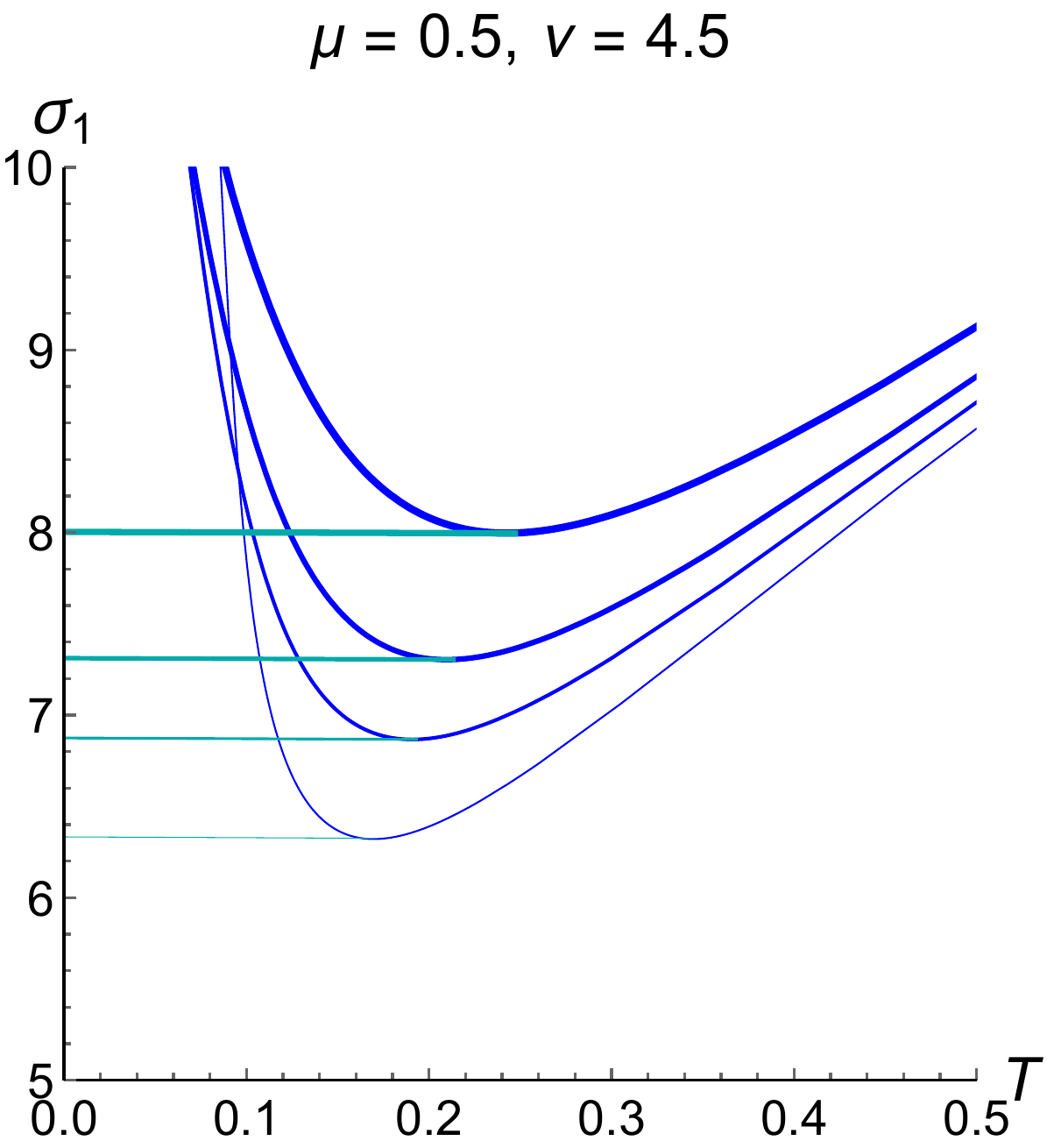}\\
  A \hspace{40 mm} B \hspace{40 mm} C \\ \ \\
  \includegraphics[scale=0.8]{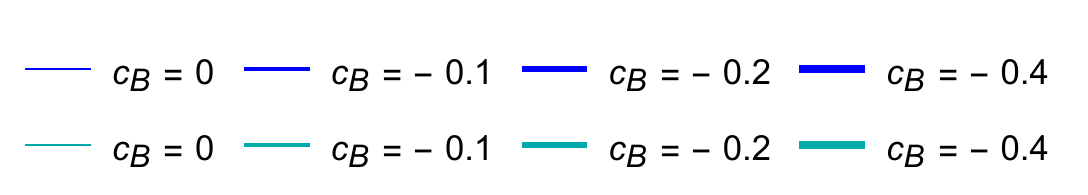}
  \caption{Dependence of $\sigma_1$ from Fig.\ref{Fig:sigma-1-nu45} in
    larger scale. Plot legends are the same for all plots.
  }
  \label{Fig:sigma-1-nu45-large}
\end{figure}

\begin{figure}[h!]
  \centering
  \includegraphics[scale=0.45]{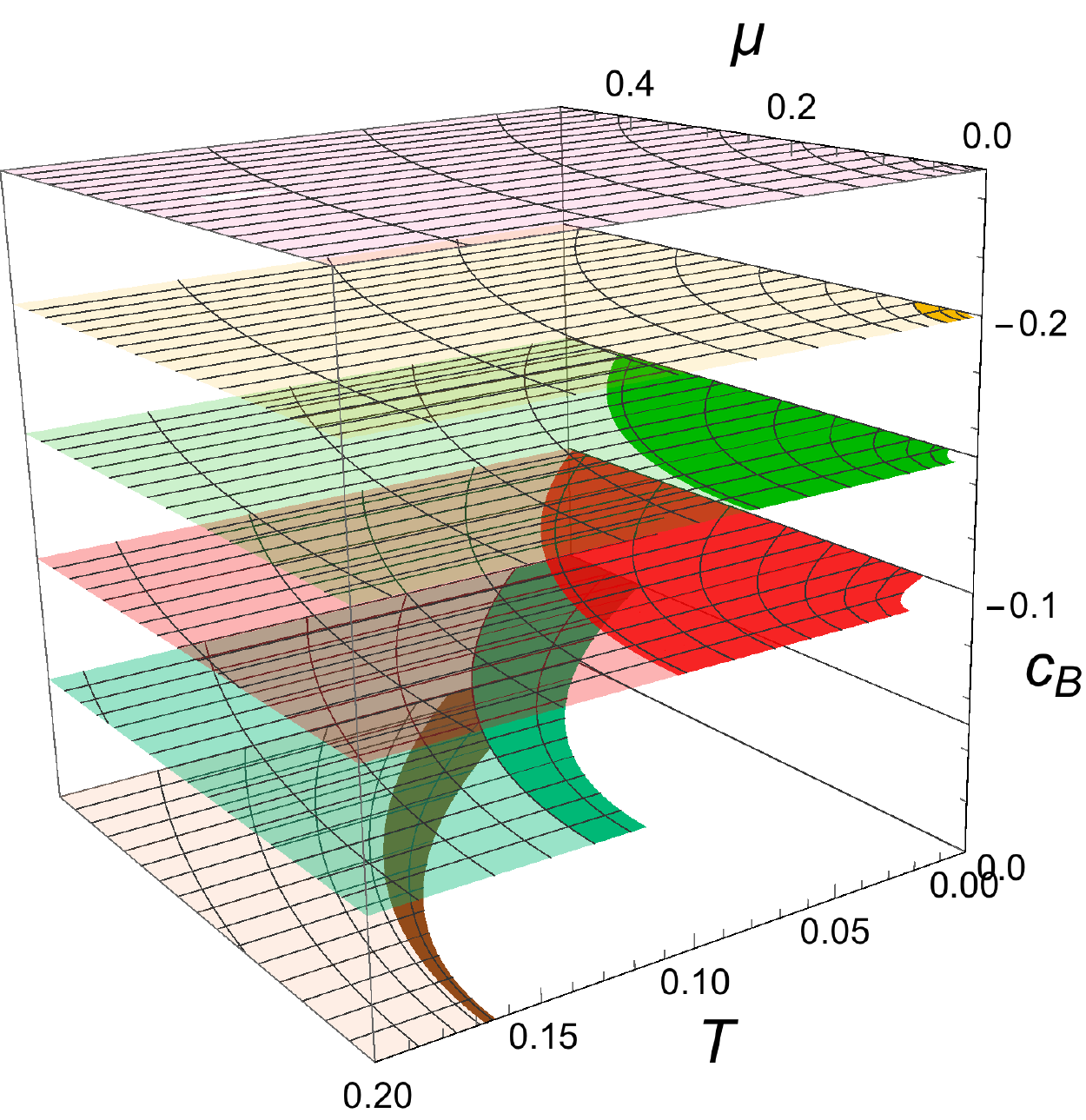} \qquad
  \includegraphics[scale=0.45]{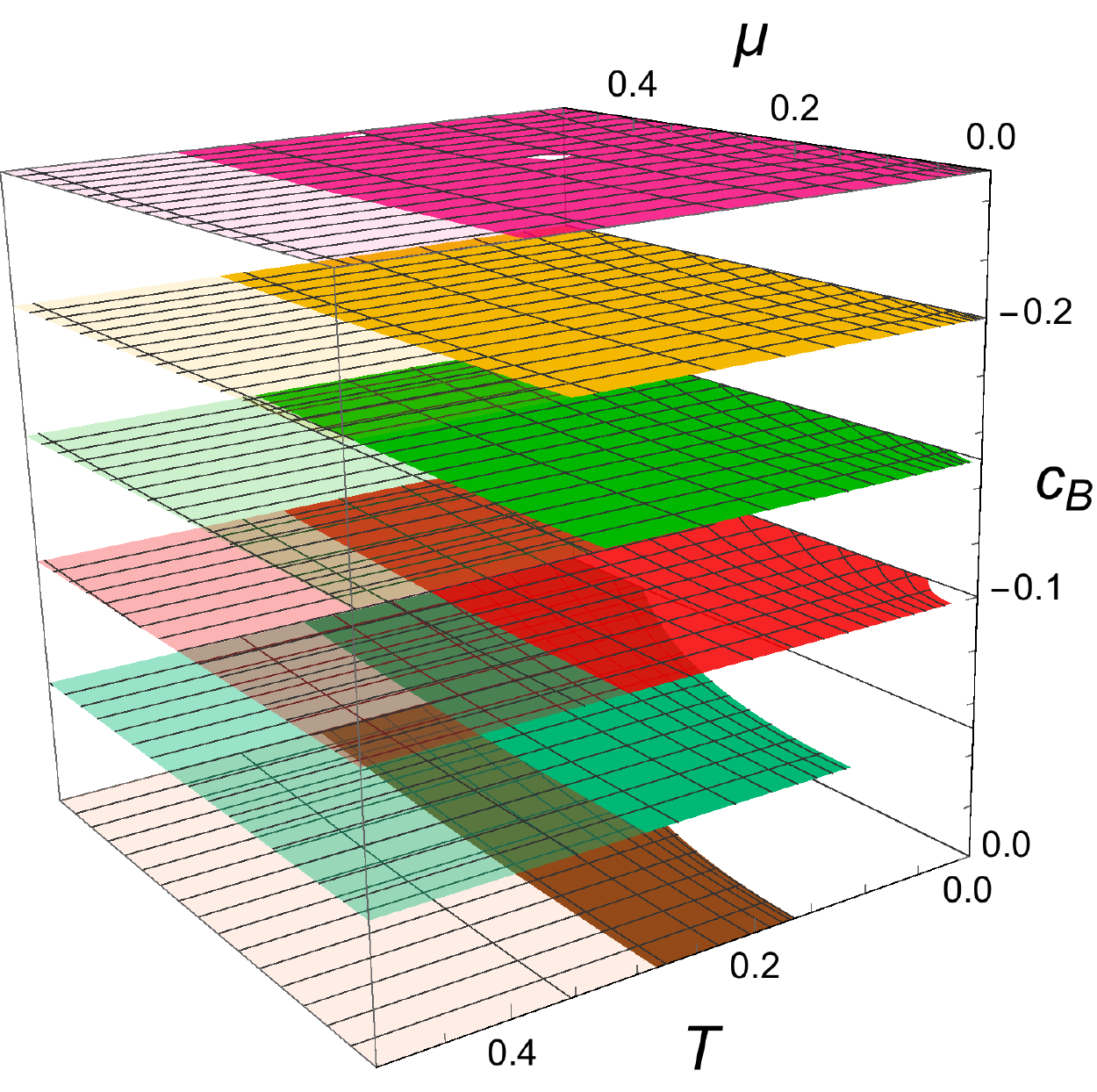} \\
  A \hspace{70 mm} B
  \caption{Transparent surfaces show the configurations touching the
    horizon, and less transparent ones -- the configurations touching
    the dynamic wall for $\nu = 1$ (A) and $\nu = 4.5$ (B). The phase
    transition corresponds to the transition from the DW configuration
    to the horizon configuration and is located at boundaries between
    light and dark surfaces. 
  }
  \label{Fig:PT-V1-surfaces-1}
\end{figure}

\begin{figure}[h!]
  \centering
  \includegraphics[scale=0.45]{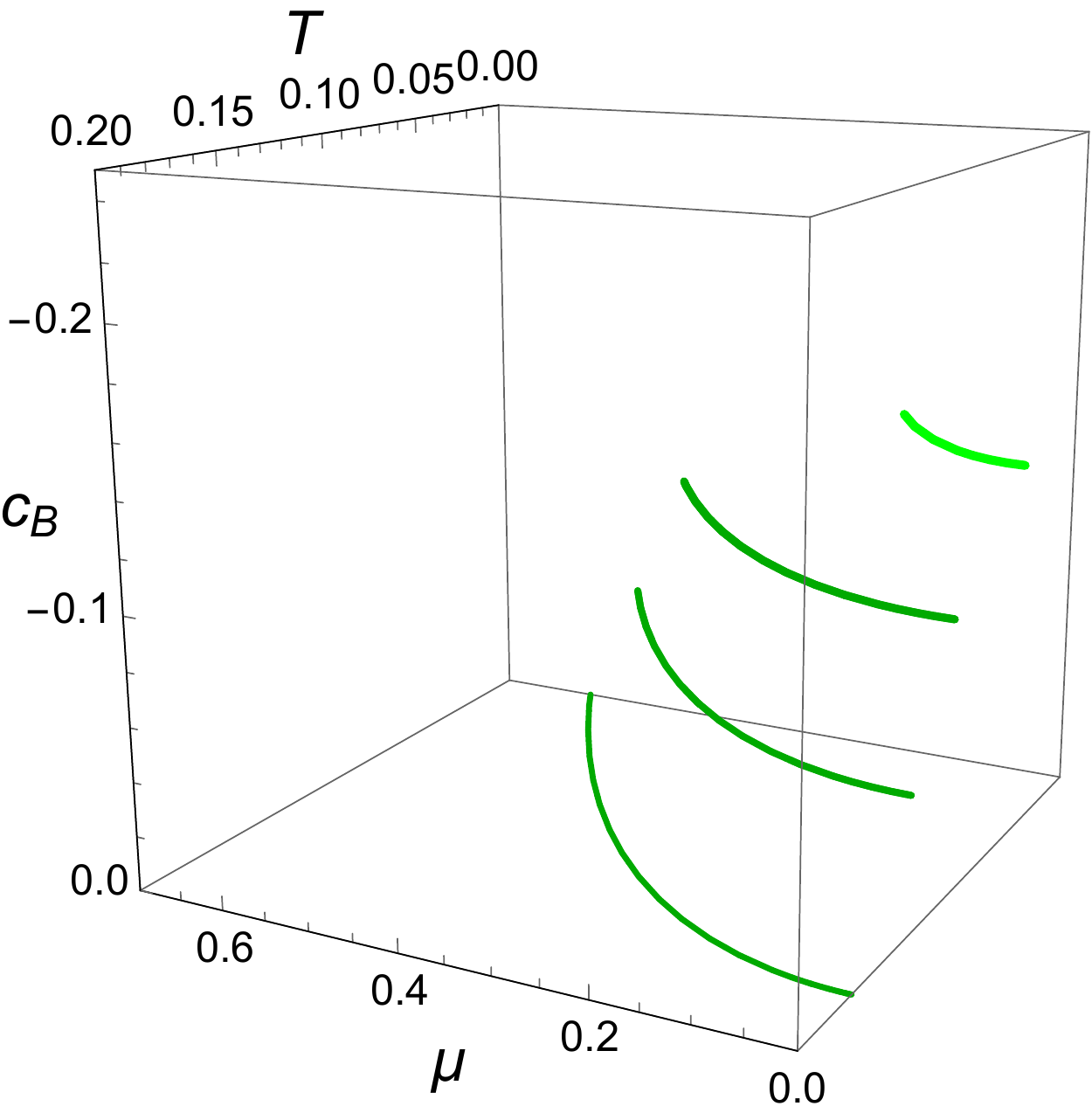}\qquad
  \includegraphics[scale=0.45]{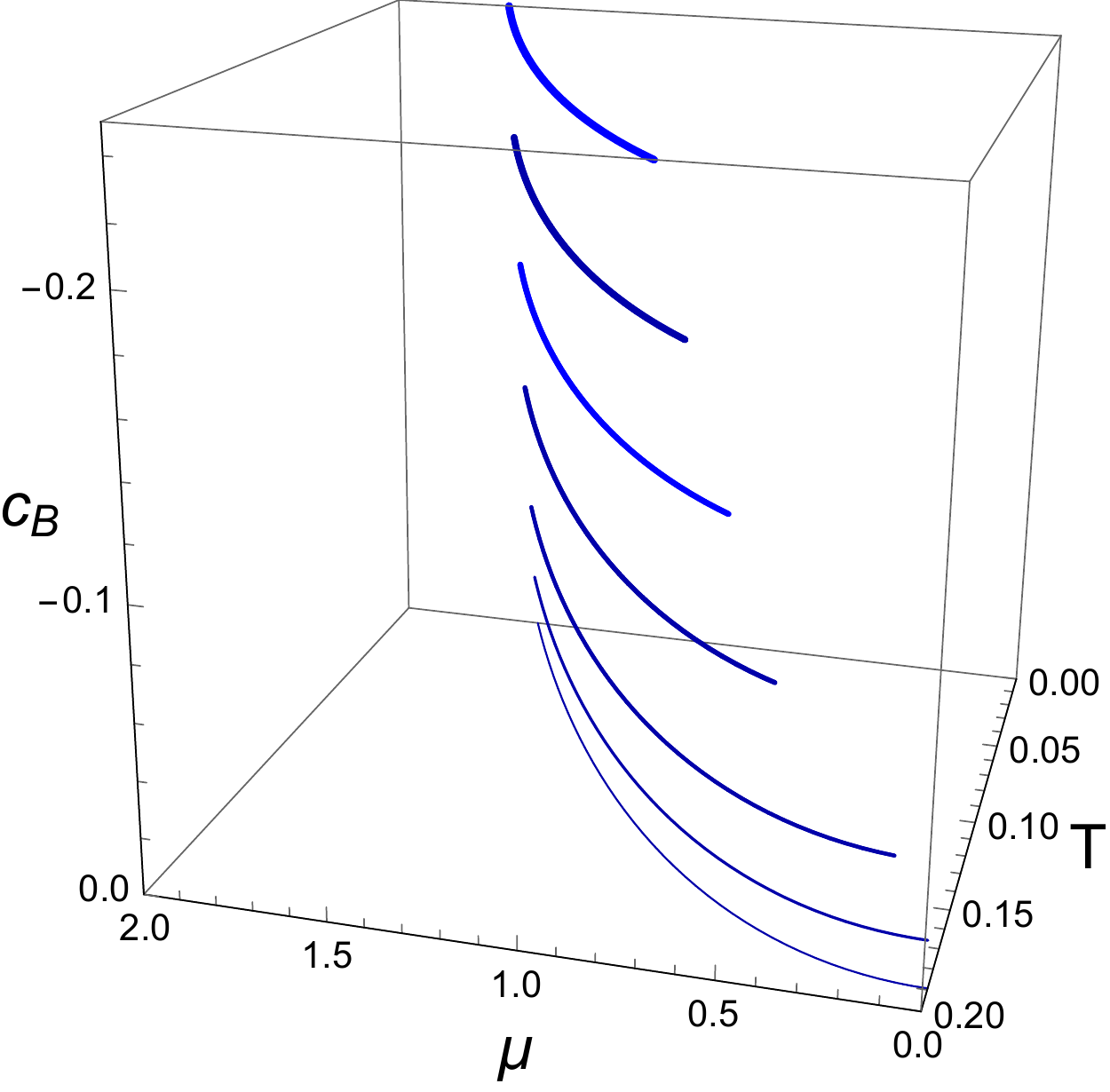}\\
  A \hspace{70 mm} B
  \caption{Locations of phase transition of $\sigma_1$ corresponding
    to the transition from the DW configuration to the horizon
    configuration for $\nu = 1$ (A) and $\nu = 4.5$ (B). 
  }
  \label{Fig:PT-V1-surfaces-2}
\end{figure}

To get the string tension of the spatial Wilson loop we have to find
the minimal value of effective potential at the dynamical wall and at 
the horizon. We show both values of effective potential in following
plots. In Fig.\ref{Fig:sigma-1-nu1} and
Fig.\ref{Fig:sigma-1-nu1-large} dependence of~$\sigma_1$ on
temperature for different values $c_B$ and $\mu$ is presented in
primary isotropic case. Green lines depict the dependence of
$\sigma(z_h)$ on temperature, cyan lines depict the dependence of
$\sigma(z_{DW})$ on temperature. In Fig.\ref{Fig:sigma-1-nu45} and
Fig.\ref{Fig:sigma-1-nu45-large} dependence of $\sigma_1$ on
temperature for different values $c_B$ and $\mu$ is presented in
primary anisotropic case ($\nu = 4.5$). Blue lines depict the
dependence of $\sigma(z_h)$ on temperature, cyan lines depict the
dependence of $\sigma(z_{DW})$ on temperature. We see the phase
transition between two connected string configuration with different
values of string tension: $\sigma(z_{DW})$ and $\sigma(z_h)$ (see also
Sect.\ref{Sect:BI}). For this phase transition the second derivative
of string tension $\partial^2\sigma/\partial T^2$ undergoes a jump. In
Fig.\ref{Fig:PT-V1-surfaces-1} transparent surfaces correspond to the
configurations touching the horizon, and less transparent ones to the
configurations touching the dynamic wall. The phase transition
corresponds to the transition from the DW configuration to the horizon
configuration and is located at boundaries between light and dark
surfaces. In Fig.\ref{Fig:PT-V1-surfaces-2} locations of phase
transition of $\sigma_1$ is presented for $\nu = 1$ and for $\nu =
4.5$.

\begin{figure}[h!]
  \centering
  \includegraphics[scale=0.45]{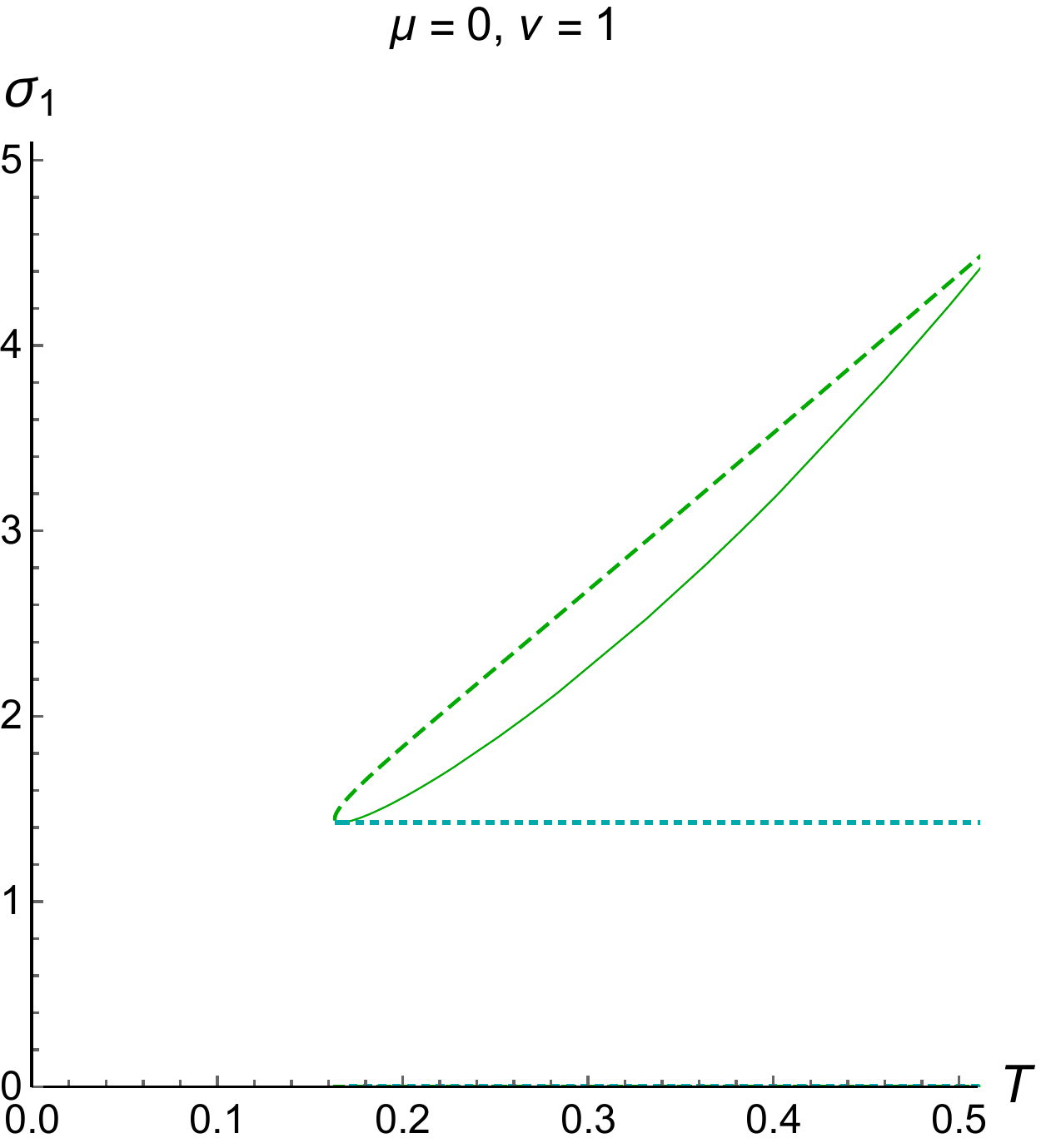}\qquad\qquad
  \includegraphics[scale=0.45]{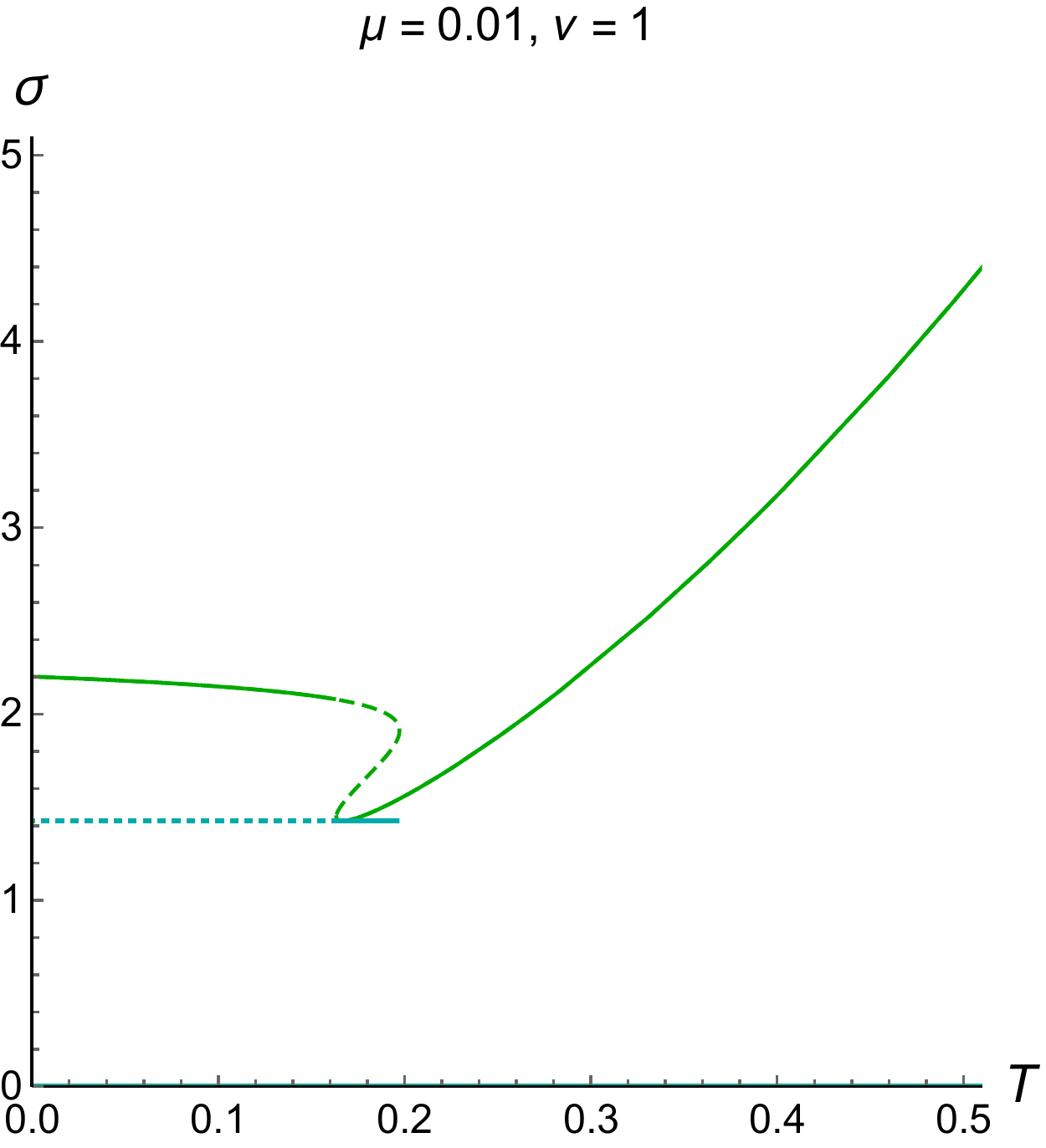}\\
  A \hspace{70 mm} B
  \caption{Phase transitions in details for $\mu = 0$ (A) and $\mu =
    0.01$ (B); $c_B = 0$.
  }
  \label{Fig:SW-PT}
\end{figure}

\newpage
$$\,$$
\newpage
\newpage
$$\,$$
\newpage

\subsection{Wilson loop $W_{xY_{2}}$}
\begin{figure}[h!]
  \centering
  \includegraphics[scale=0.44]{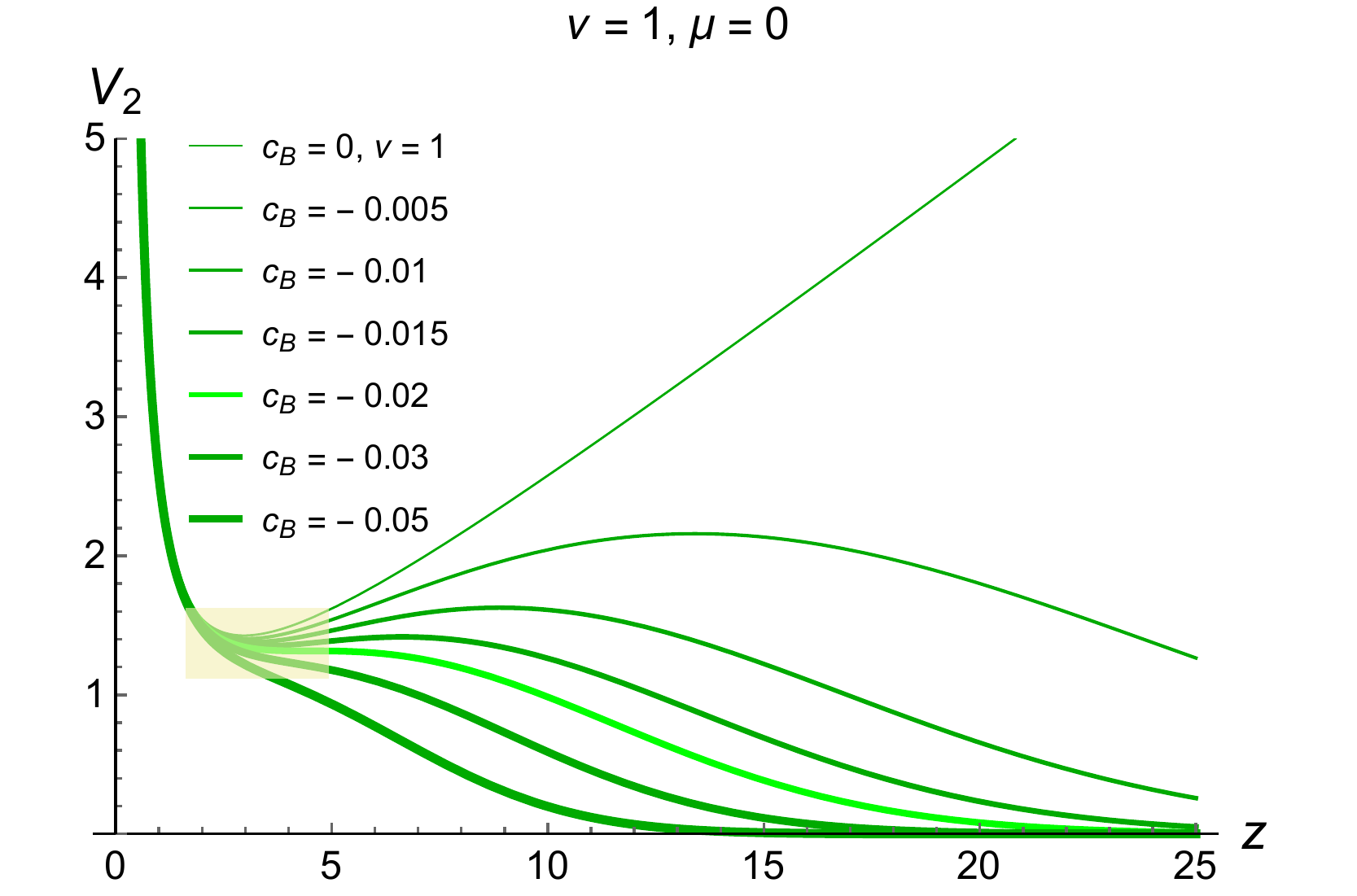}
  \includegraphics[scale=0.44]{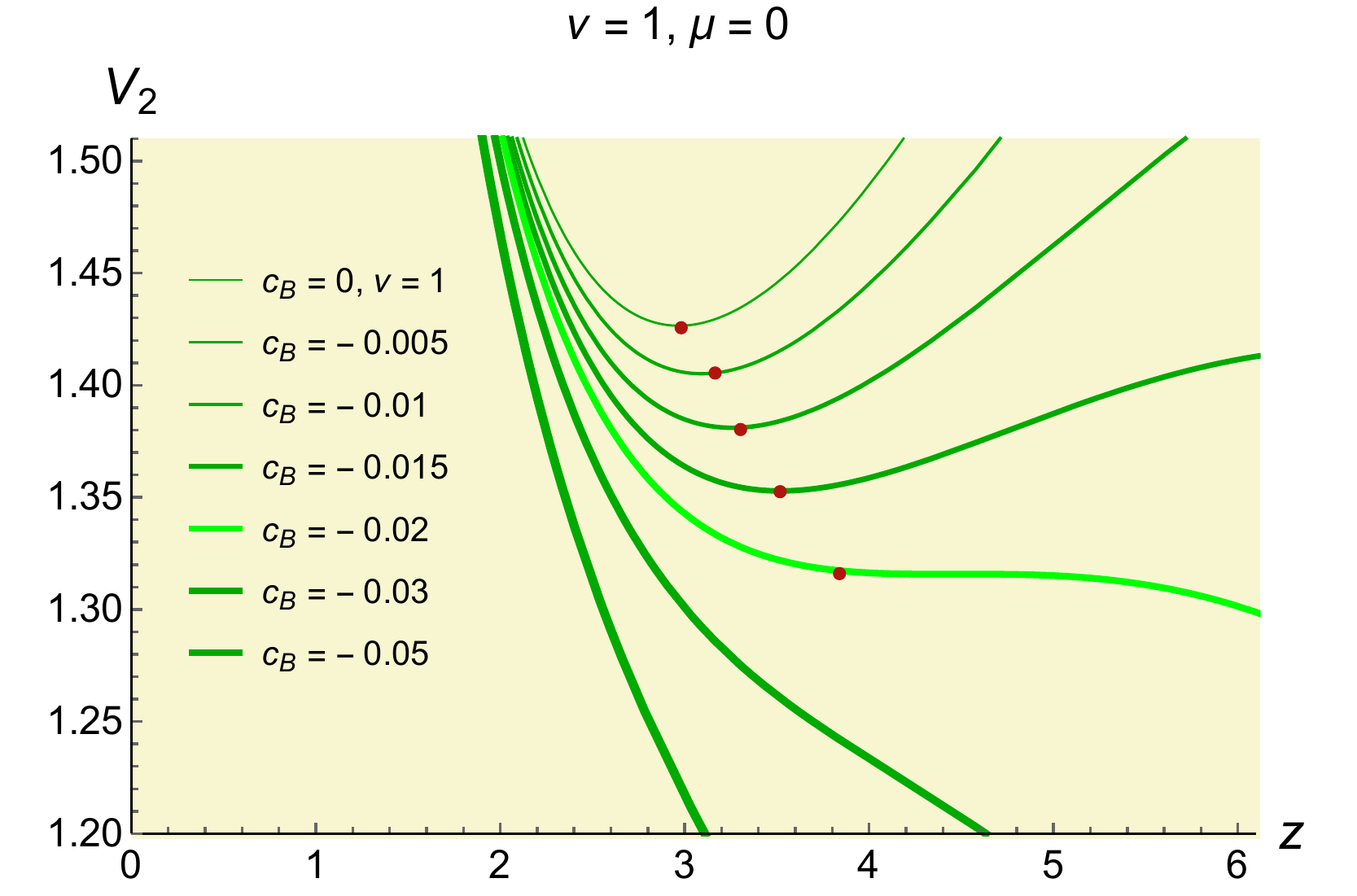}\\
  A\\
  \includegraphics[scale=0.44]{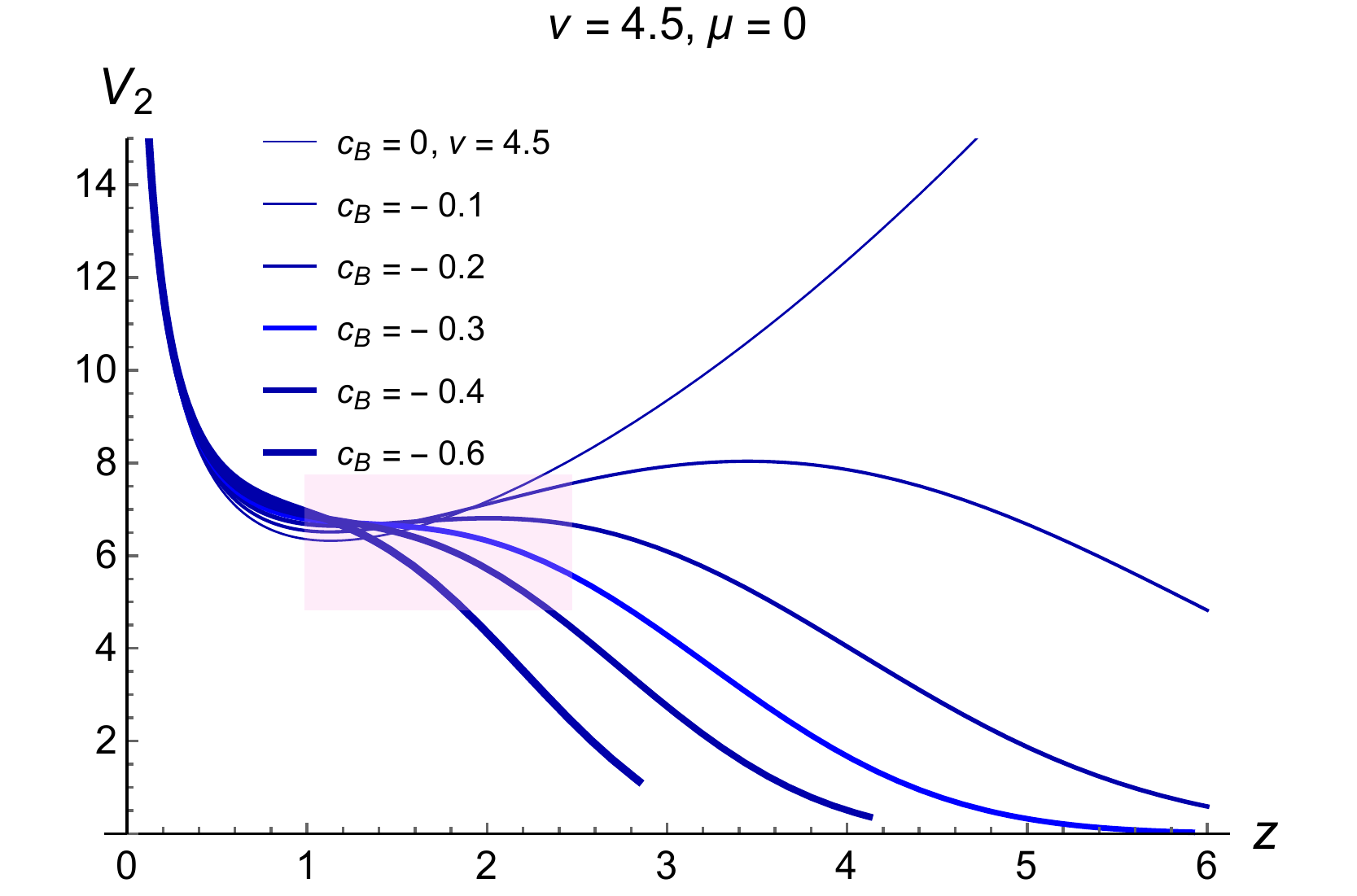}
  \includegraphics[scale=0.42]{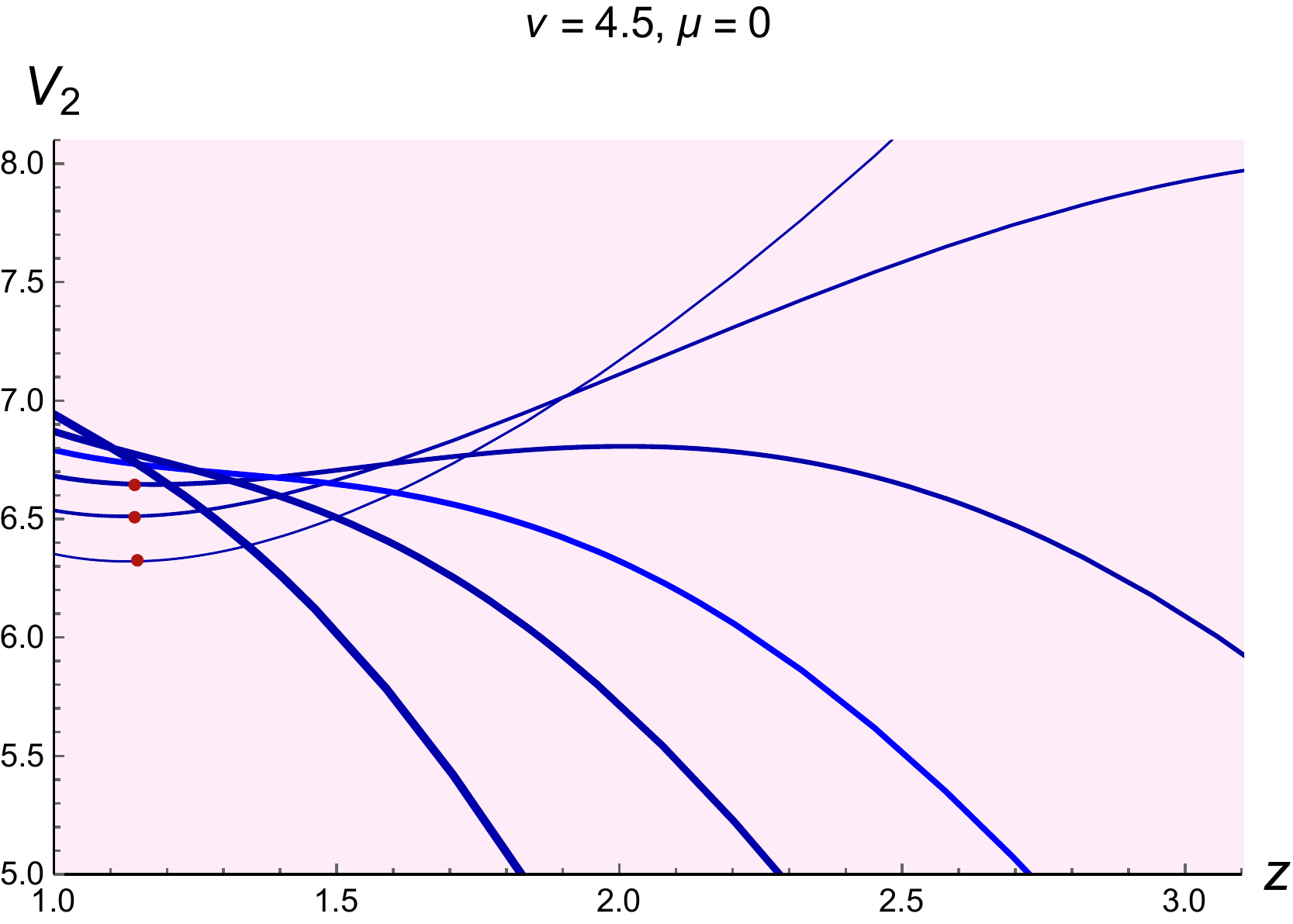}\\
  B
  \caption{Effective potential ${\cal V}_2$ as function of $z$ for
    different $c_B$; $\nu = 1$ (A) and $\nu = 4.5$ (B). Locations of
    DW are indicated by dots. We see that for $\mu = 0$ at large $c_B$
    DW disappears. Here $c_{B,DWcr} \approx 0.2$ for $\nu = 1$ and
    $c_{B,DWcr} \approx 0.3$ for $\nu = 4.5$. 
  }
  \label{Fig:V2}
\end{figure}

In Fig.\ref{Fig:V2} the effective potential ${\cal V}_2$ is presented
as function of $z$ for different $c_B$ in primary isotopic ($\nu = 1$)
and primary anisotropic ($\nu = 4.5$) cases. 

\begin{figure}[h!]
  \centering
  \ \\ \ \\
  \includegraphics[scale=0.35]{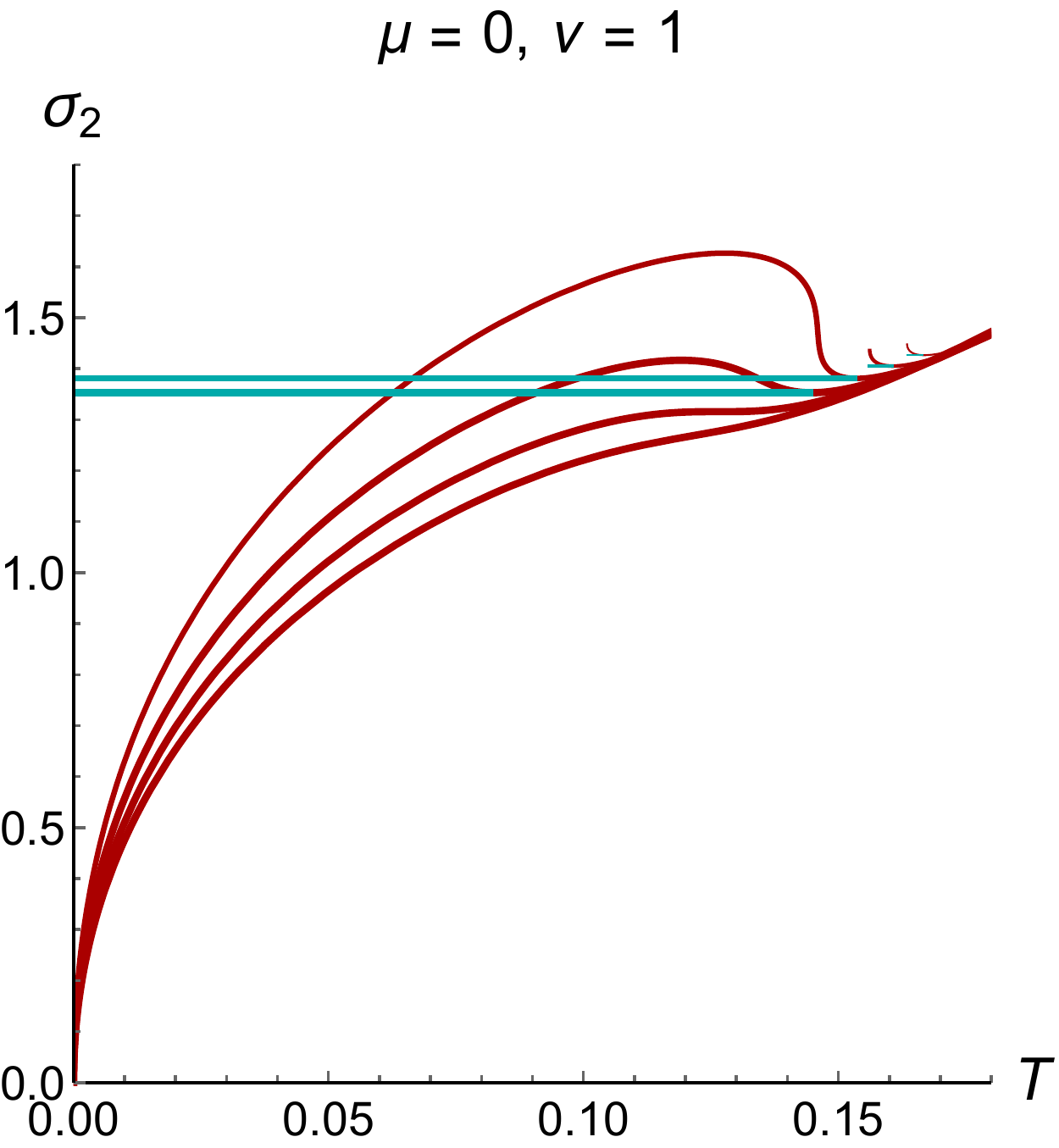}
  \includegraphics[scale=0.35]{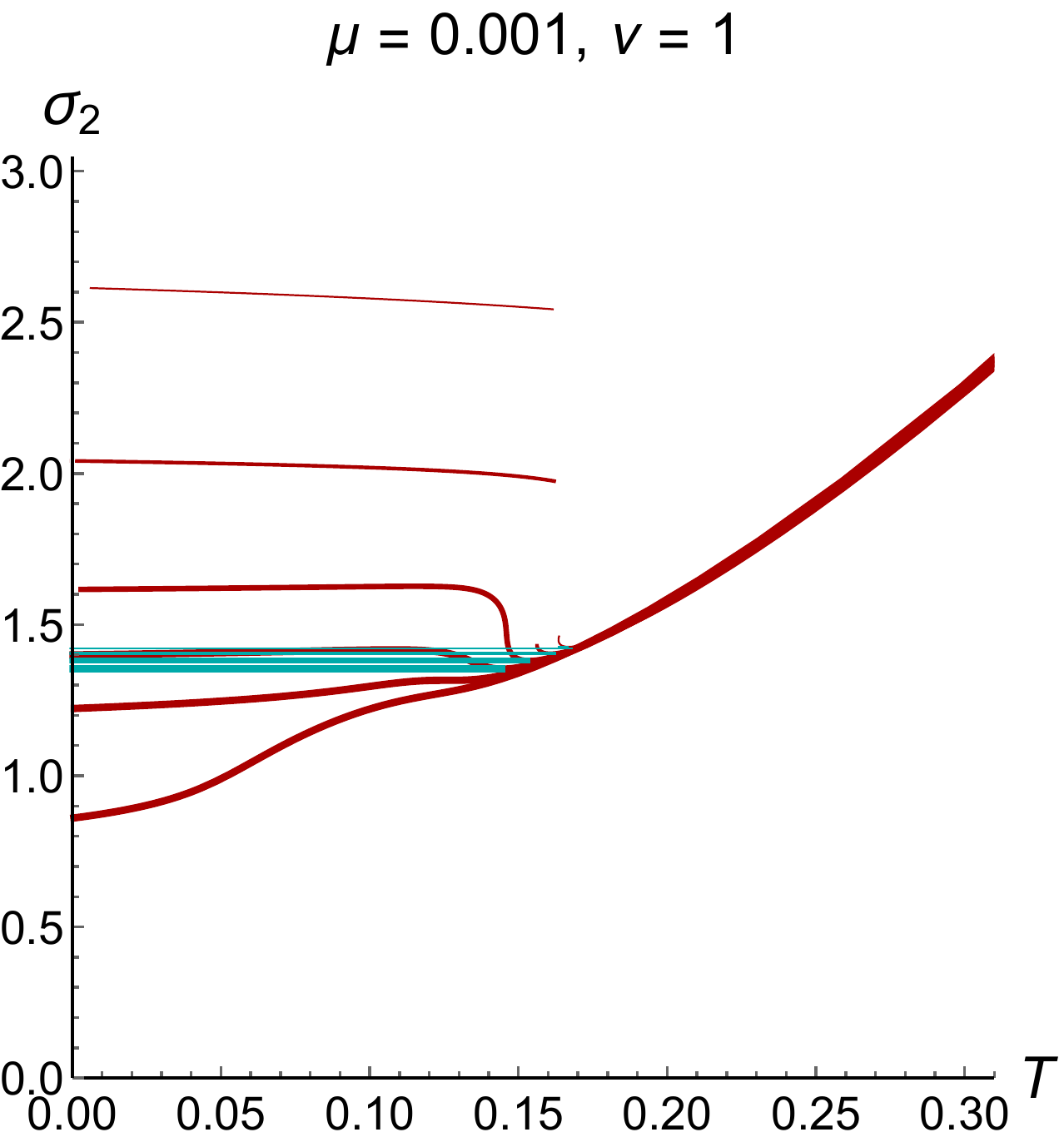}
  \includegraphics[scale=0.35]{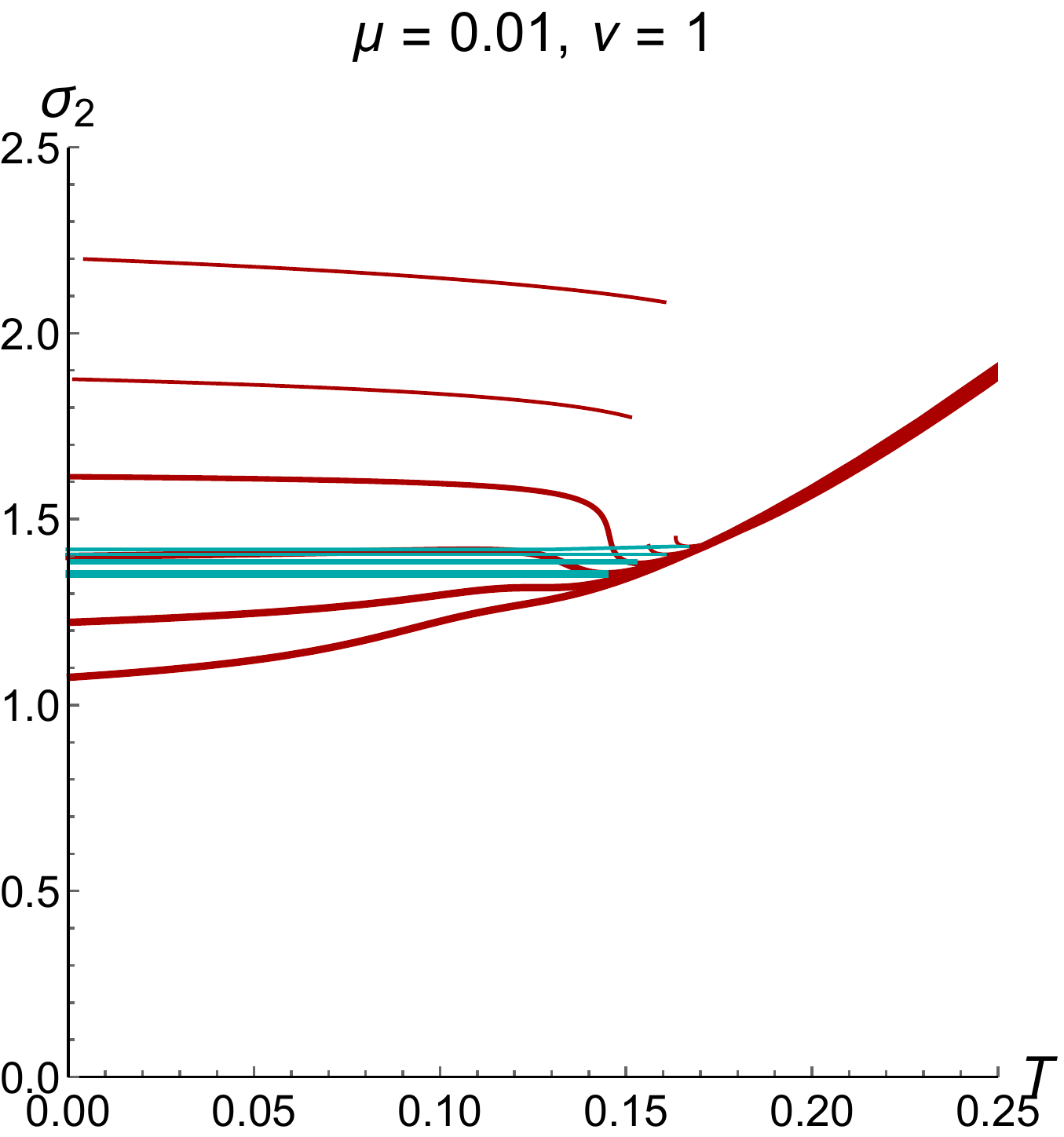}\\
  A \hspace{40 mm} B \hspace{40 mm} C \\
  \hspace{80 mm}$\,$\\
  \includegraphics[scale=0.35]{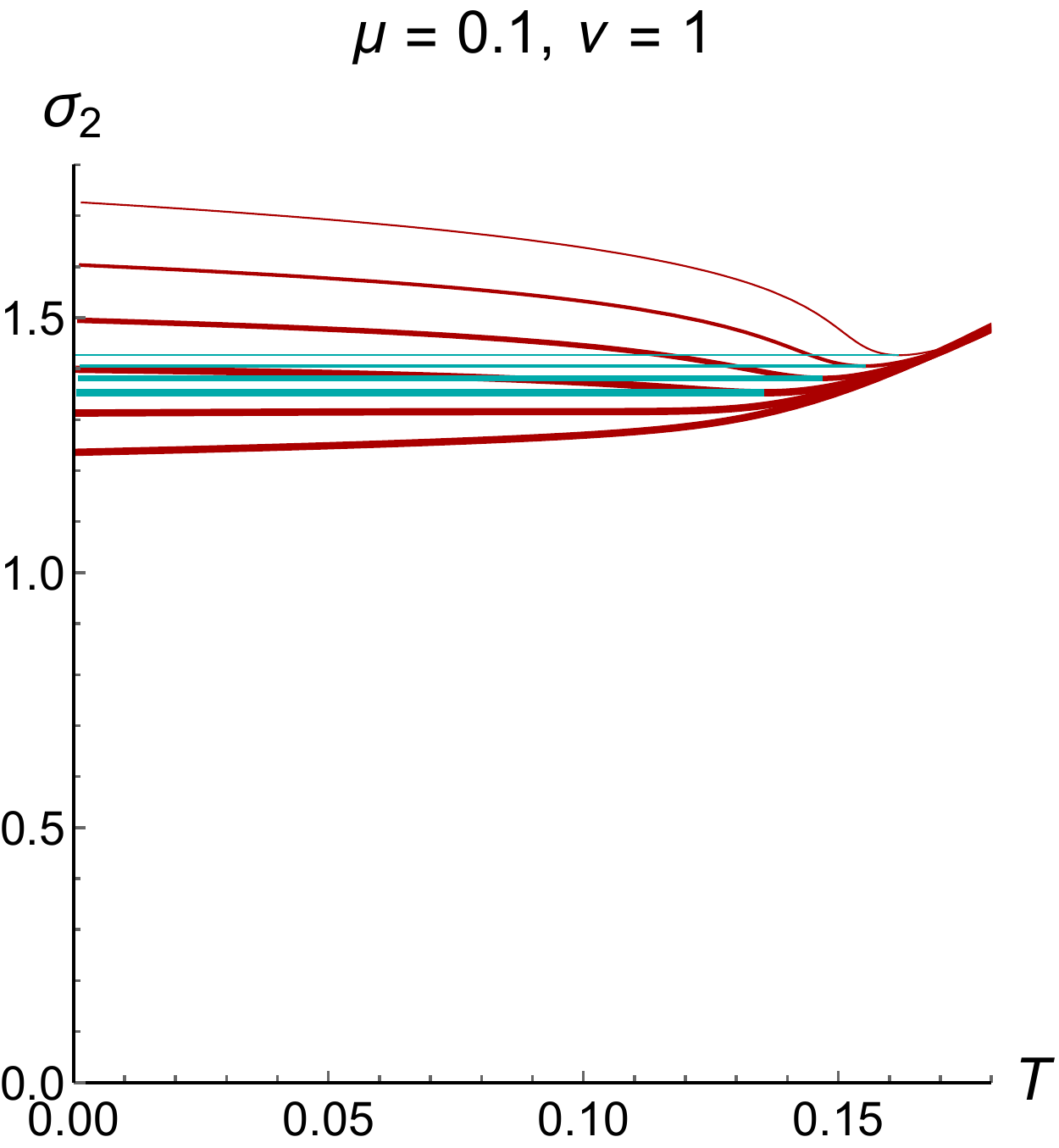}
  \includegraphics[scale=0.35]{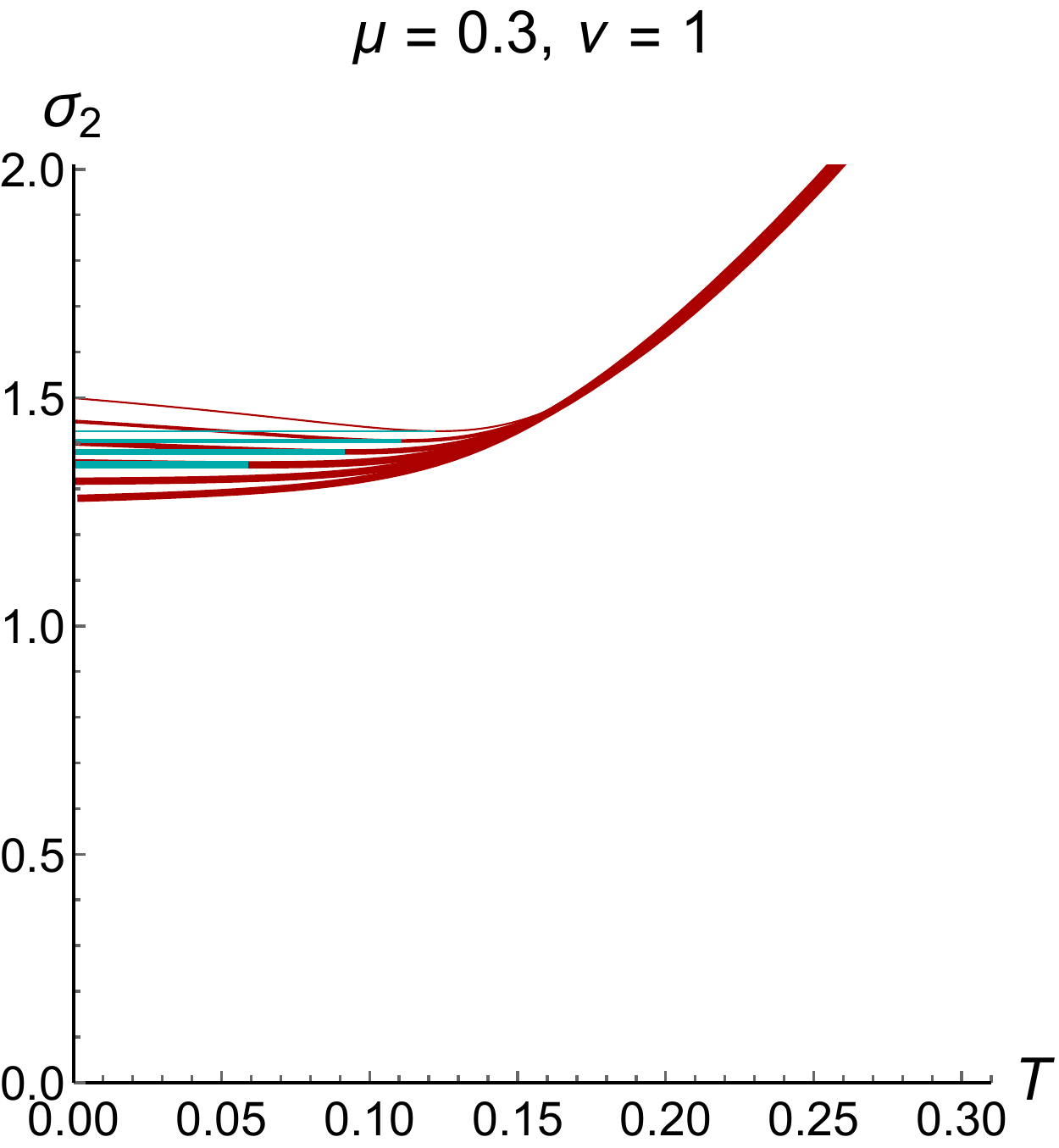}
  \includegraphics[scale=0.35]{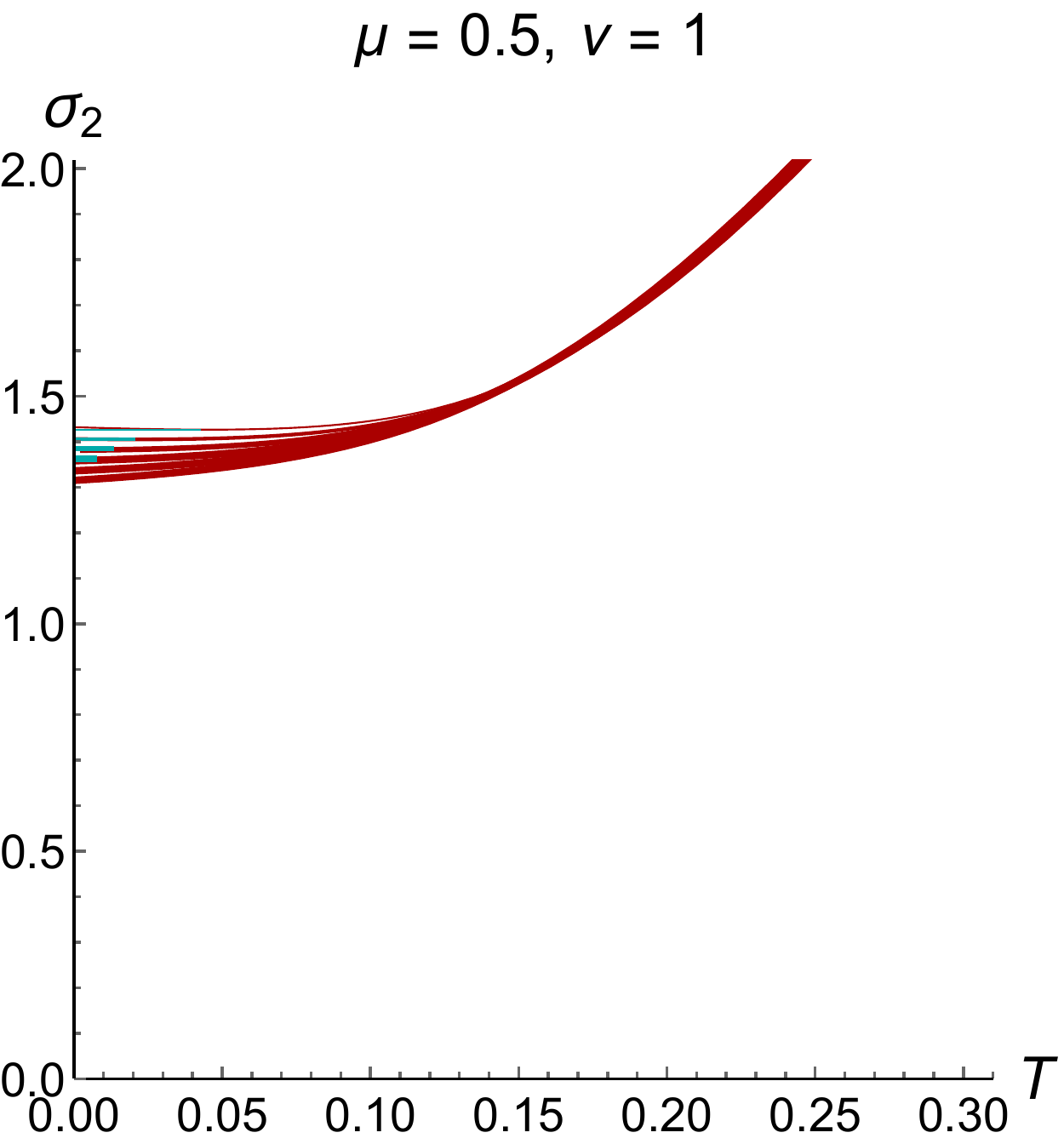}\\
  D \hspace{40 mm} E \hspace{40 mm} F \\ \ \\
  \includegraphics[scale=0.7]{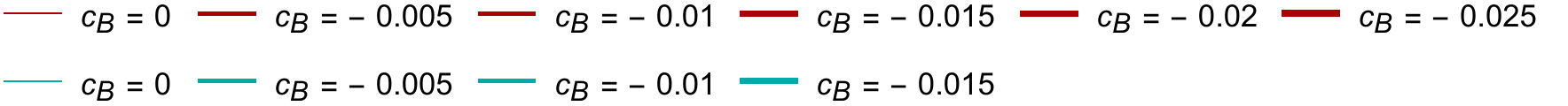}
  \caption{Dependence of $\sigma_2$ on temperature and $c_B$ for $\mu
    = 0$ (A), $0.001$ (B), $0.01$ (C), $0.1$ (D), $0.3$ (E) and $0.5$
    (F); $\nu = 1$. Plot legends are the same for all plots.
   }
   \label{Fig:sigma-2small}
\end{figure}

\begin{figure}[h!]
  \centering
  \includegraphics[scale=0.35]{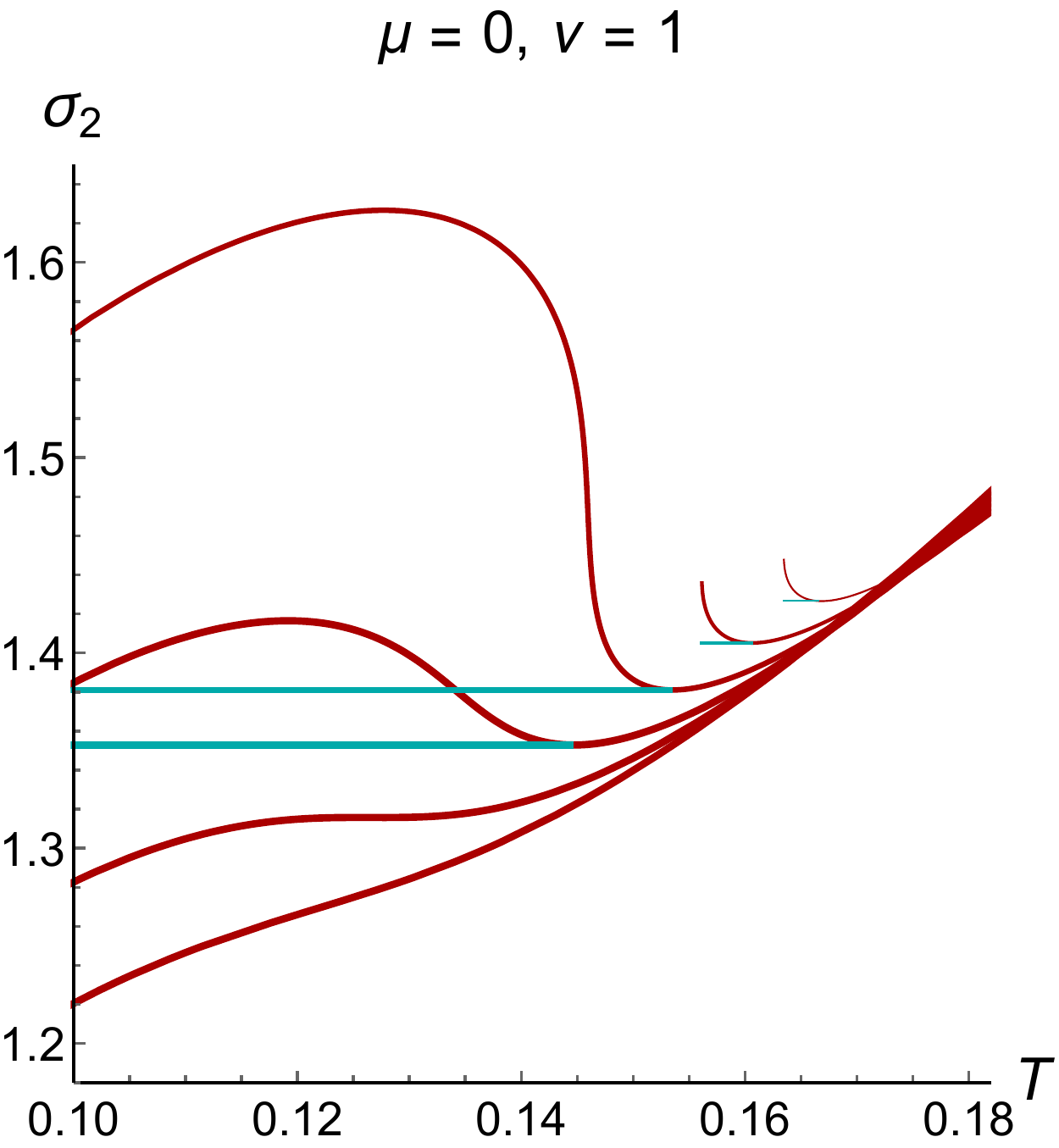}\quad
  \includegraphics[scale=0.35]{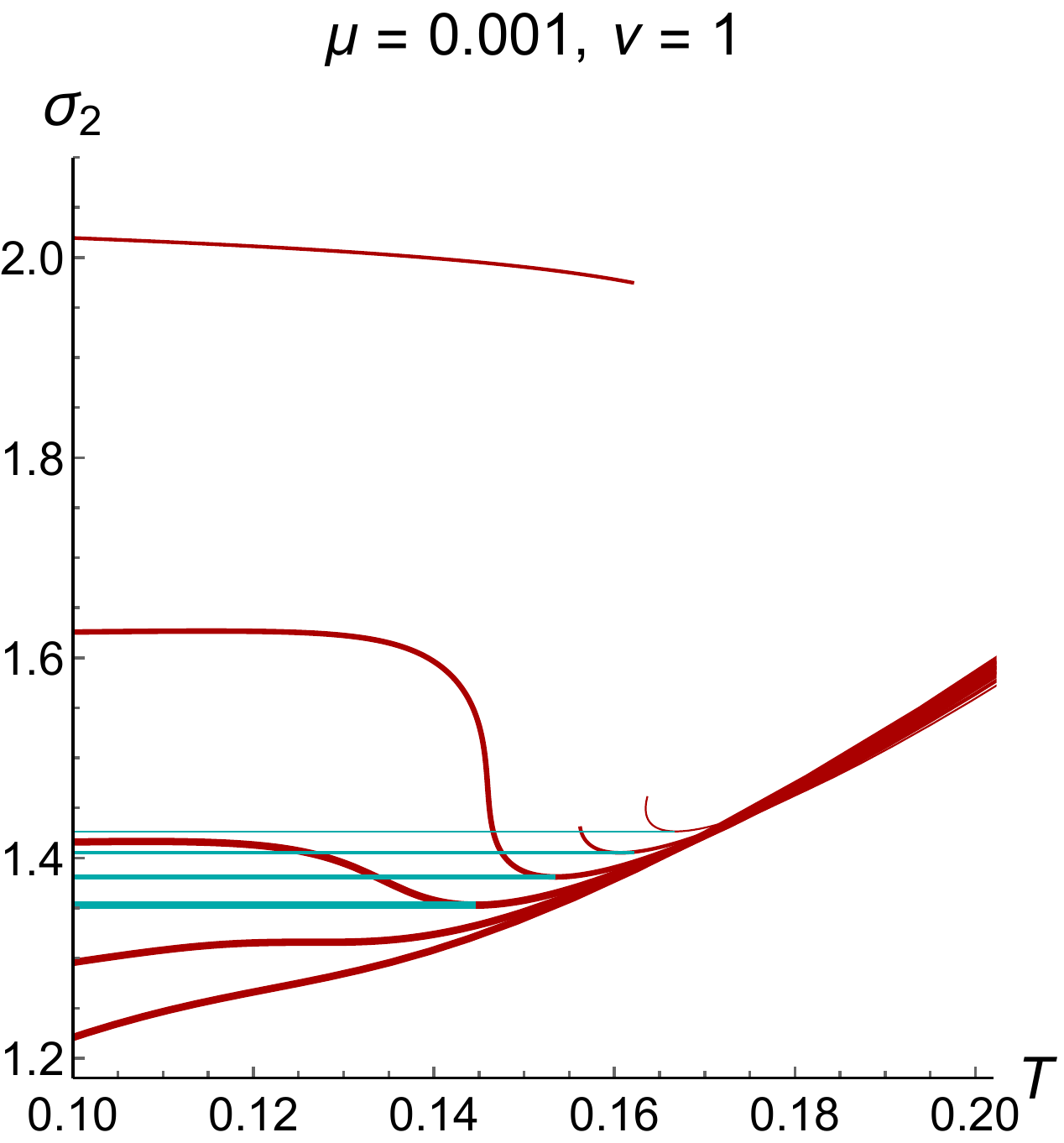}\quad
  \includegraphics[scale=0.35]{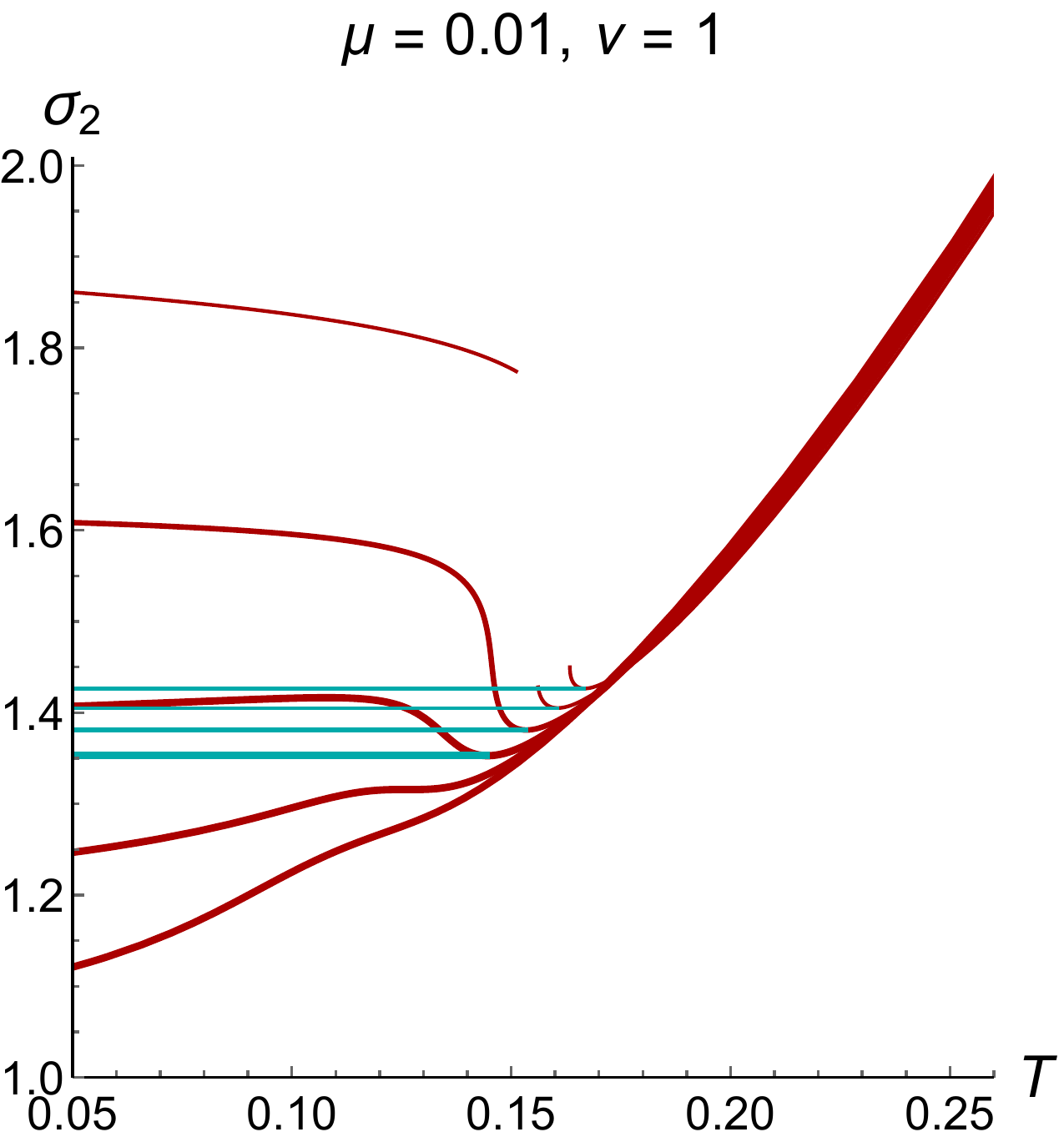} \quad \\
  A \hspace{40 mm} B \hspace{40 mm} C\\
  \includegraphics[scale=0.35]{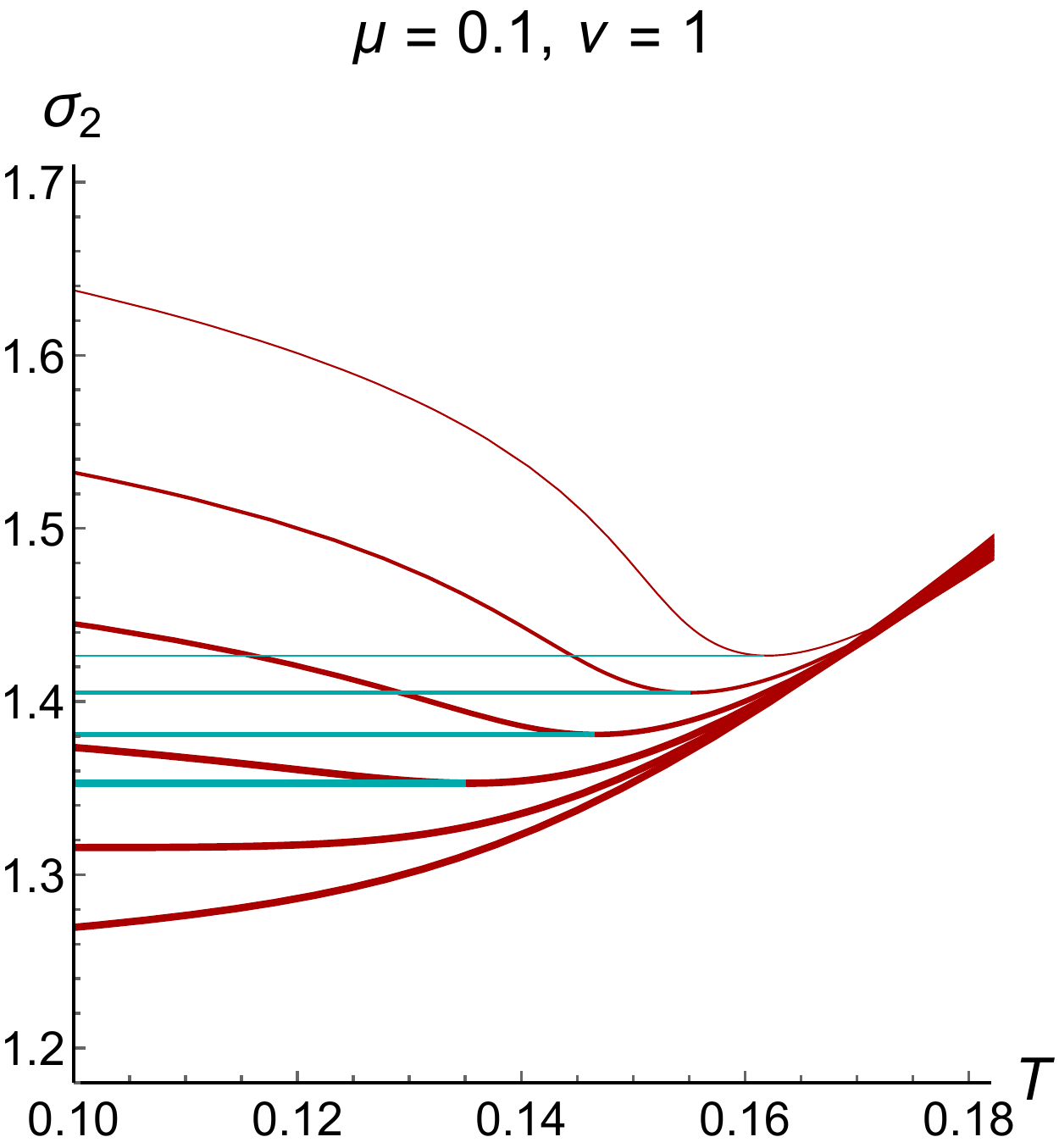}\quad
  \includegraphics[scale=0.35]{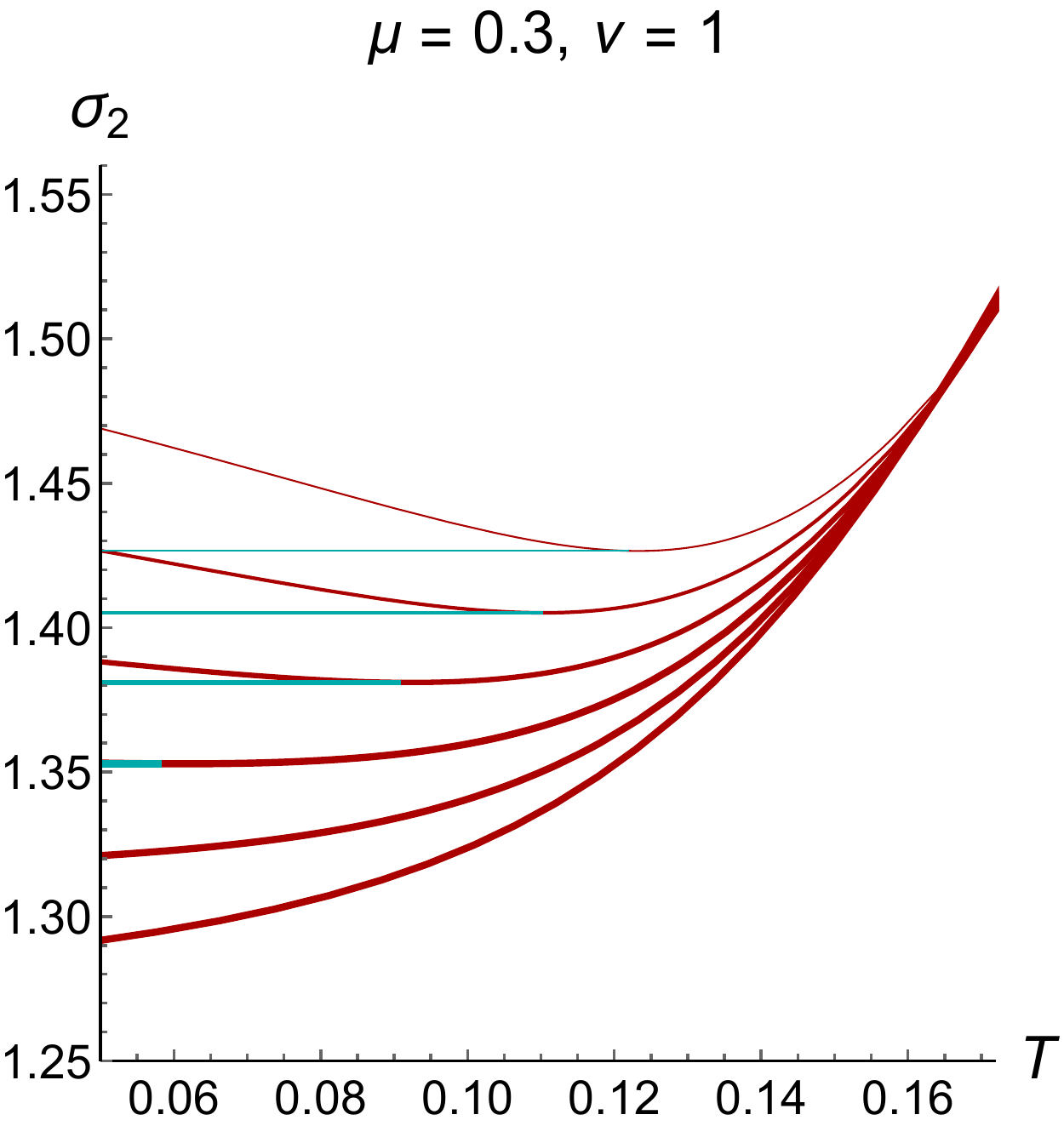}\quad
  \includegraphics[scale=0.35]{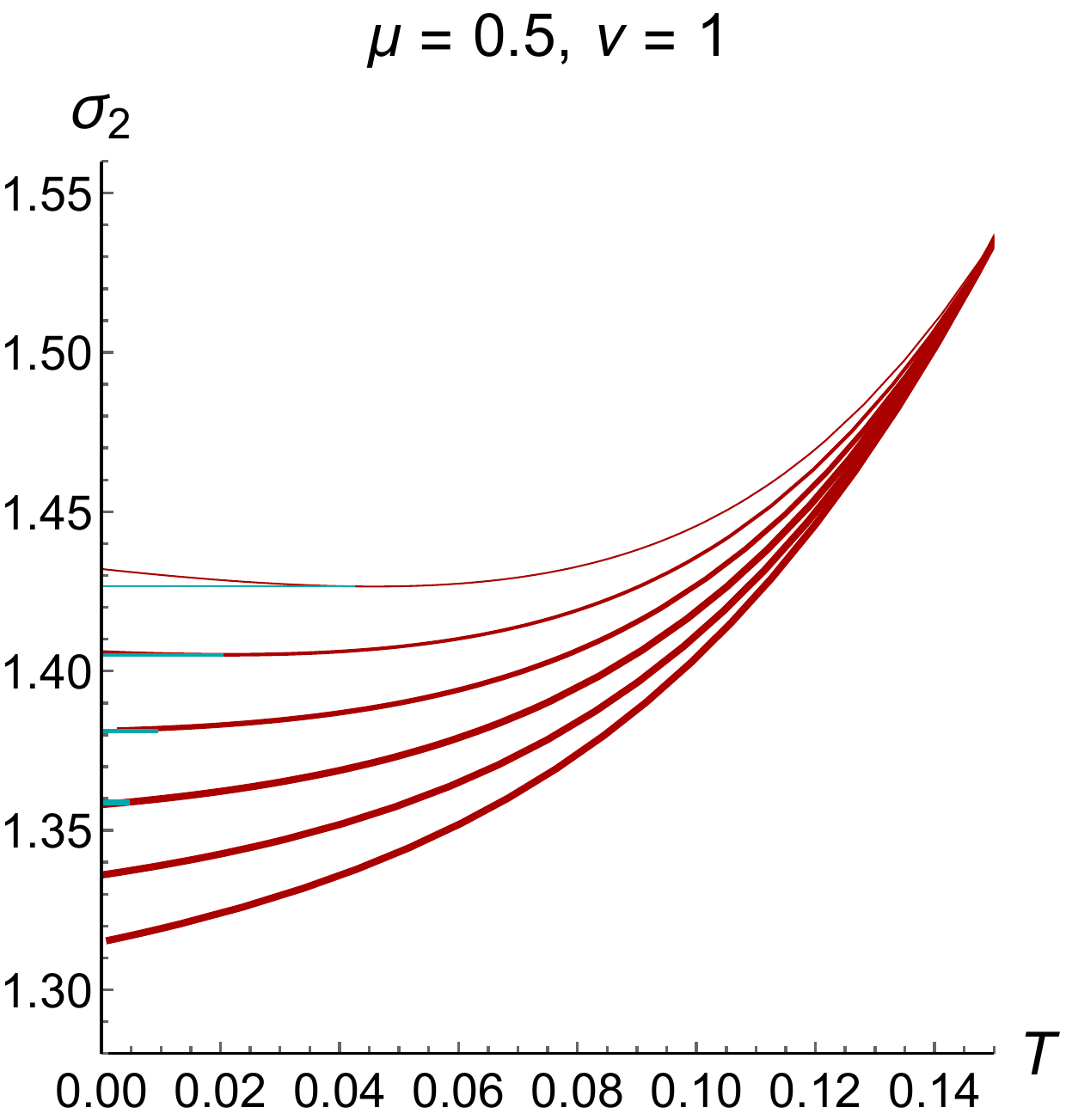}  \quad \\
  D \hspace{40 mm} E \hspace{40 mm} F\\ \ \\
  \includegraphics[scale=0.7]{Plots/sigma2-nu1-leg.pdf}
  \caption{Dependence of $\sigma_2$ from Fig.\ref{Fig:sigma-2small} in larger
    scale. Plot legends are the same for all plots.
  }
  \label{Fig:sigma-2large}
\end{figure}

\begin{figure}[h!]
  \centering
  \includegraphics[scale=0.45]{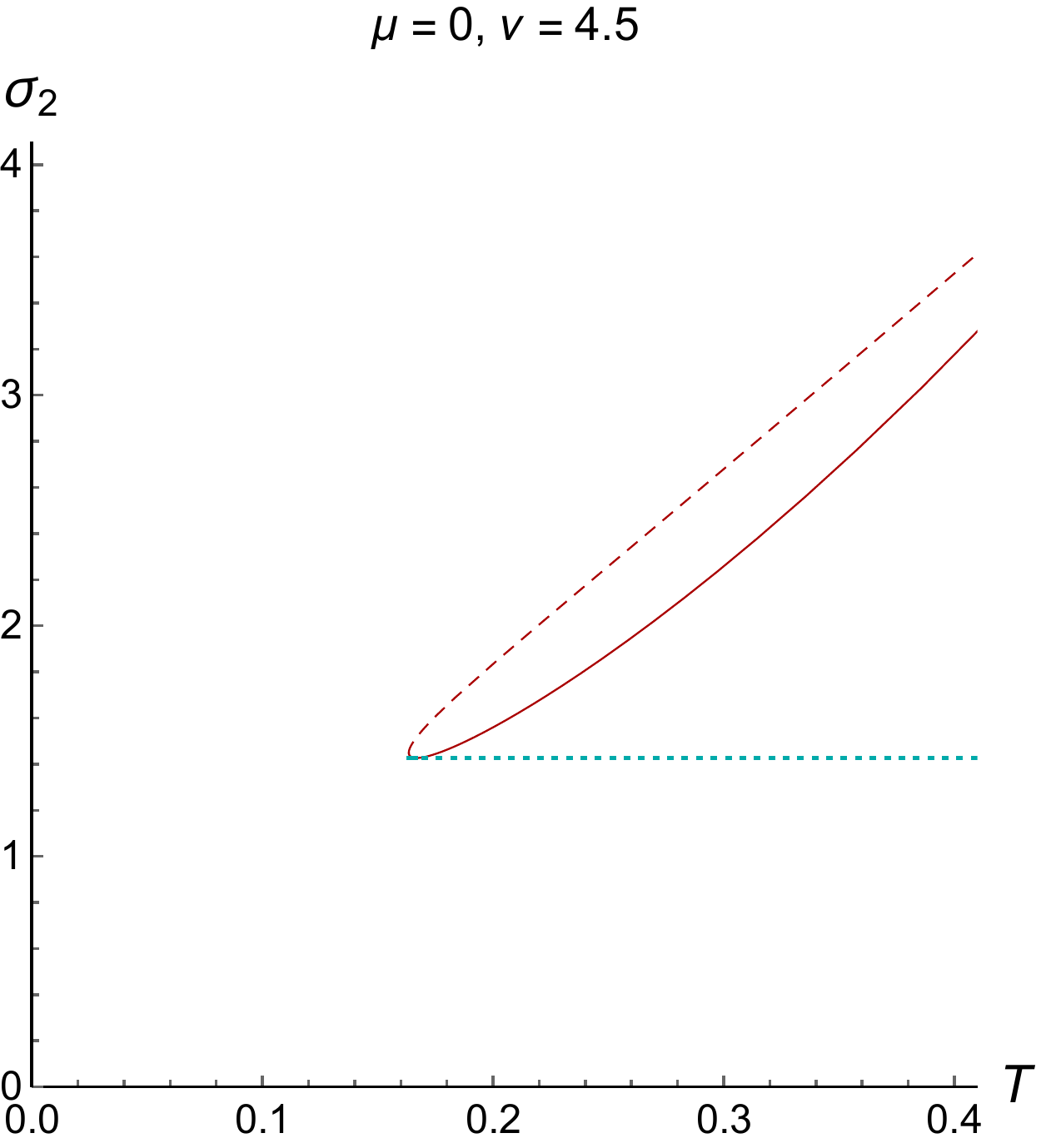}\qquad\qquad
  \includegraphics[scale=0.45]{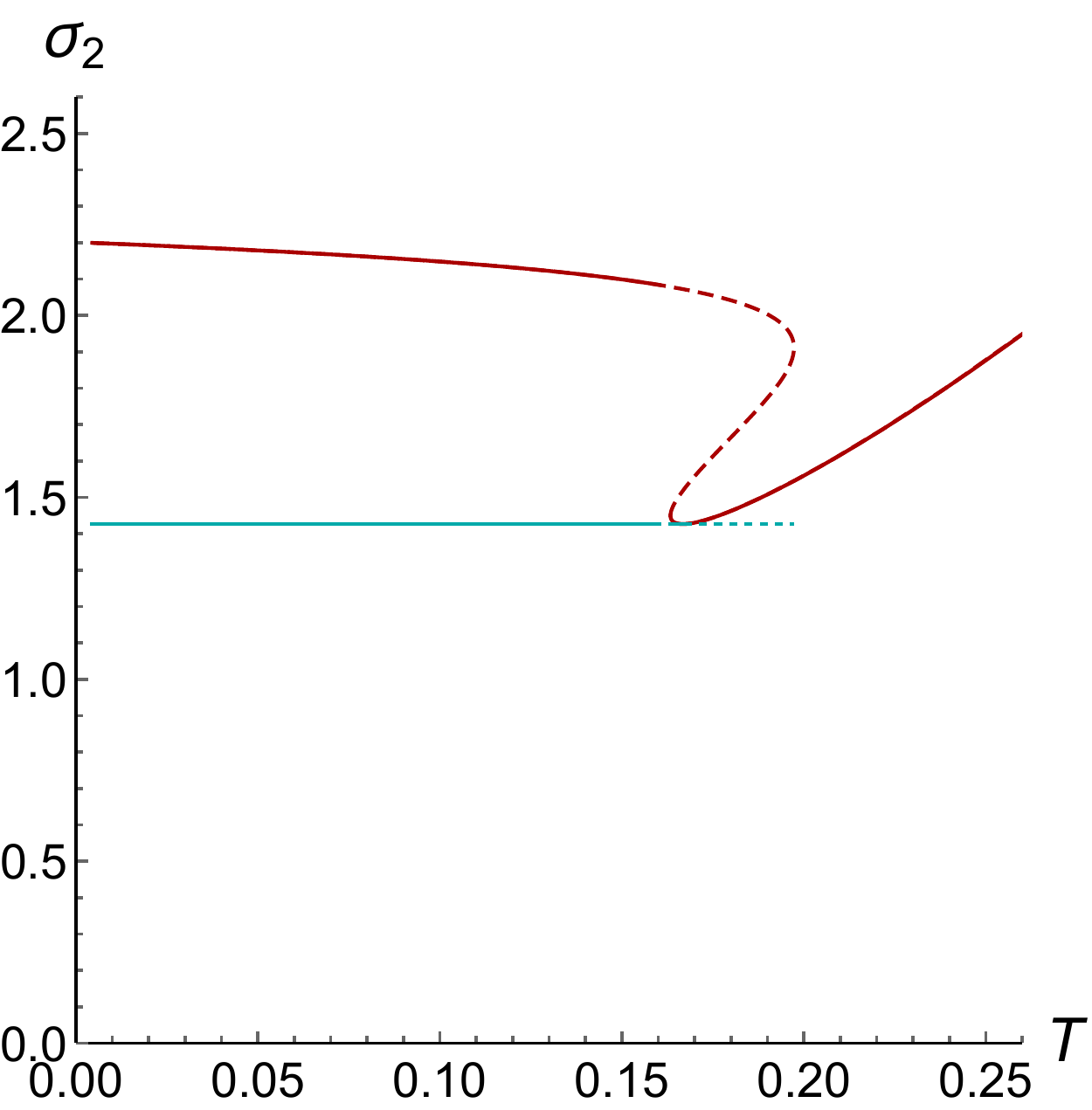} \\
  A \hspace{70 mm} B
  \caption{Phase transitions in details for $\mu = 0$ (A) and $\mu =
    0.01$ (B); $c_B = 0$.
  }
  \label{Fig:SW-PT-1}
\end{figure}

\begin{figure}[h!]
  \centering
  \includegraphics[scale=0.35]{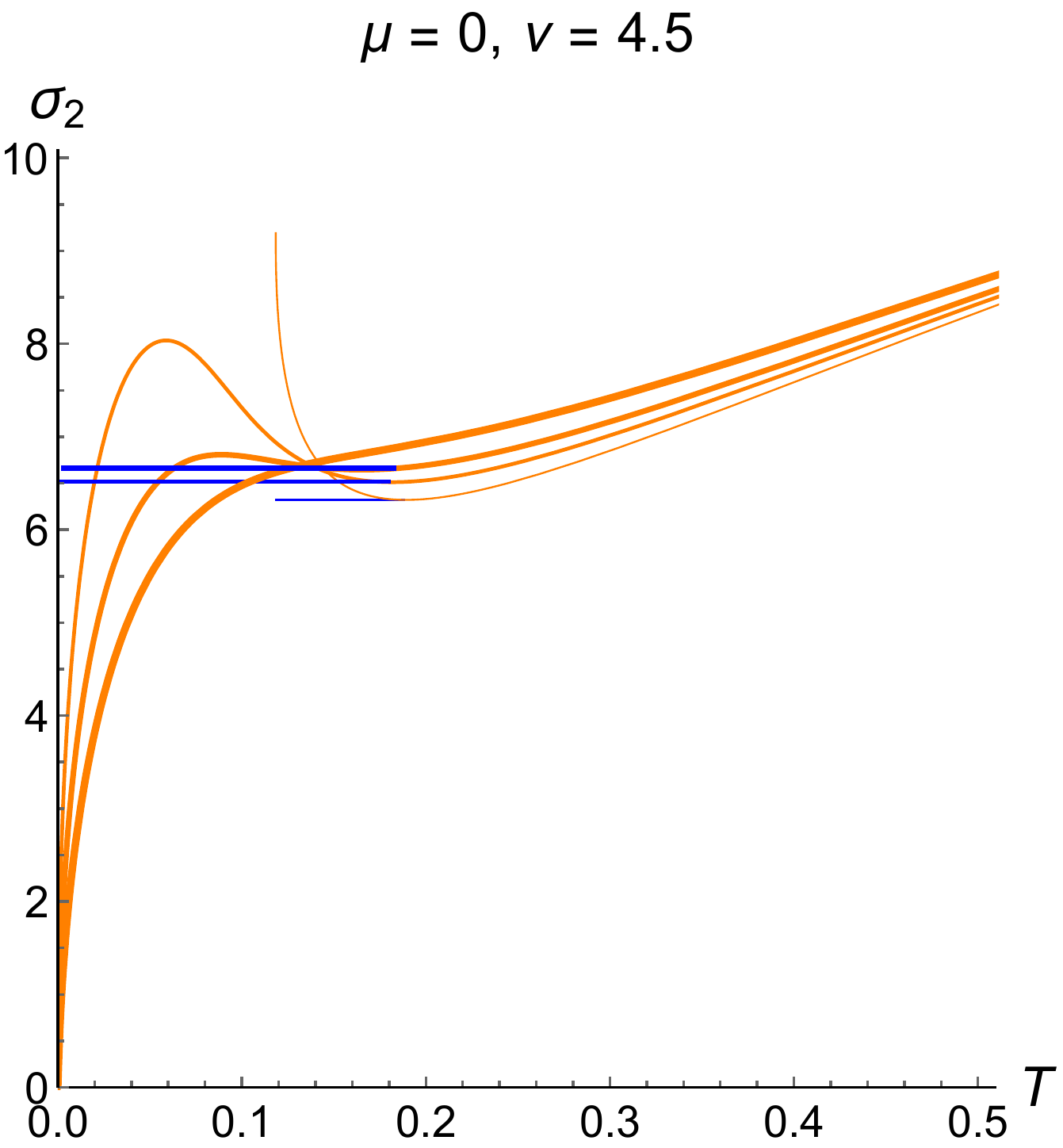}
  \includegraphics[scale=0.35]{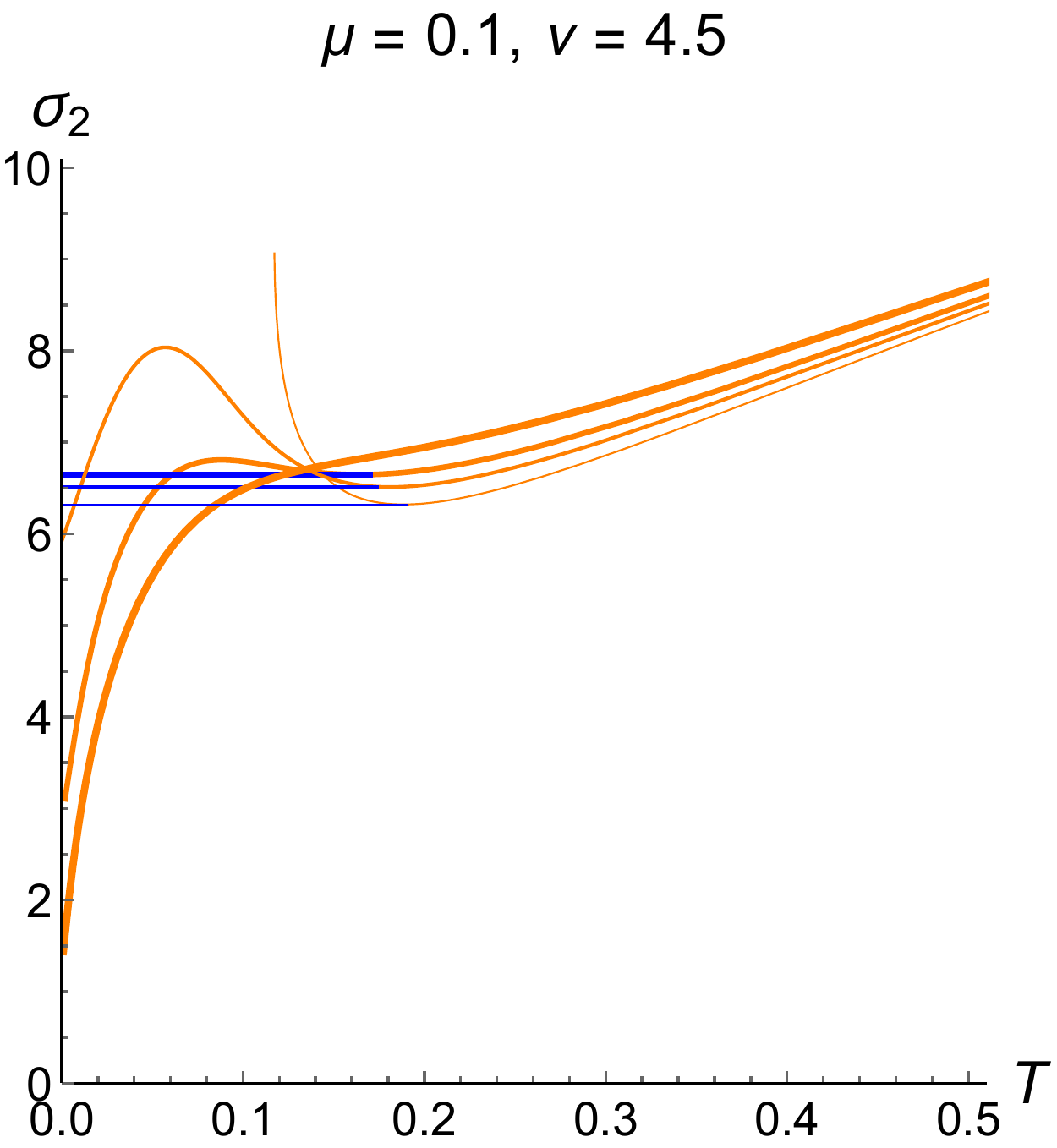}
  \includegraphics[scale=0.35]{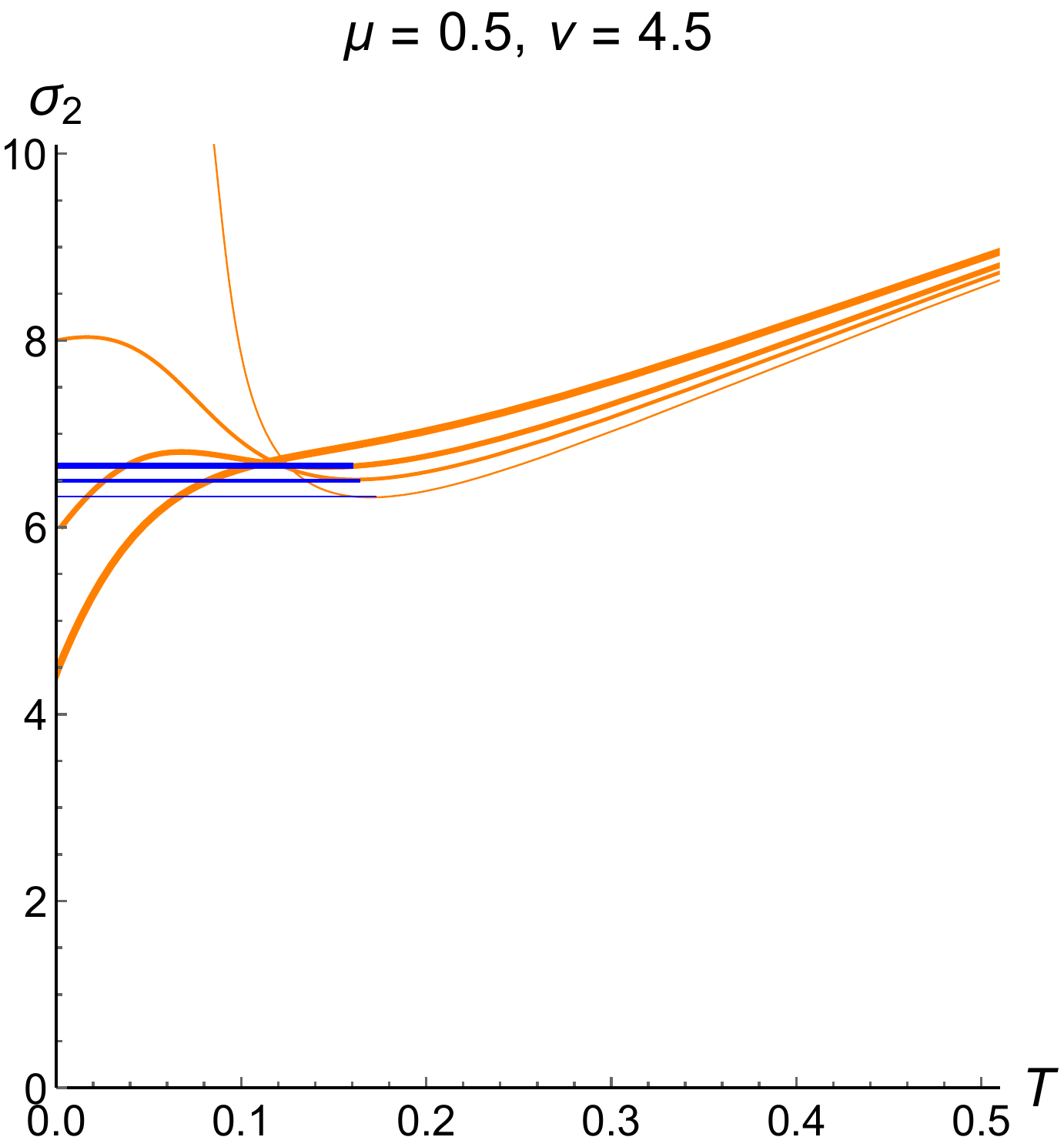}\\
  A \hspace{40 mm} B \hspace{40 mm} C \\ \ \\
  \includegraphics[scale=0.7]{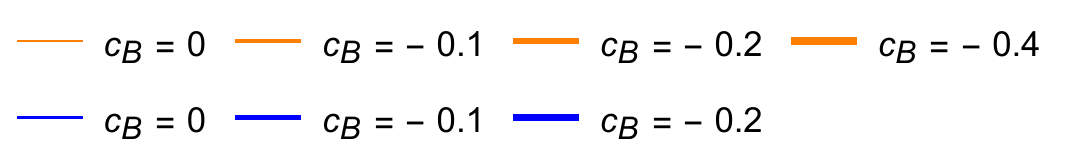}
  \caption{Dependence of $\sigma_2$ on temperature and $c_B$ for $\mu
    = 0$ (A), $0.1$ (B) and $0.5$ (C); $\nu = 4.5$. Plot legends are
    the same for all plots.
  }
  \label{Fig:sigma-2-nu45small}
\end{figure}

\begin{figure}[h!]
  \centering
  \includegraphics[scale=0.35]{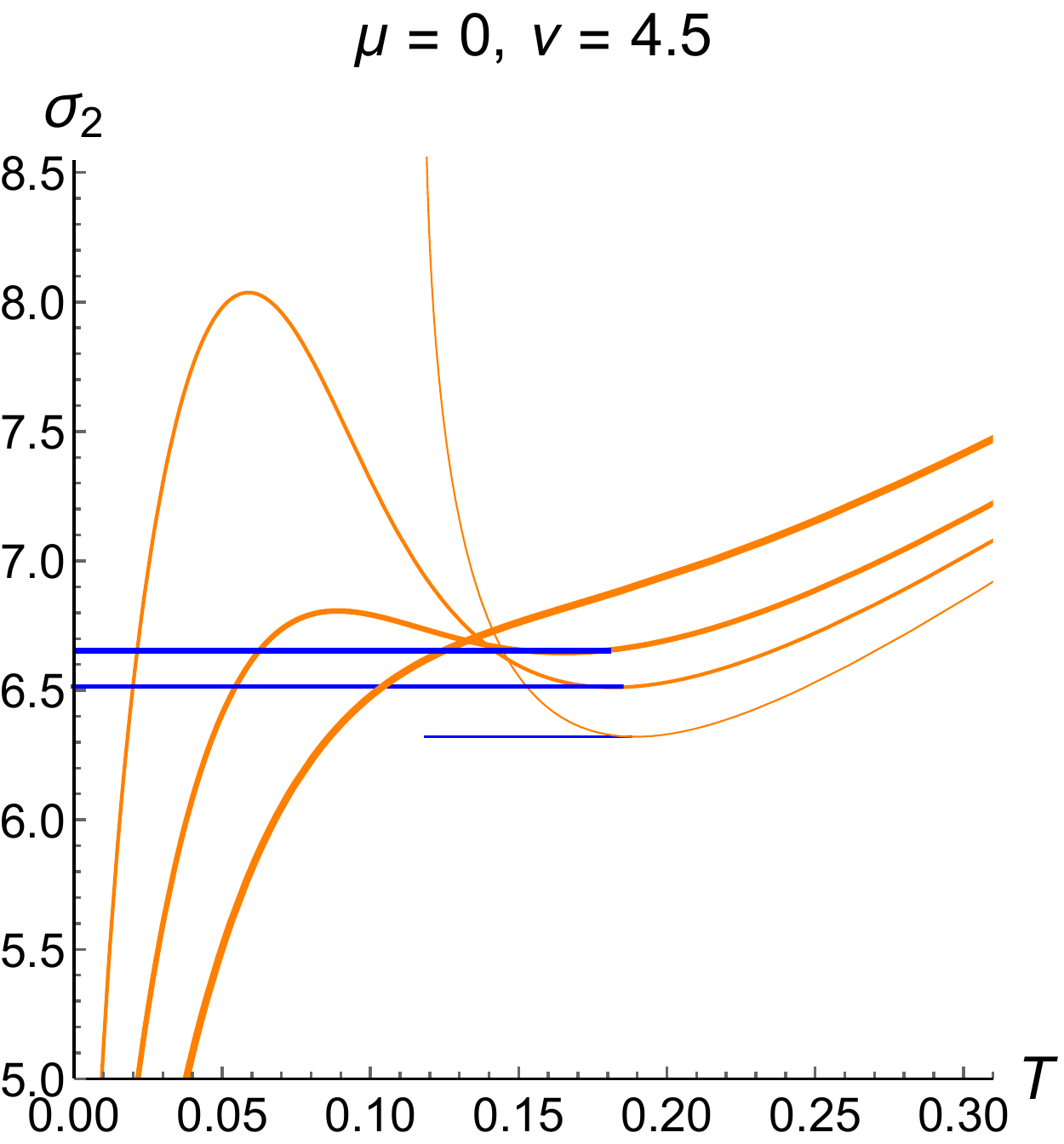}
  \includegraphics[scale=0.35]{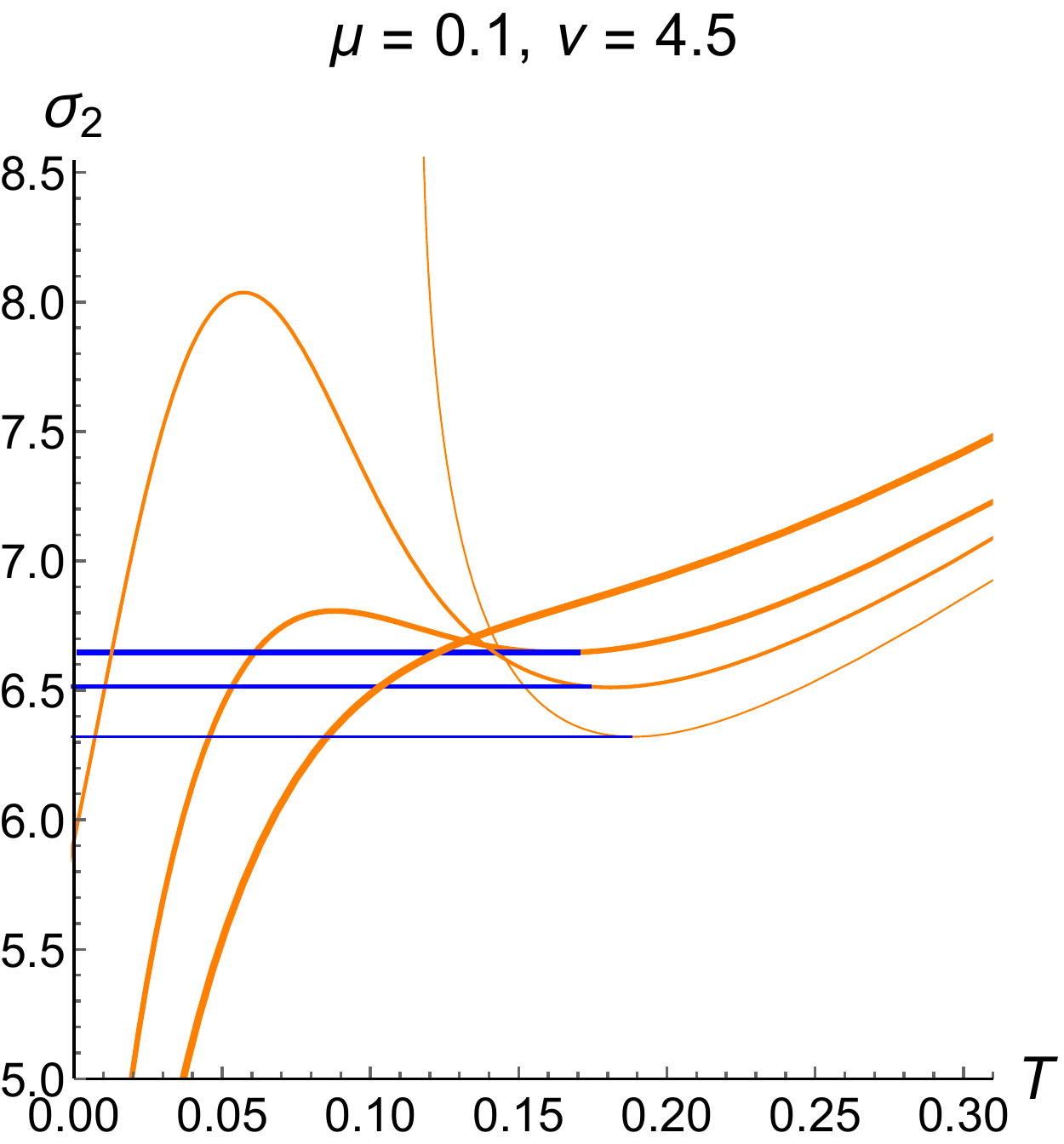}
  \includegraphics[scale=0.35]{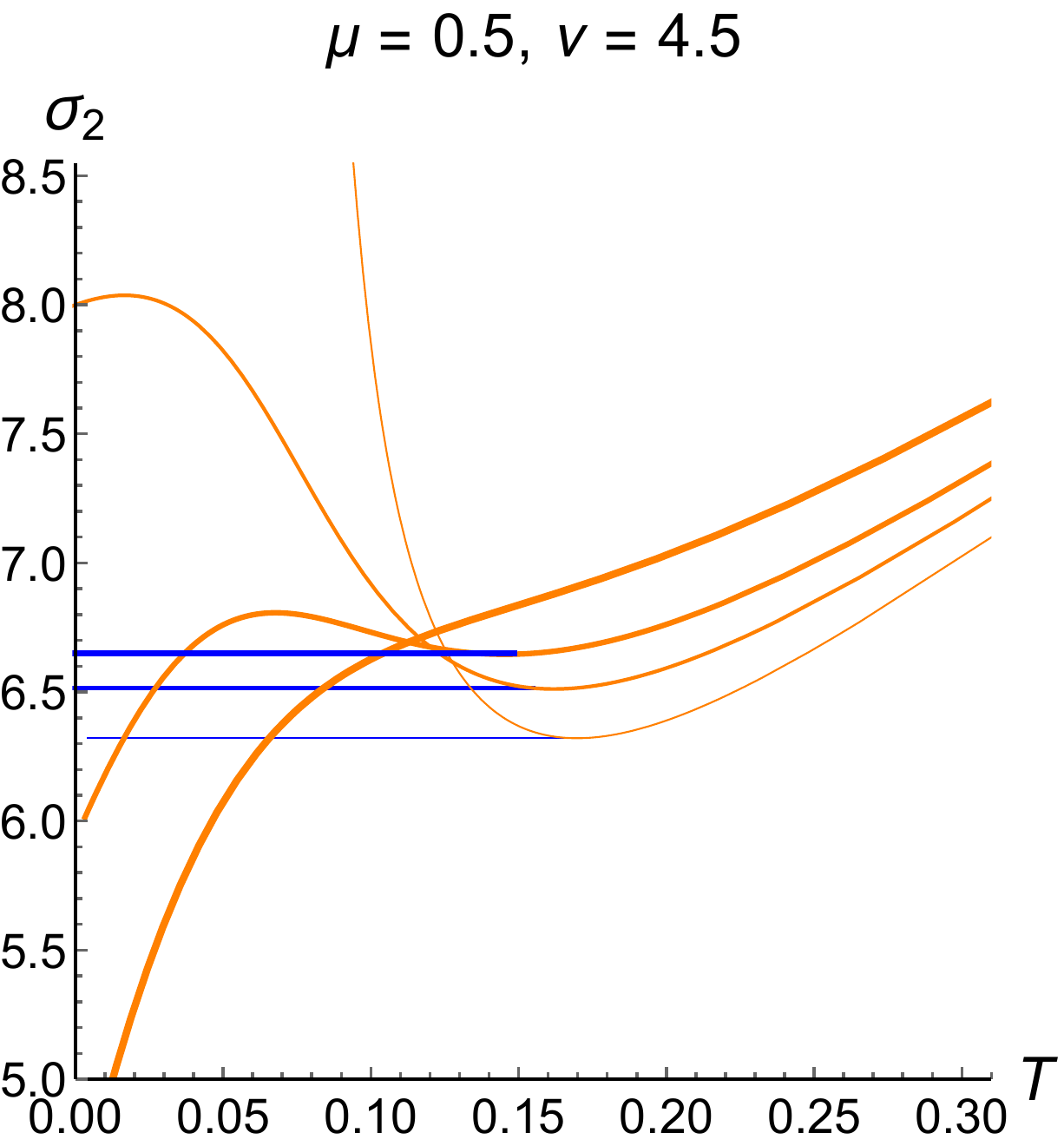}\\
  A \hspace{40 mm} B \hspace{40 mm} C \\ \ \\
  \includegraphics[scale=0.7]{Plots/sigma2-nu45-leg.pdf}
  \caption{Dependence of $\sigma_2$ from
    Fig.\ref{Fig:sigma-2-nu45small} in larger scale. Plot legends are
    the same for all plots.
  }
  \label{Fig:sigma-2-nu45large}
\end{figure}

In Fig.\ref{Fig:sigma-2small} and Fig.\ref{Fig:sigma-2large}
dependence of $\sigma_2$ on temperature is presented for different
values of chemical potentials $\mu$ and $c_B$ in primary isotropic
case. In Fig.\ref{Fig:SW-PT-1} more detailed pictures with indications of thermodynamically 
    unstable phases are shown.  In Fig.\ref{Fig:sigma-2-nu45small} and
Fig.\ref{Fig:sigma-2-nu45large} dependence of $\sigma_2$ on
temperature is presented for different values of chemical potentials
$\mu$ and $c_B$ in primary anisotropic case ($\nu = 4.5$). In both
cases we can see that there are two values of $\sigma$ for some
temperature values. This is the consequence of a multi-valued
dependence of temperature on the size of horizon and two-valued
dependence of sigma on the size of horizon for some sets of parameters
$c_B$, $\nu$ and $\mu$. In  Fig.\ref{Fig:SW-PT-1} the situation with
phase transitions is depicted in details for $\mu = 0$, $\mu = 0.01$,
$c_B = 0$ in both cases. The result for ${\cal V}_2$ is qualitatively
similar to results for ${\cal V}_1$. Dynamic wall positions are
presented in Tables~\ref{tab:V2nu1} and~\ref{tab:V2nu45}.

\begin{table}[h!]
\centering
\begin{tabular}{|l|c|l|l|l|}
\hline
\multicolumn{1}{|c|}{${\cal V}_2$} & \multicolumn{4}{c|}{$\nu=1$} \\ \hline
$-\,c_B$ & 0 & 0.05 & 0.1 & 0.15 \\ \hline
$z_{DW}$ & \multicolumn{1}{l|}{2.952} & 3.110 & 3.238 & 3.465 \\ \hline
\end{tabular}
\caption{Locations of DW for ${\cal V}_2$ at $\nu=1$}
\label{tab:V2nu1}
\end{table}

\begin{table}[h]
\centering
\begin{tabular}{|l|c|l|l|l|l|l|l|}
\hline
\multicolumn{1}{|c|}{${\cal V}_2$} & \multicolumn{7}{c|}{$\nu=4.5$} \\ \hline
$-\,c_B$ & 0 & 0.02 & \multicolumn{1}{c|}{0.05} & \multicolumn{1}{c|}{0.1} & \multicolumn{1}{c|}{0.15} & \multicolumn{1}{c|}{0.21} & \multicolumn{1}{c|}{0.265} \\ \hline
$z_{DW}$ & \multicolumn{1}{l|}{1.130} & 1.126 & 1.124 & 1.124 &1.142 & 1.178 & 1.192 \\ \hline
\end{tabular}
\caption{Locations of DW for ${\cal V}_2$ at $\nu=4.5$}
\label{tab:V2nu45}
\end{table}

$$\,$$
\newpage
$$\,$$
\newpage
$$\,$$

\newpage
$$\,$$
\begin{figure}[h!]
  \centering
  \includegraphics[scale=0.35]{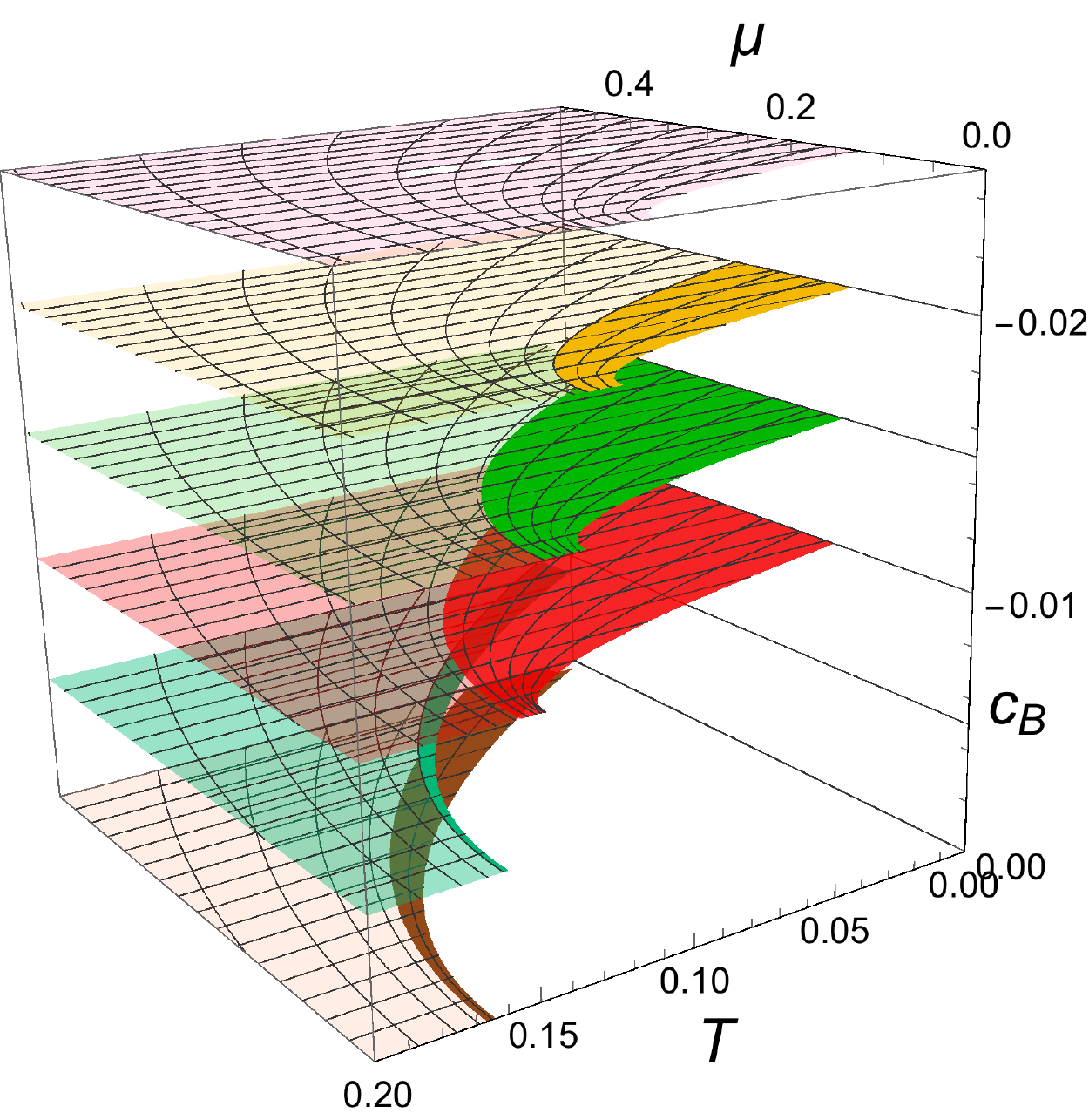}\qquad
  \includegraphics[scale=0.35]{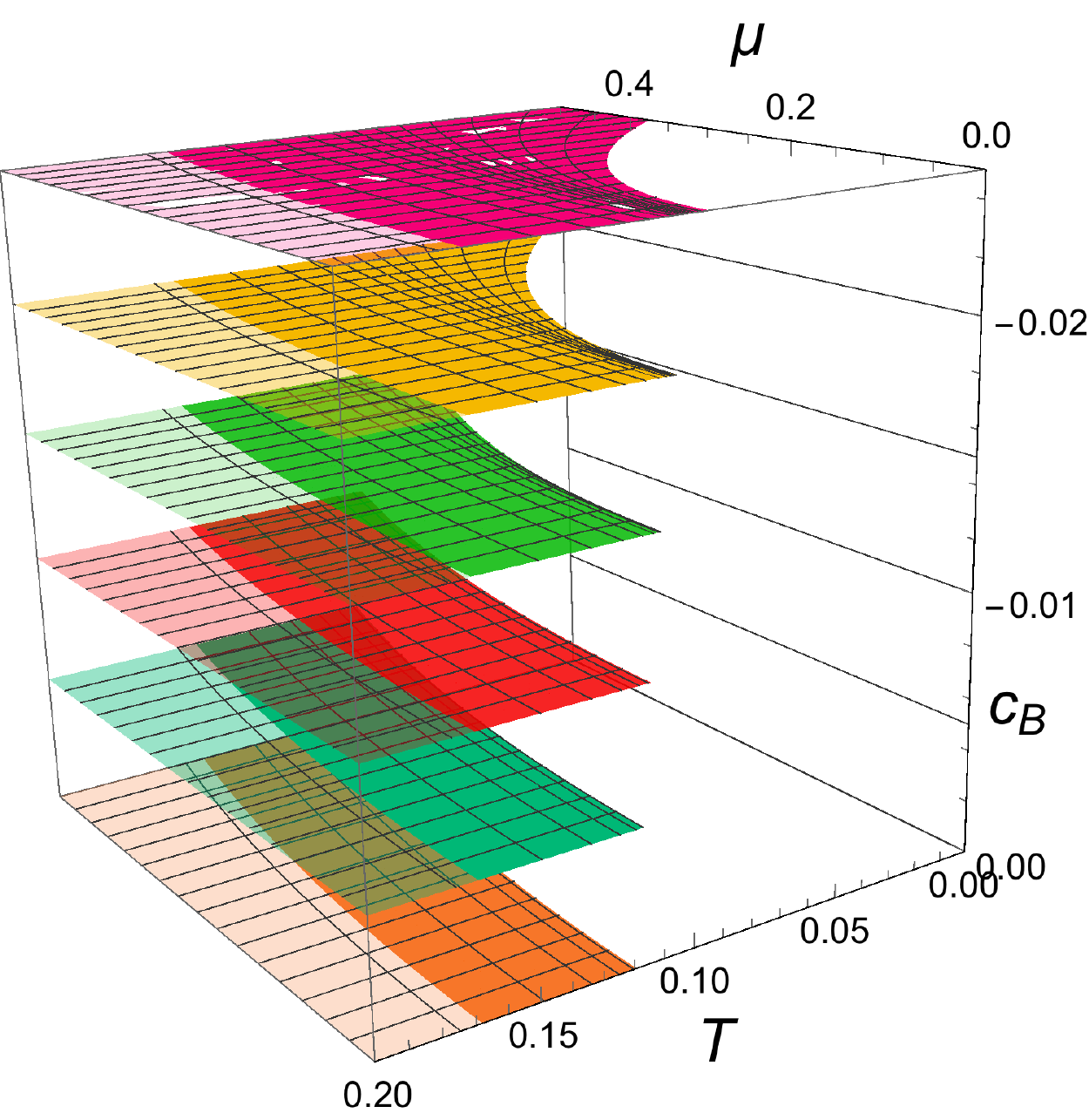}
  \includegraphics[scale=0.35]{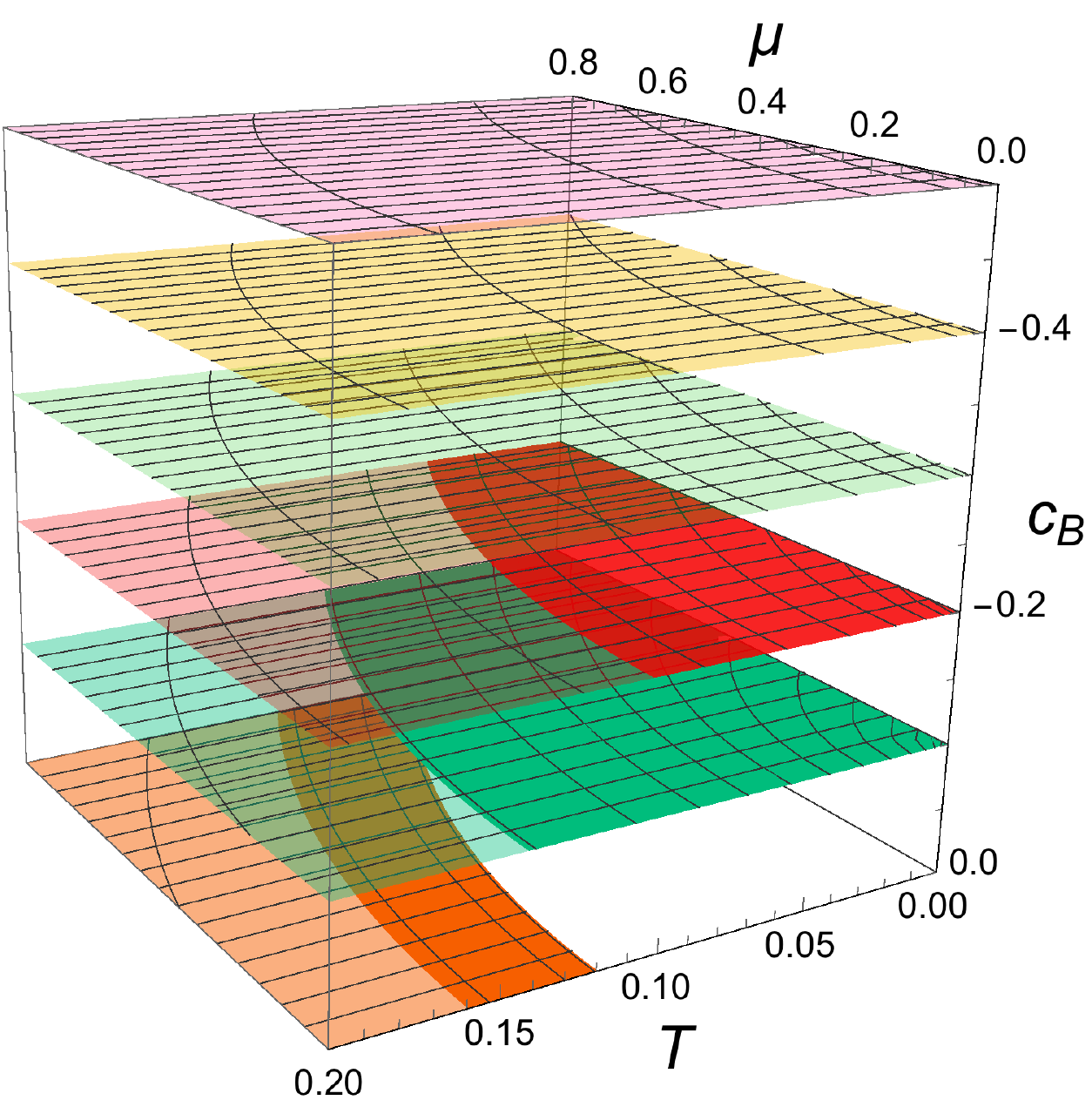}\\
  A \hspace{40 mm} B \hspace{40 mm}C
  \caption{Phase transitions for $\nu = 1$ (A) and $\nu = 4.5$ (B,C).
  }
  \label{Fig:SW-PT}
\end{figure}

\begin{figure}[h!]
  \centering
  \includegraphics[scale=0.45]{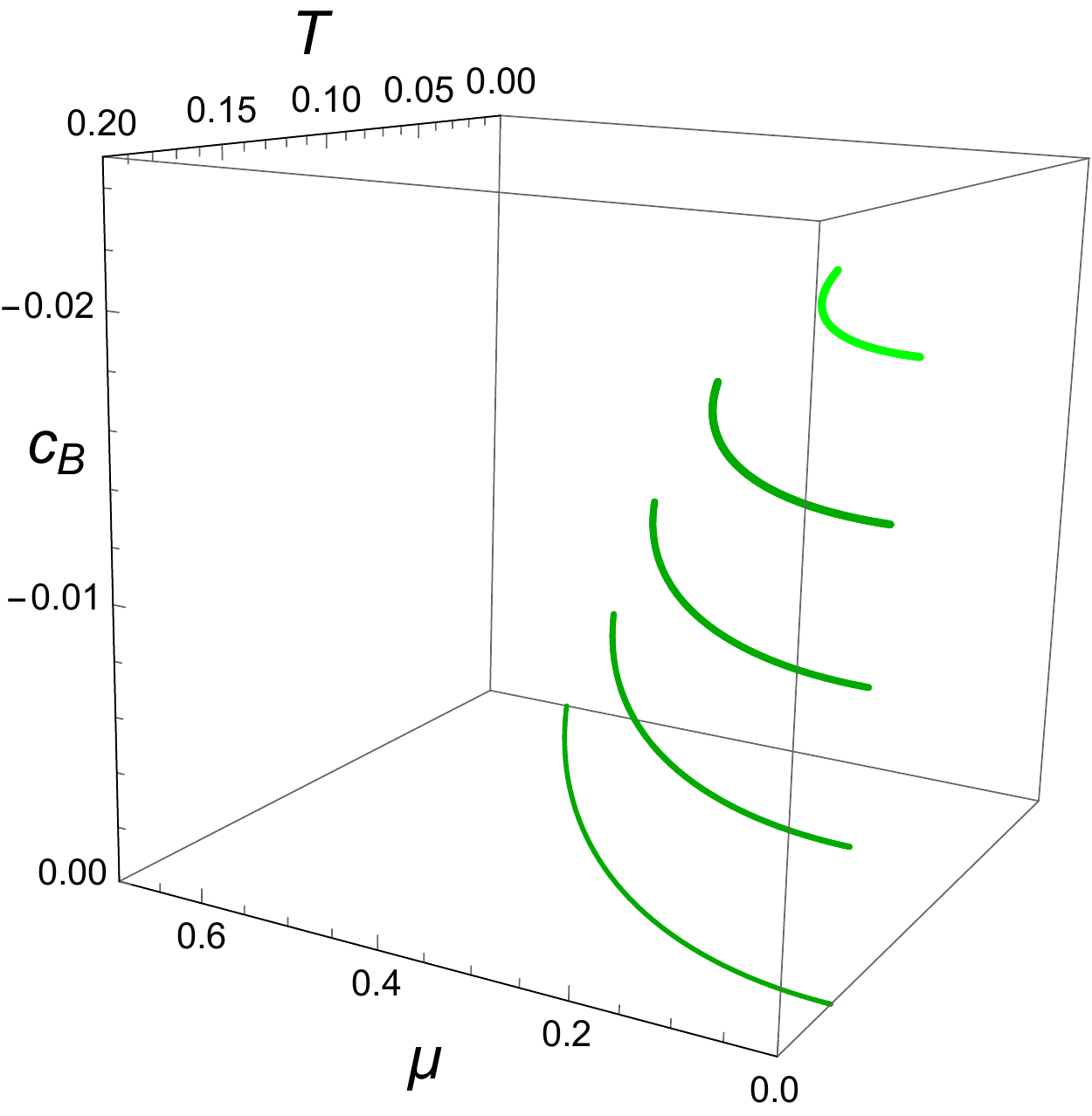}\qquad
  \includegraphics[scale=0.45]{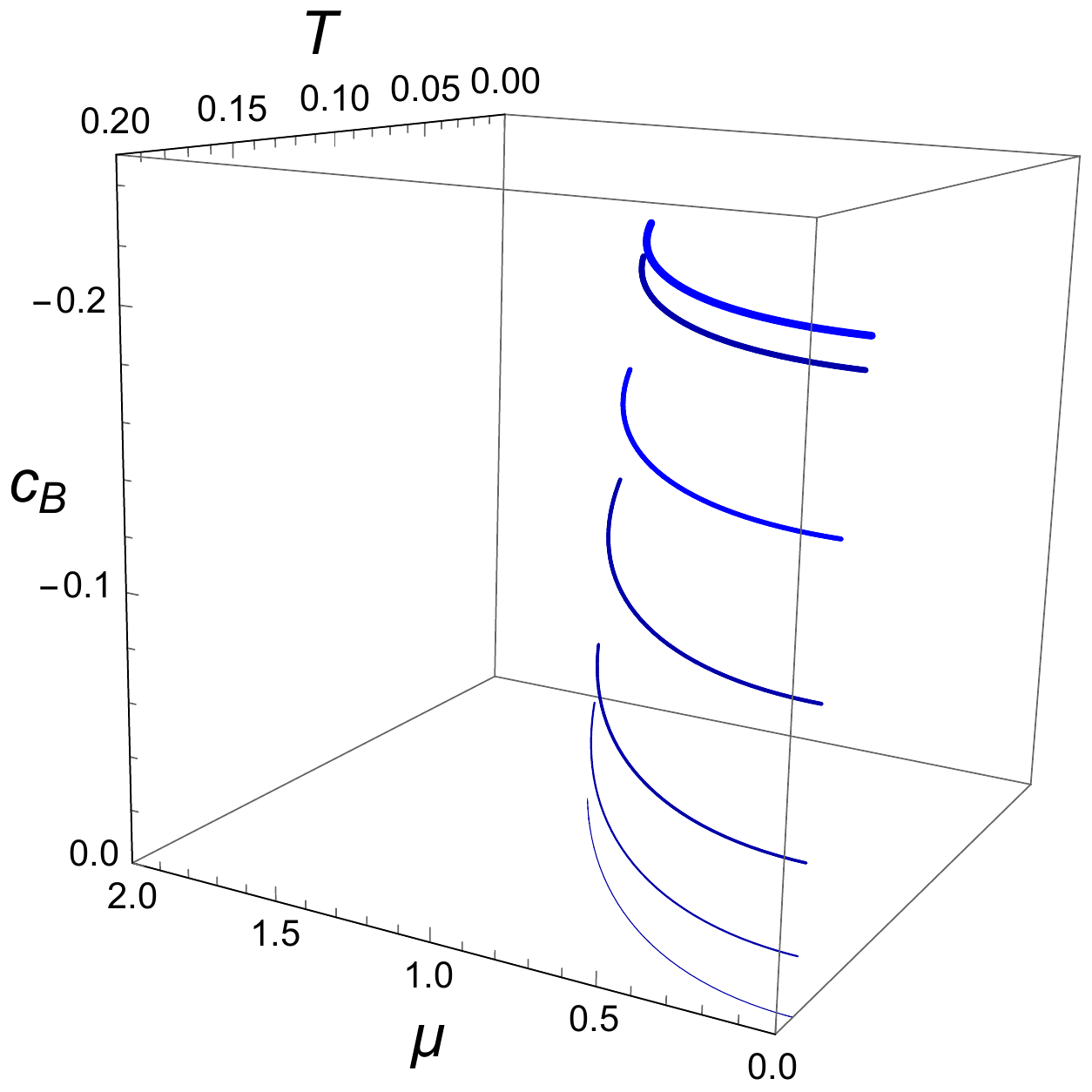}\\
  A \hspace{60 mm} B \hspace{40 mm}
  \caption{Phase transitions of ${\cal V}_2$ for $\nu = 1$ (A) and
    $\nu = 4.5$ (B). Light line represents the set of parameters where
    phase transitions disappear.
  }
\label{Fig:SW-PT-2}
\end{figure}

Finally, we get the surfaces that depict phase transition presented at
Fig.\ref{Fig:SW-PT} and Fig.\ref{Fig:SW-PT-2}. These results can be
compared with ${\cal V}_1$ consideration from previous subsection. 

\newpage

\subsection{Wilson loop $W_{y_1Y_2}$}

\begin{figure}[h!]
  \centering
  \includegraphics[scale=0.5]{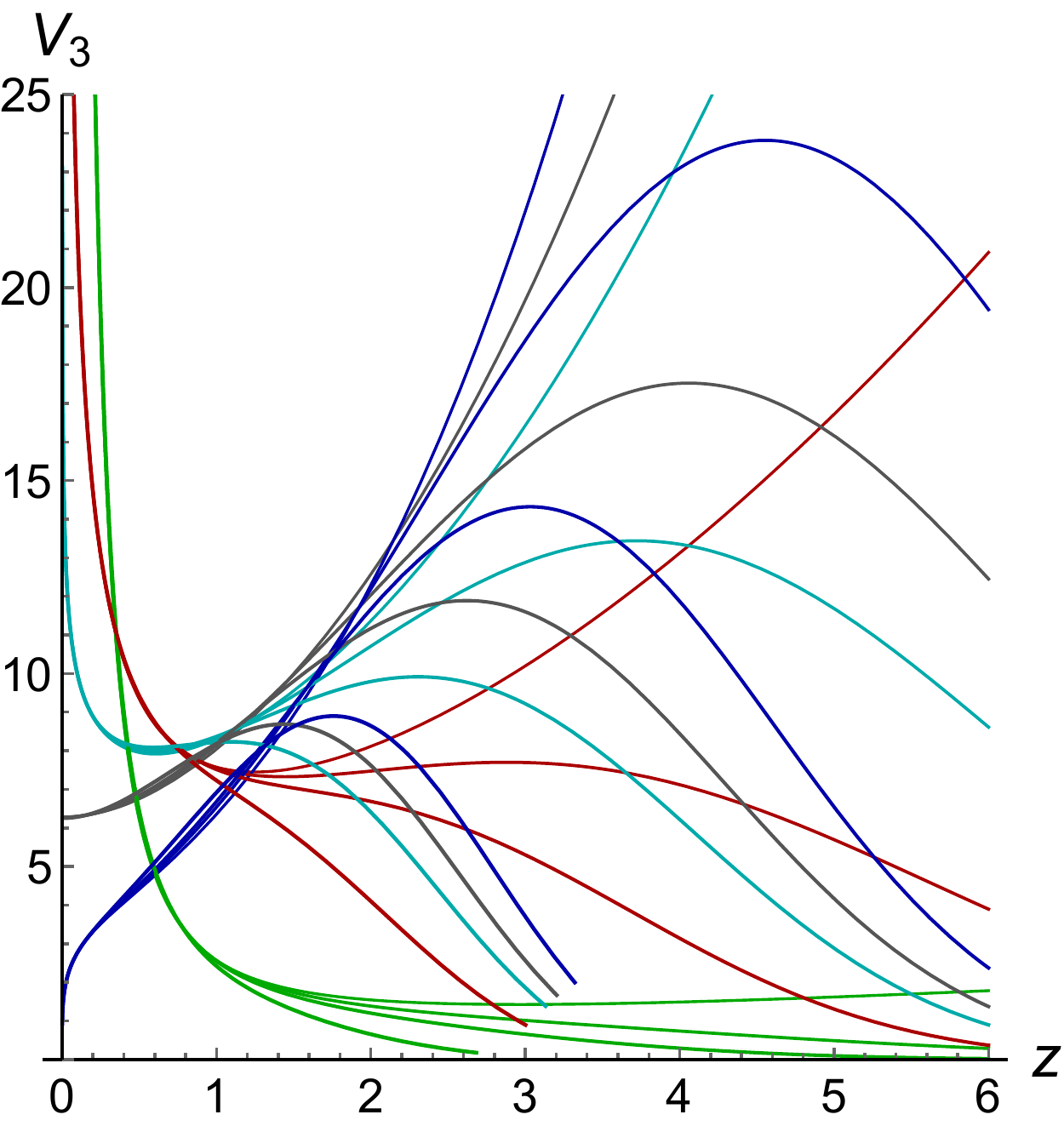}\\
  \caption{Effective potential ${\cal V}_3$ for $c_B = 0, 0.05, 0.1,
    0.2, 0.5$ (line's thickness increases with increasing $c_B$) for
    $\nu = 1$ (green), $1.5$ (red), $2$ (cyan), $2.48$ (gray), $4.5$
    (blue).  
  }
  \label{Fig:V3mix1}
\end{figure}

\begin{figure}[h!]
  \centering
  \includegraphics[scale=0.5]{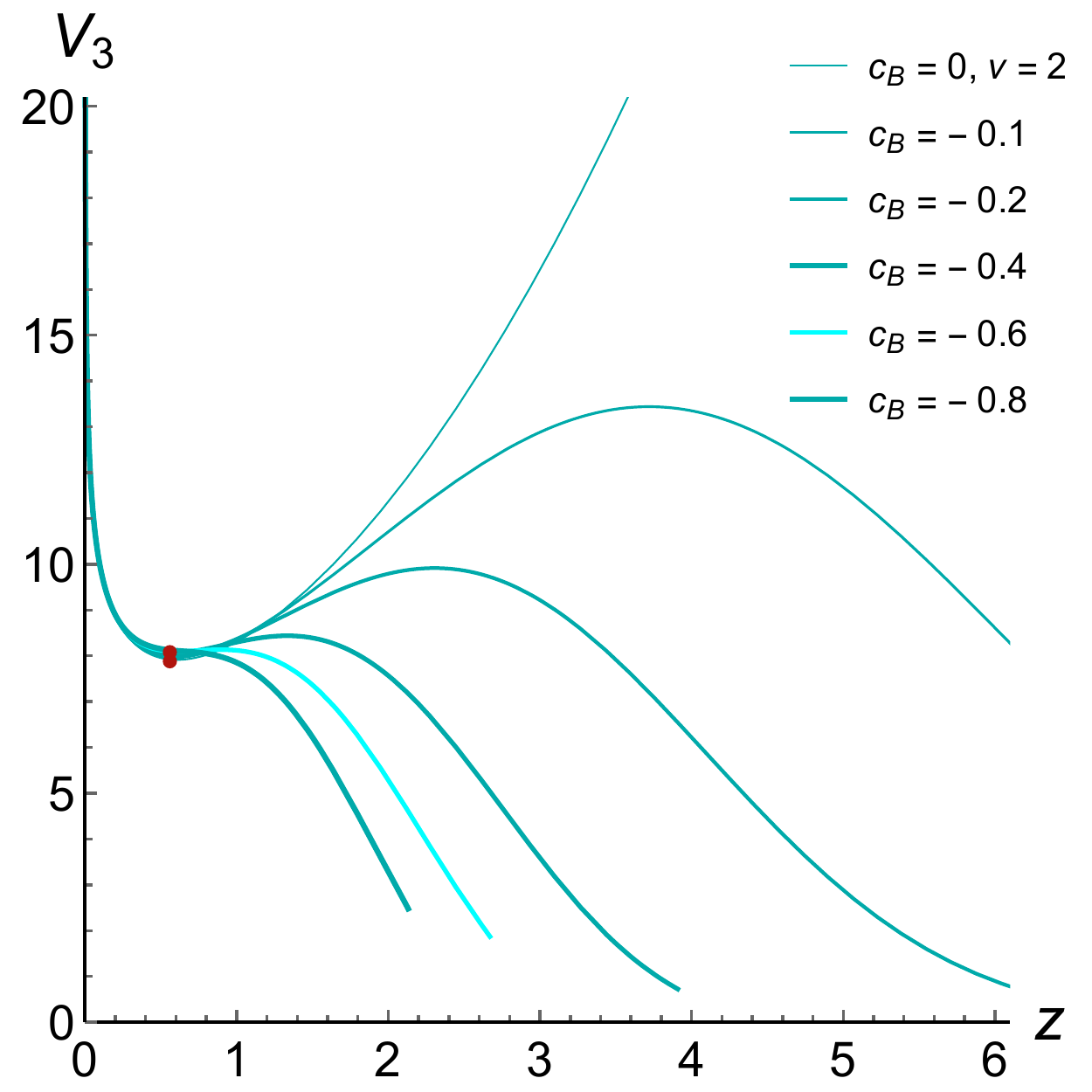}\qquad
  \includegraphics[scale=0.5]{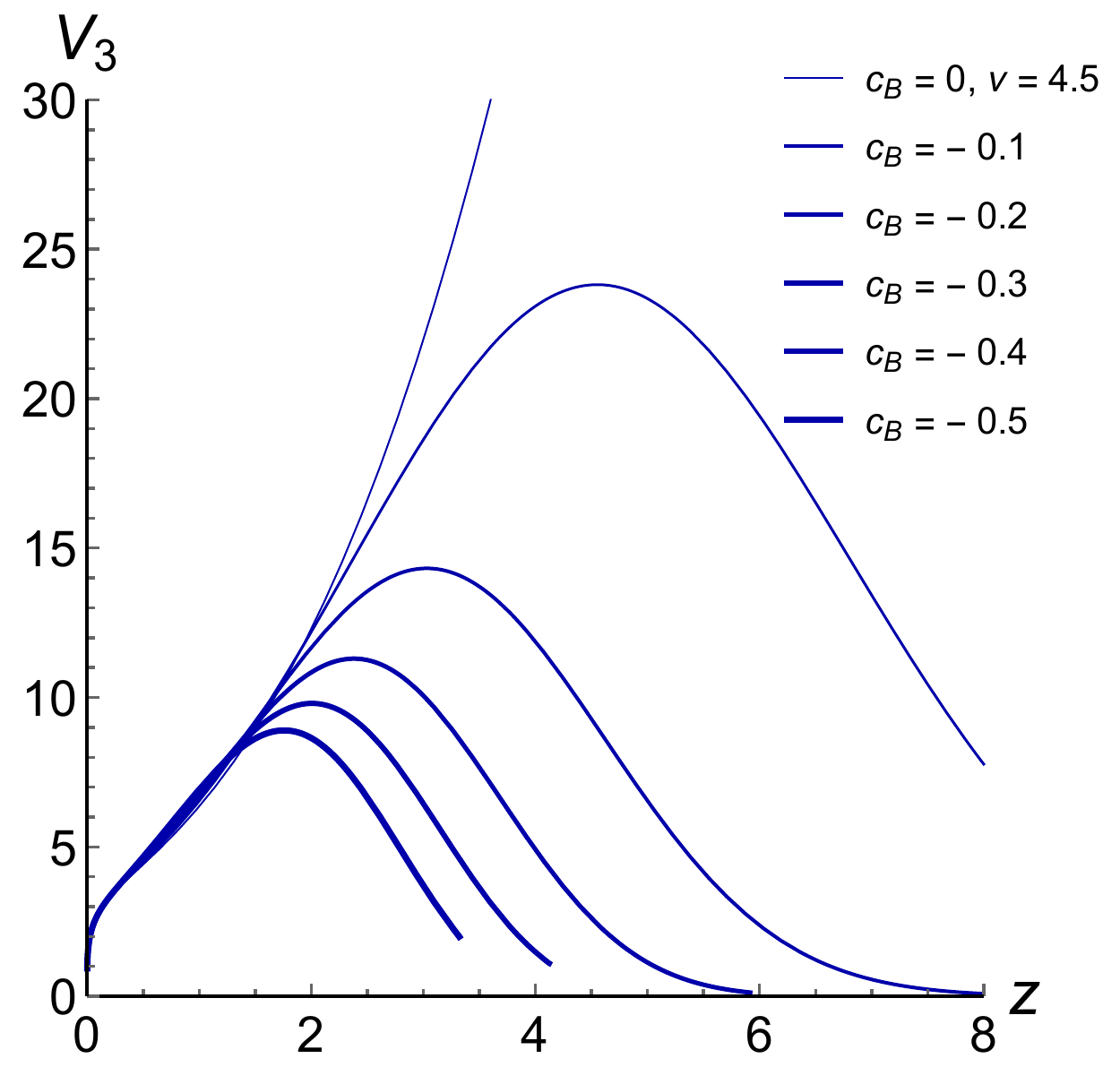}\\
  A \hspace{80 mm} B
  \caption{Effective potential ${\cal V}_3$ for different $c_B$
    values, $\nu = 2$ (A) and $\nu = 4.5$ (B).
  }
  \label{Fig:V3-2}
\end{figure}

In Fig.\ref{Fig:V3mix1} the effective potential dependence on $z$ is
presented. The form of the potential ${\cal V}_3$ depends on $c_B$ and
$\nu$. The DW disappears with increasing $c_B$. The
effective potential changes its behavior due to the dilation field
asymptotic on the boundary $z = \epsilon$, so that ${\cal V}(0) =
\infty$ for $\nu <\nu_{cr}$ and ${\cal V}(0) = 0$ for $\nu > \nu_{cr}$, $\nu_{cr}= 2.48$. This
fact is demonstrated also in Fig.\ref{Fig:V3-2}, where the effective
potential dependence on $z$ is presented for $\nu = 2$ and $\nu =
4.5$. This behavior of the effective potential means that for large
values of anisotropic parameter $\nu$ the connected string
configuration disappears in the considered background.

\begin{figure}[h!]
  \centering
  \includegraphics[scale=0.5]{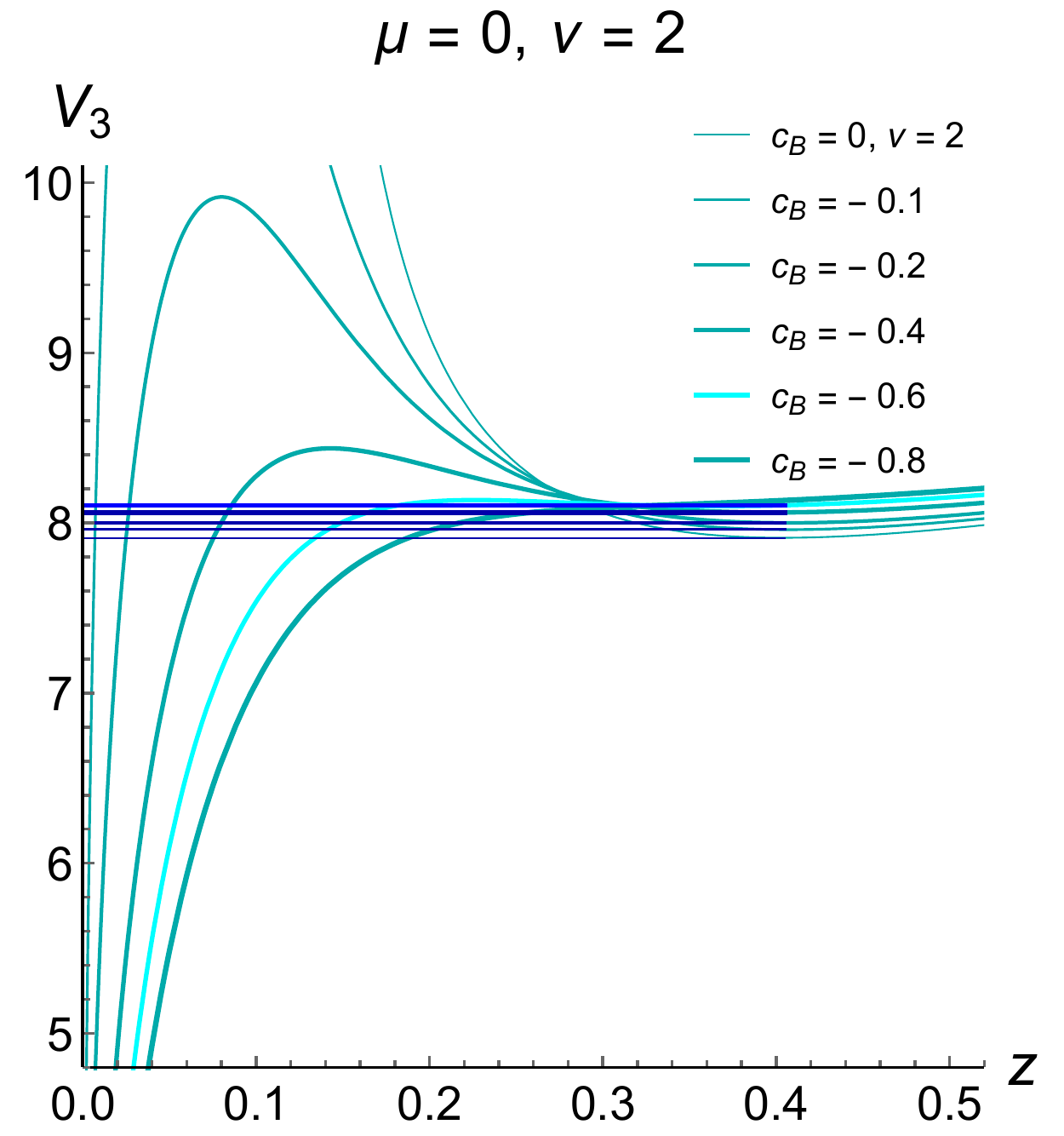}\\
  \caption{
    Dependence of $\sigma_3$ on temperature for different $c_B$
    values. Blue lines depict $\sigma_3$ calculated on the DW.}
  \label{Fig:V3}
\end{figure}

In Fig.\ref{Fig:V3}.A the effective potential ${\cal V}_3$ is
presented for different values of $c_B$ and $\nu = 2$. The DW exists
for all considered $c_B$ values. In Fig.\ref{Fig:V3}.B $\sigma_3$
dependence on temperature $T$ is presented. Blue lines depict the
$\sigma(z_{DW})$. Dynamical wall positions for ${\cal V}_3$ are listed in
Table~\ref{tab:V3nu1}.

\begin{figure}[h!]
  \centering
  \includegraphics[scale=0.5]{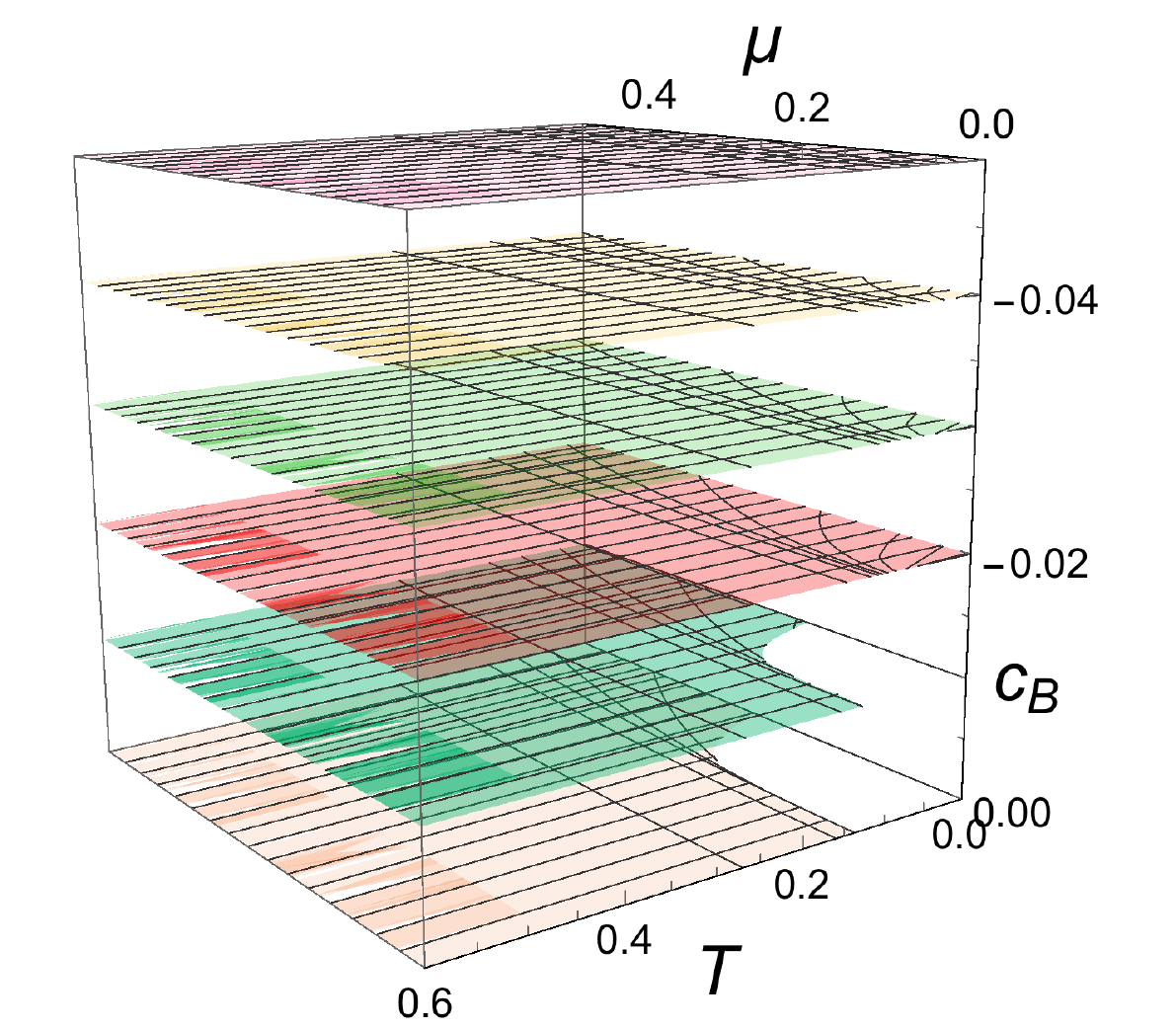}\qquad
  \includegraphics[scale=0.41]{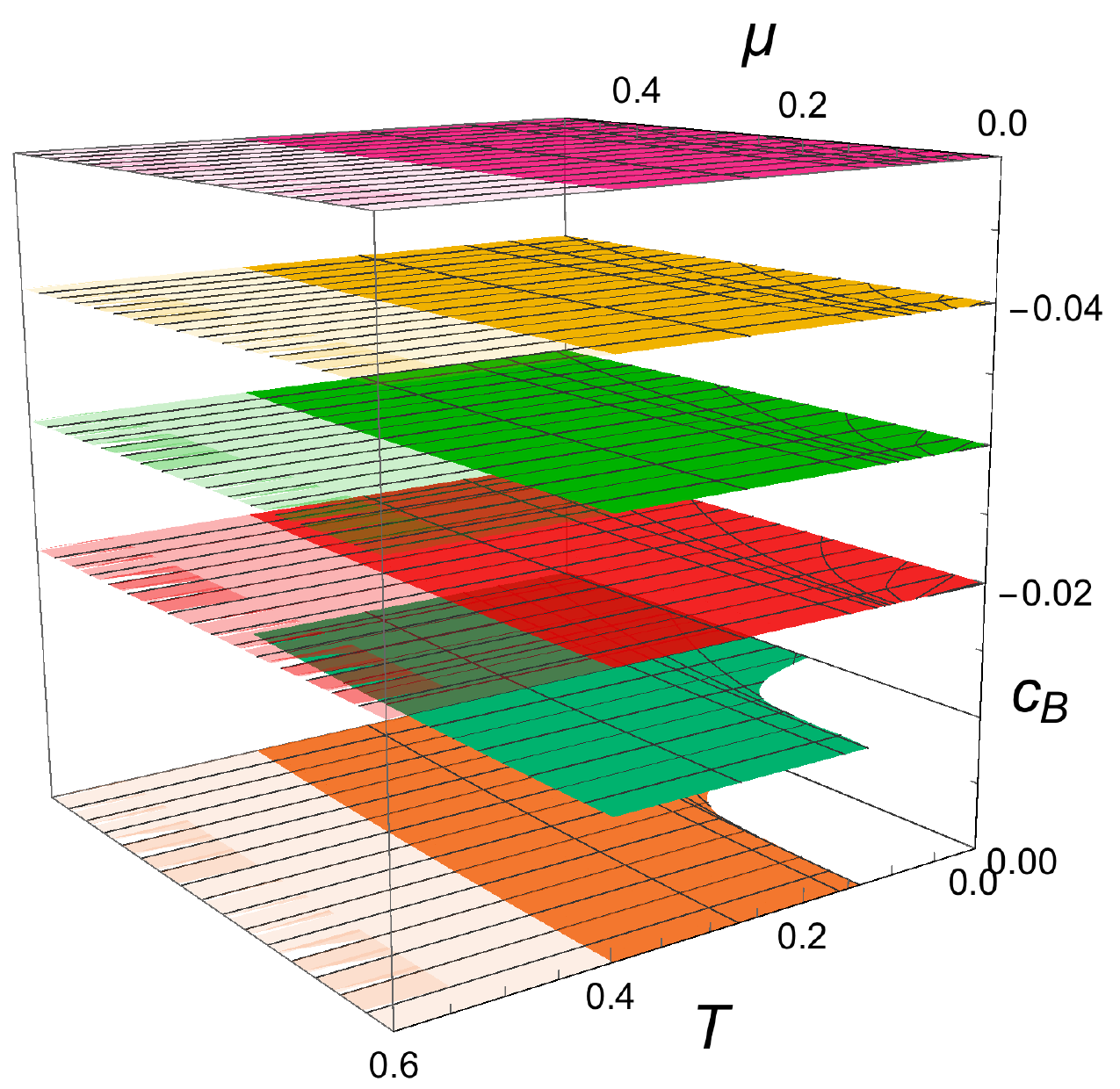}\\
  A \hspace{70 mm} B
  \caption{The domain of acceptable physical parameters $T$, $\mu$,
    $c_B$; $\nu = 2$ (A). Phase transition for $\sigma_3$ for different
    $c_B$ values; $\nu = 2$ (B). 
  }
  \label{Fig:V3mix2}
\end{figure}

We can also see the phase transition like it was in previous cases for
${\cal V}_1$ and ${\cal V}_2$. In Fig.\ref{Fig:V3mix2}.A the domain of
acceptable physical parameters $T$, $\mu$, $c_B$ at $\nu = 2$ is
depicted. In Fig.\ref{Fig:V3mix2}.B phase transition is presented for
$\nu = 2$ and different values of $c_B$. These results can be compared
with ${\cal V}_1$ and ${\cal V}_2$ cases with previous subsections. 

\begin{table}[h!]
\centering
\begin{tabular}{|l|c|l|l|l|}
\hline
\multicolumn{1}{|c|}{${\cal V}_3$} & \multicolumn{3}{c|}{$\nu=1$} \\ \hline
$-\,c_B$ & 0 & 0.01 & 0.02 \\ \hline
$z_{DW}$ & \multicolumn{1}{l|}{3.030} & 3.352 & 3.978 \\ \hline
\end{tabular}
\caption{Locations of DW for ${\cal V}_3$ at $\nu=1$}
\label{tab:V3nu1}
\end{table}

\newpage
$$\,$$
\newpage

\section{Conclusion}\label{Sect:conclusion}

In this work  expressions for the string tension $\sigma$  describing
differently oriented SWL in the fully anisotropic background are
obtained. Dependence of $\sigma$ for three particular orientations of
SWL, $\sigma_i$, $i = 1, 2, 3$, on the anisotropy and external
magnetic field parameters are studied. The considered orientations
correspond to drag forces acting on heavy quarks moving along
different directions.

\begin{figure}[b!]
  \centering
  \includegraphics[scale=0.45]{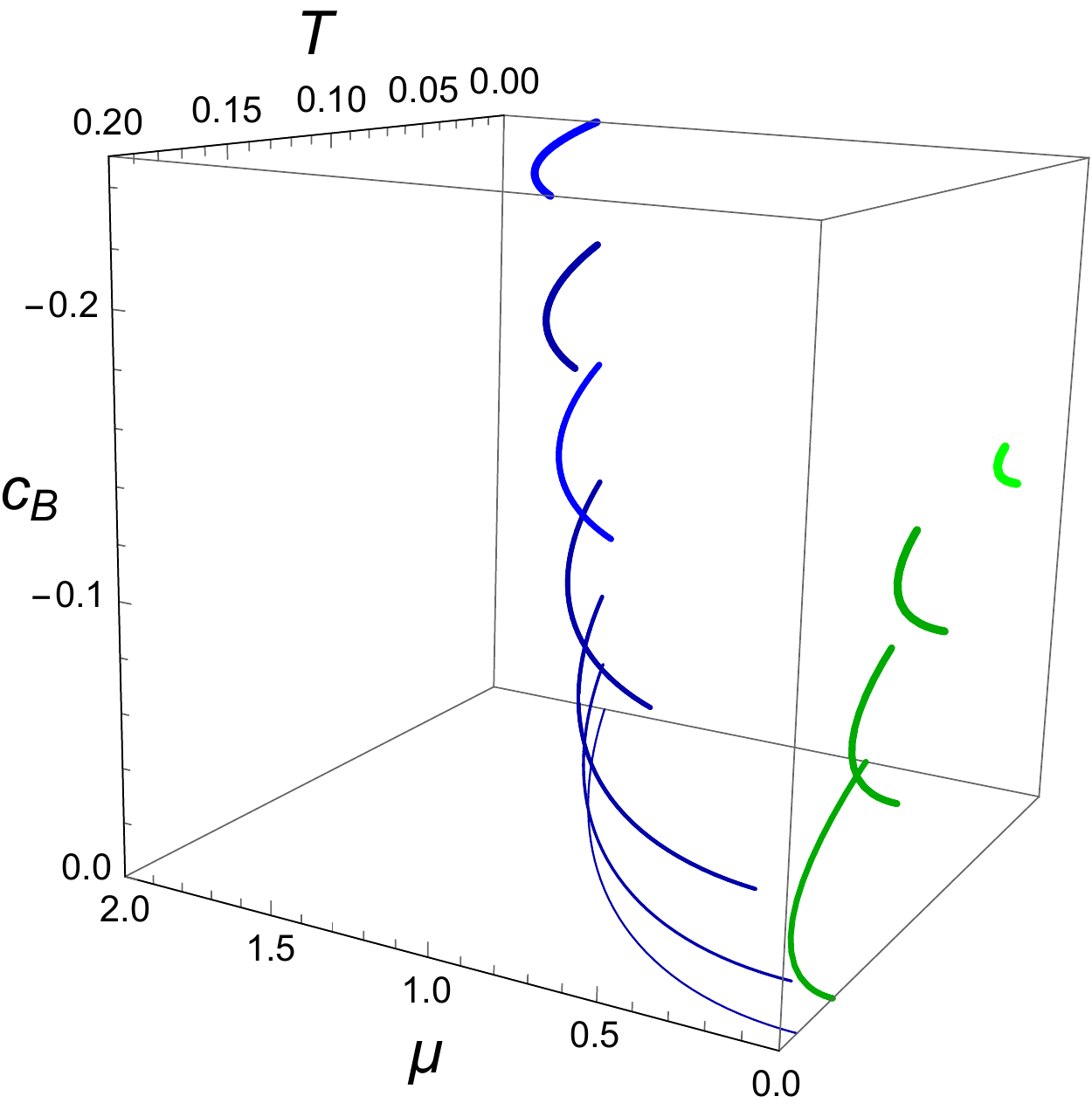}\qquad
  \includegraphics[scale=0.45]{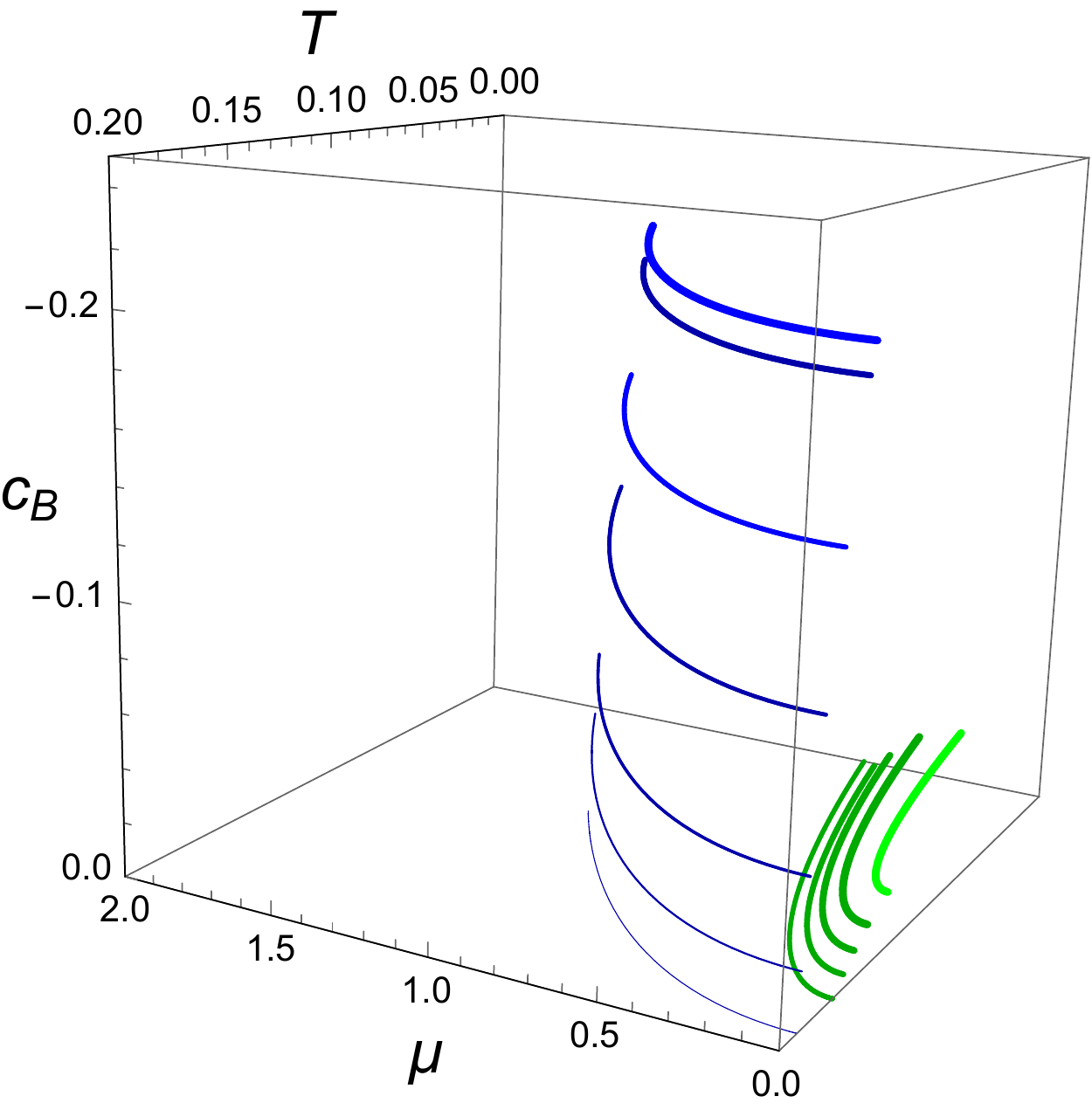}\\
  A \hspace{60 mm} B
  \caption{Phase transitions of ${\cal V}_1$ (A) and ${\cal V}_2$ (B) for $\nu = 1,
    4.5$.
  }
  \label{Fig:PTV2-nu1-45}
\end{figure}

Under variation of thermodynamic parameters -- temperature $T$,
chemical potential $\mu$ and magnetic field -- the string tensions
$\sigma_i$ undergo the phase transition. Their phase transitions are related
 to the appearance/disappearance of
dynamical walls for these effective potentials. Existence of the DW strongly depends on the parameters of
anisotropy. The disappearance of DW means that we should calculate
$\sigma$ at the horizon only. In primary isotropic case, i.e. when the
only source of anisotropy is the magnetic field, the DW exists for all
values of $z_h$, so we have $\sigma(z_{DW})$ and $\sigma(z_h)$ for all
$z_h$ and the configuration with the minimal effective potential is
realized. In fully anisotropic case the DW exists for $\nu < \nu_{cr}$ and $|c_B|<|c_{B,cr}|$. The DW can disappear at the point
$z_{DW}(z_h,\nu,c_B)$, reached with  $\nu =
\nu_{cr}$, $c_B = c_{B,cr}$. Values of $\nu_{cr}$ and $ c_{B,cr}$ depend on the orientation. With a slight increasing of
anisotropy near $\nu = \nu_{cr}$ and $c_B = c_{B,cr}$ the DW
disappears and the string tension is calculated at the horizon,
therefore the phase transition between two configurations disappears too.  Phase transition planes in the three dimensional space of physical parameters $T,\mu,c_B$ for $\sigma_1$ and $\sigma_2$ are presented in Fig.\ref{Fig:PTV2-nu1-45}
by green and blue lines for $\nu=1$ and $\nu=4.5$ respectively. The light green and blue lines represent to the end of the phase transitions.

It is interesting to consider SWLs  and drag forces in the fully
anisotropic case for different warp  factors  $b_s$, including factor
corresponding light quarks, as well as the hybrid factors
\cite{2009.05562}, and study the influence of the DW and anisotropy on
the drag forces. Also it would be interesting to study the same problem in other anisotropic models \cite{2010.04578,2011.09474}.

$$\,$$
\section{Acknowledgments}

This work is supported by Russian Science Foundation grant \textnumero
20-12-00200.

$$\,$$
\newpage

\appendix
\section*{Appendix}

\section{EOM solution}\label{appendixA}

The EOM system corresponding to action \eqref{eq:2.01} and metric
\eqref{eq:2.03} has the solution
\begin{gather}
  A_t = \mu \ \cfrac{e^{\frac{1}{4}(c-2c_B)z^2} -
    e^{\frac{1}{4}(c-2c_B)z_h^2}}{1 -
    e^{\frac{1}{4}(c-2c_B)z_h^2}} \, , \label{eq:3.06} \\
  \begin{split}
    g &= e^{c_B z^2} \left\{ 1
      - \cfrac{\Gamma\left(1 + \frac{1}{\nu} \ ; 0\right) -
        \Gamma\left(1 + \frac{1}{\nu} \ ; \frac{3}{4} (2 c_B - c)
          z^2\right)}{\Gamma\left(1 + \frac{1}{\nu} \ ; 0\right) - 
        \Gamma\left(1 + \frac{1}{\nu} \ ; \frac{3}{4} (2 c_B - c)
          z_h^2 \right)} \right. - \\ 
    &\qquad \qquad \quad - \cfrac{\mu^2 \ (2 c_B -
      c)^{-\frac{1}{\nu}}}{4 L^2 \left( 1 -
        e^{(c-2c_B)\frac{z_h^2}{4}} \right)^2} \
    \left(\Gamma\left(1 + \frac{1}{\nu} \ ; 0\right) -
      \Gamma\left(1 + \frac{1}{\nu} \ ; \frac{3}{4} \ (2 c_B - c) \
        z^2\right)\right) \times \\
    &\left. \times \left[ 1
        - \cfrac{\Gamma\left(1 + \frac{1}{\nu} \ ; 0\right) -
          \Gamma\left(1 + \frac{1}{\nu} \ ; \frac{3}{4} \ (2 c_B - c)
            z^2\right)}{\Gamma\left(1 + \frac{1}{\nu} \ ; 0\right) -  
          \Gamma\left(1 + \frac{1}{\nu} \ ; \frac{3}{4} \ (2 c_B - c)
            z_h^2 \right)} \ 
        \cfrac{\Gamma\left(1 + \frac{1}{\nu} \ ; 0\right) -
          \Gamma\left(1 + \frac{1}{\nu} \ ; (2 c_B - c)
            z_h^2\right)}{\Gamma\left(1 + \frac{1}{\nu} \ ; 0\right) -
          \Gamma\left(1 + \frac{1}{\nu} \ ; (2 c_B - c) z^2
          \right)}\right] \right\}, \label{eq:3.07}
  \end{split} \\
  f_B = - \ 2 \left( \cfrac{z}{L} \right)^{-\frac{2}{\nu}}
  e^{-\frac{1}{2}cz^2} \ \cfrac{c_B z}{q_B^2} \ g \left( \cfrac{3 c
      z}{2} + \cfrac{2}{\nu z} - c_B z - \cfrac{g'}{g}
  \right), \label{eq:3.08} \\
  f_2 = 4 \left( \cfrac{z}{L} \right)^{2-\frac{4}{\nu}}
  e^{-\frac{1}{2}(c-2c_B)z^2} \ \cfrac{\nu - 1}{q^2 \nu z} \ g \left(
    \cfrac{\nu + 1}{\nu z} + \cfrac{3 c - 2 c_B}{4} \ z -
    \cfrac{g'}{2g} \right), \label{eq:3.09} \\
  \begin{split}
    \phi &= \cfrac{\sqrt{2}}{4 \nu} \left\{
      2 \sqrt{2} \ \sqrt{\nu - 1} \left[ \log\left( \cfrac{z^2}{z_0^2}
        \right) - \log\biggl(
        8(\nu - 1) + 4 \nu c_B z^2 + 3 (3 c - 2 c_B)
        \nu^2 z^2 \right. + \right. \\
    &\qquad \ + \ 2 \sqrt{2} \ \sqrt{\nu - 1} \sqrt{8 (\nu - 1) + 8 c_B
      \nu z^2 + 6 (3 c - 2 c_B) \nu^2 z^2 + (3 c^2 - 4 c_B^2) \nu^2
      z^4} \biggr) \ + \\
    &+ \ \log\biggl( 8(\nu - 1) + 4 \nu c_B z_0^2 + 3 (3 c - 2 c_B)
    \nu^2 z_0^2 \ + \\
    &\qquad \ + \left. 2 \sqrt{2} \ \sqrt{\nu - 1} \ \sqrt{8 (\nu - 1) +
        8 c_B \nu z_0^2 + 6 (3 c - 2 c_B) \nu^2 z_0^2 + (3 c^2 - 4
        c_B^2) \nu^2 z_0^4} \biggr) \right] + \\
    &+ \ \sqrt{8 (\nu - 1) + 8 c_B \nu z^2 + 6 (3 c - 2 c_B)
      \nu^2 z^2 + (3 c^2 - 4 c_B^2) \nu^2 z^4} \ - \\
    &- \ \sqrt{8 (\nu - 1) + 8 c_B \nu z_0^2 + 6 (3 c - 2 c_B)
      \nu^2 z_0^2 + (3 c^2 - 4 c_B^2) \nu^2 z_0^4} \ + \\
    &+ \ \cfrac{4 c_B + 3 (3 c - 2 c_B) \nu}{\sqrt{3
        c^2 - 4 c_B^2}} \ \times \\
    \hspace{-20pt} \times \log &\left. \left( \text{\small{$\frac{4
              c_B + 3 (3 c - 2 c_B) \nu z^2 
              \sqrt{3 c^2 - 4 c_B^2} \ \sqrt{8 (\nu - 1) + 8 c_B \nu z^2 + 6
                (3 c - 2 c_B) \nu^2 z^2 + (3 c^2 - 4 c_B^2) \nu^2 z^4}}{4
              c_B + 3 (3 c - 2 c_B) \nu^2 z_0^2 + \sqrt{3 c^2 - 4 c_B^2}
              \sqrt{8 (\nu - 1) + 8 c_B \nu z_0^2 + 6 (3 c - 2 c_B) \nu^2
                z_0^2 + (3 c^2 - 4 c_B^2) \nu^2 z_0^4} } $}} \right) \!
    \right\}, 
  \end{split} \label{eq:3.10} 
\end{gather}
\begin{gather}
  \begin{split}
    V = - \ \cfrac{e^{\frac{1}{2}cz^2}}{4 L^2 \nu^2} \ &\biggl\{
    \bigl[ 
    8 (1 + 2 \nu) (1 + \nu) + 2 (3 + 2 \nu) (3 c - 2 c_B) \nu z^2 + (3
    c - 2 c_B)^2 \nu^2 z^4 \bigr] g \ - \\
    &\quad - \left[ 2 (4 + 5 \nu) + 3 (3 c - 2 c_B) \nu z^2 \right] g'
    + 2 g'' \nu^2 z^2 \biggr\}. \label{eq:3.11}
  \end{split}
\end{gather}
Here $z_0$ is the point specifying the boundary condition for the dilaton field,
$\phi(z_0)=0$ \cite{ARS-2019qfthep, 2009.05562}.

\end{document}